\documentclass[]{aa}
\usepackage{lmodern}
\usepackage{graphicx}
\usepackage{natbib}
\bibpunct{(}{)}{;}{a}{}{,} 
\usepackage{multicol}
\usepackage{multirow}
\usepackage{float}
\usepackage{amsmath}
\usepackage{adjustbox}
\usepackage{txfonts}
\usepackage{url}

\makeatletter
\renewcommand*\aa@pageof{, page \thepage{} of \pageref*{LastPage}}
\makeatother

\usepackage[pdfencoding=auto, psdextra,colorlinks=true,allcolors=blue]{hyperref}

\usepackage{bm}		

\begin{document}
 
\title{The imprint of magnetic fields on absorption spectra from circumgalactic wind-cloud systems}

\author{Benedetta Casavecchia$^{1,3,4}$, Wladimir E. Banda-Barrag\'an$^{2,3}$, Marcus Br\"uggen$^3$, Fabrizio Brighenti$^4$, and Evan~Scannapieco$^{5}$}

\institute{Max-Planck-Institut f\"ur Astrophysik, Karl-Schwarzschild-Strasse 1, D-85748 Garching b. M\"unchen, Germany
\\
              \email{benecasa@mpa-garching.mpg.de}
\and Escuela de Ciencias F\'isicas y Nanotecnolog\'ia, Universidad Yachay Tech, Hacienda San Jos\'e S/N, 100119 Urcuqu\'i, Ecuador
\and Hamburger Sternwarte, University of Hamburg, Gojenbergsweg 112, 21029 Hamburg, Germany
\and Dipartimento di Fisica e Astronomia, Universit\'{a} di Bologna, Via Gobetti 93/2, 40122, Bologna, Italy
\and School of Earth and Space Exploration, Arizona State University, Tempe AZ, USA}

\authorrunning{Casavecchia et al.}
\titlerunning{Imprints of magnetic fields on absorption spectra}

\date{Accepted ???. Received ???; in original form ???}

  \abstract
   {Galactic winds determine how stellar feedback regulates the mass and metallicity of galaxies. Observational studies show that galactic winds are multi-phase and magnetised. In the local Universe, the dense phase is traced by emission and absorption lines, which reveal the presence of fast-moving clouds embedded in hot streams. Simulations of such streams indicate to us that magnetic fields can shield such clouds and help to delay their disruption, but observational effects are rarely discussed.
   }
   {Using a suite of 3D magnetohydrodynamical simulations, we studied the influence of two orientations of the magnetic field (aligned and transverse to the wind) on the cloud morphology, temperature and density structure, mixing fraction, ion kinematics, column densities, and absorption spectra.}
   {We numerically studied supersonic wind-cloud systems with radiative processes, and developed a framework to extract ion column density maps and synthetic absorption spectra. The framework relies on studying ion populations and creating down-the-barrel spectra via an interface that links our PLUTO simulations to TRIDENT using the yt-package infrastructure, CLOUDY, and STARBURST99.}
   {We find that the transverse initial magnetic field makes the cloud asymmetric, shields and protects dense cold gas, and reduces mixing fractions compared to the aligned case. Ions can reach higher velocities in the transverse field case. The imprints of the initial orientation of the field on the synthetic spectra can be described as follow: a) in the cold phase, we find no signature of C\,{\sc ii} and Si\,{\sc ii} when the field is aligned; b) in the intermediate phase traced by C\,{\sc iv} and Si\,{\sc iv}, we find broader lines in the transverse case; and c) in the warm phase, we find deeper lines for O\,{\sc vi} and N\,{\sc v} in the aligned case, but they are less sensitive overall to the field orientation.}
   {Magnetic fields significantly affect the absorption spectra of cold clouds. Intermediate ions are the most sensitive to the magnetic field orientation and can potentially yield information about magnetic field topology.}

   \keywords{Galaxy: halo -- Magnetic fields -- Magnetohydrodynamics (MHD) -- quasars: absorption lines
               }
   \maketitle

\section{Introduction}
\label{intro}

Starburst galaxies are observed to host galactic winds in both the local Universe and at high redshifts (\citealt{2005ARA&A..43..769V}, \citealt{2017arXiv170109062H} \citealt{2020A&ARv..28....2V}, \citealt{2023ARA&A..61..131F}). The large-scale winds spread metals, mass, and energy throughout the surrounding circumgalactic medium (CGM), driven by the collective action of supernovae (SNe) (\citealt{2001ApJ...557..605S}, \citealt{2005ARA&A..43..769V}, \citealt{2020MNRAS.491.2855V}) and winds from OB stars  (\citealt{2012A&A...547A...3V}, \citealt{2021ApJ...921...91D}, \citeyear{2022ApJ...937...68D}). The feedback from galactic winds plays a fundamental role in the evolution of galaxies, as it strongly influences baryon exchange, star formation (SF), the metallicity distribution, and the configuration of large-scale magnetic fields (\citealt{2004ApJ...613..898T}, \citealt{2010MNRAS.406.2325O}, \citealt{2015ApJ...810L..14L},  \citealt{2021ApJ...914...24L}). \par

On a large scale, galactic winds usually present a bi-conical shape that develops from star-forming regions in the galaxy's bulge and disc. This structure has been observed, for instance, in M82 (\citealt{1998ApJ...493..129S}, \citealt{2013Natur.499..450B}, \citealt{2015ApJ...814...83L}) or our own Galaxy (\citealt{2003ApJ...582..246B,2010ApJ...724.1044S,2013Natur.493...66C}), and has also been reproduced through galactic-disc wind simulations that capture whole outflow lobes (\citealt{2008ApJ...674..157C}, \citealt{2018ApJ...862...56S}) or large-scale vertical slices (\citealt{2015MNRAS.454..238W}, \citealt{2018ApJ...853..173K}). These outflows display a multi-phase structure, observed both in nearby starburst galaxies (\citealt{2005ApJS..160..115R}, \citealt{2006A&A...448...43T}, \citealt{2017arXiv170109062H}) and in galaxies that have experienced intense star formation activity in the past, such as the Milky Way (MW) (\citealt{2013ApJ...770L...4M}, \citealt{2018ApJ...855...33D}, \citealt{2019Natur.573..235H,2023A&A...674L..15V}).\par

Spectroscopy plays a key role in understanding the complex dynamics and structure of multi-phase galactic winds. In nearby galaxies, outflow patterns have been observed in both emission and absorption (\citealt{2008A&A...487..583B,2017A&A...607A..48R}), but for distant systems, spectroscopy is typically limited to absorption lines arising from cold gas around galaxies (see \citealt{2017ARA&A..55..389T}, \citealt{2018MNRAS.474.1688C}, \citealt{2022ApJ...927..147T}). These are detected either down-the-barrel with the stellar continuum of the galaxy in the background (e.g. \citealt{2014ApJ...794..156R}) or transversely along the lines of sight of distant quasars (QSOs) (see \citealt{2015ApJ...804...79L}, \citealt{2022ApJ...936..171R}). In the CGM, the existence of a hot stream with $T \geq 10^6$ K, ionised gas with $T = 10^{4}-10^{6}$ K, and atomic clouds with $T \leq 10^4$ K, can be deduced from the combination of different ionisation states detected via down-the-barrel absorption-line spectroscopy, in the same velocity range along the line of sight (see \citealt{2018Galax...6..114Z} and \citealt{2020A&ARv..28....2V} for reviews). Gas with temperatures below $10^6$ K is commonly studied via absorption lines of H, C, Fe, N, Ne, Mg, O, S, and Si. As shown in Fig. 6 of \citealt{2017ARA&A..55..389T}, the different ionisation states of these tracers cover temperatures down to $10^4$ K. Especially, in the Local Group, some of the ions frequently observed are: O\,{\sc vi} and N\,{\sc v} in the warm phase; C\,{\sc iv}, Si\,{\sc iii}, and Si\,{\sc iv} in the intermediate-temperature phase; and C\,{\sc ii}, C\,{\sc iii}, Si\,{\sc ii}, and H\,{\sc i} in the cold phase (\citealt{2006ApJ...646..951K,2015ApJ...804...79L,2019ApJ...884...53F}).
In our Galaxy, absorption spectra of these ions reveal a complex multi-phase CGM (\citealt{2017A&A...607A..48R}), and neutral atomic clouds have also been detected in H\,{\sc i} emission, reaching distances from the Galactic plane of $\sim 1$ kpc and velocities of $ \sim 200-500\,\rm km\,s^{-1}$ (\citealt{2013ApJ...770L...4M}).\par


How neutral atomic and ionised gas clouds can be observed at such velocities is still an open question, and numerical simulations play a crucial role in helping to understand the observations. Two alternative models point to either cloud acceleration or dense-gas precipitation. The former suggests that fast-moving cold gas could be accelerated directly from the outflow through ram pressure (e.g. \citealt{2008ApJ...674..157C}, \citeyear{2009ApJ...703..330C}, \citealt{2012MNRAS.421.3522H}, \citealt{2016MNRAS.455.1309B}, \citeyear{2018MNRAS.473.3454B}), cosmic ray pressure (e.g. \citealt{2008ApJ...674..258E,2020ApJ...905...19B}), or radiation pressure (e.g. \citealt{2011ApJ...735...66M}, \citealt{2012MNRAS.424.1170Z}). The latter suggests that atomic and molecular gas clouds could be generated by the precipitation of the hot medium ejected by stellar wind (e.g. \citealt{2016MNRAS.455.1830T}, \citealt{2018MNRAS.480L.111G}, \citealt{2020MNRAS.499.2173B}, \citeyear{2021MNRAS.506.5658B}). In this context, many wind-cloud simulations have investigated the effects of magnetic fields (e.g. \citealt{2016MNRAS.455.1309B}, \citeyear{2018MNRAS.473.3454B}, \citealt{2020ApJ...892...59C}, \citealt{2020MNRAS.499.4261S}), radiative cooling (\citealt{2009ApJ...703..330C}, \citealt{2015ApJ...805..158S}, \citealt{2019MNRAS.482.5401S}), and thermal conduction on the survival of cold clouds in the hot flow (e.g. \citealt{2016ApJ...822...31B}). Observational evidence for magnetic fields in the CGM of galaxies come for example, from Faraday rotation, for example (\citealt{2023A&A...670L..23H}). These radiative, magnetised, and thermally conductive models show that it is possible to extend the cloud lifetimes significantly, but not enough to accelerate dense gas to the observed speeds. Thus, re-condensation is currently believed to play a pivotal role in explaining the survival of cold gas in hot streams (\citealt{2020MNRAS.492.1970G,2021MNRAS.501.1143K,2021MNRAS.506.5658B}).\par

One of the most challenging aspects of distinguishing between these different possibilities is comparing the results of simulations of galactic winds with observational diagnostics, such as gas column densities, the shape and optical depth of absorption spectral lines, and their covering fractions. Producing synthetic observables from simulations of dense clouds interacting with galactic winds is essential for this purpose. Idealised, wind-cloud simulations can achieve very high resolutions and capture small-scale physics and the dynamics of ions in great detail. Previous studies have constructed ion distributions, taking into account the effects of radiative cooling, thermal conduction, the UV background from the starburst galaxy, and the impact of turbulence on the density distribution of the cloud (see \citealt{2018ApJ...864...96C,2021ApJ...919..112D}). Since magnetic fields have also been shown to alter the dynamics of wind-cloud systems (see \citealt{2015MNRAS.449....2M}, \citealt{2016MNRAS.455.1309B}, \citeyear{2018MNRAS.473.3454B}), they should also influence the ion column densities and the shape of the absorption lines produced by cold gas clouds during their evolution. The effects of magnetic fields on such observables have been investigated for C\,{\sc iv} and Mg\,{\sc ii} in \citealt{2024MNRAS.527..991D}, but additional detailed discussions are still missing in the literature.\par

In this paper, we present synthetic column density maps and absorption spectra from high-resolution wind-cloud simulations, which include the effects of radiative processes and uniform magnetic fields with different orientations. Using a framework developed by our group that links PLUTO simulations to the TRIDENT analysis code
(see \citealt{2023IAUS..362...56C}), we produce synthetic absorption lines of C\,{\sc ii} and Si\,{\sc ii} (which trace the cooler phase of material with $T = 10^4-10^{4.5}$ K), C\,{\sc iv}, Si\,{\sc iv}, (which trace intermediate-temperature material with $T = 10^{4.5}-10^{5}$), and N\,{\sc v} and O\,{\sc vi} (which trace the warmer phases with $T = 10^{5}-10^{5.5}$). Our spectra are produced by including a UV background that accounts for the star formation rate of the starburst system. The specific questions we aim to answer in this paper include: what ions we can detect in the different evolutionary stages of clouds interacting with a supersonic galactic wind; what the average abundances and typical column densities of individual ions are in such interactions; how the orientation of the underlying magnetic field affects the ion column densities and the shape of their spectra. \par

The structure of this paper is as follows. Section \ref{Methods} describes our wind-cloud simulations, and provides an overview of the analysis tools and the interface for creating synthetic spectra. In Section \ref{Results}, we describe the results of our study, including the characteristic evolution of atomic wind-swept clouds, the column densities and kinematics of different ions, and their overall spectral profiles. In Section \ref{Discussion} we comment on the time evolution of the spectra and we compare our results with the literature. Finally, Section \ref{Conclusions} summarises our conclusions.\par

\section{Methods}
\label{Methods}

\subsection{Simulation code}
\label{subsec:SimulationCode}
\begin{figure*}
\begin{center}
  \begin{tabular}{c}
    \multicolumn{1}{l}{1a. R32-AL}\vspace{-0.1cm}\\
    \includegraphics[width=0.85\textwidth]{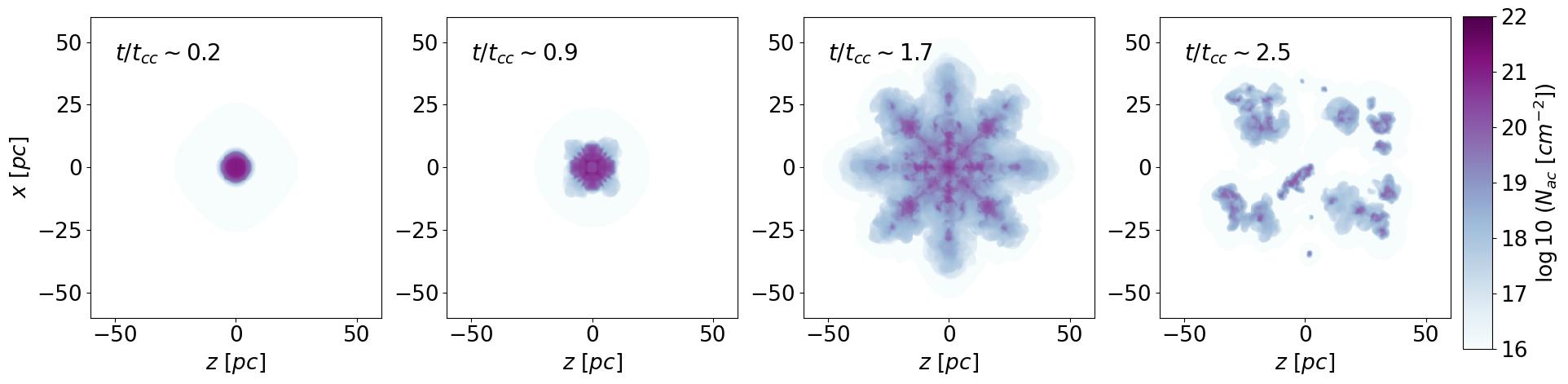}\vspace{-0.1cm}\\
    \includegraphics[width=0.85\textwidth]{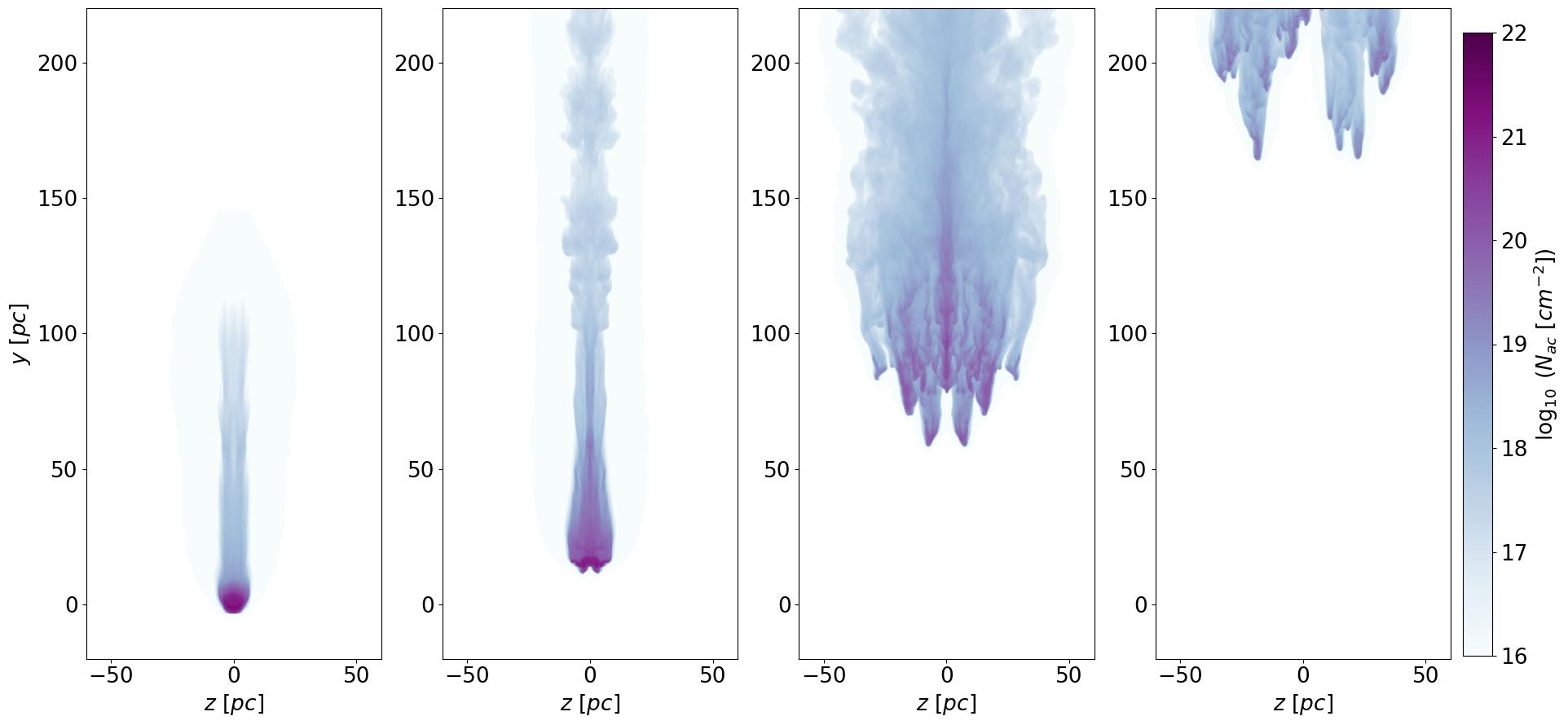}\\
    \multicolumn{1}{l}{1b. R32-TR}\vspace{-0.1cm}\\
    \includegraphics[width=0.85\textwidth]{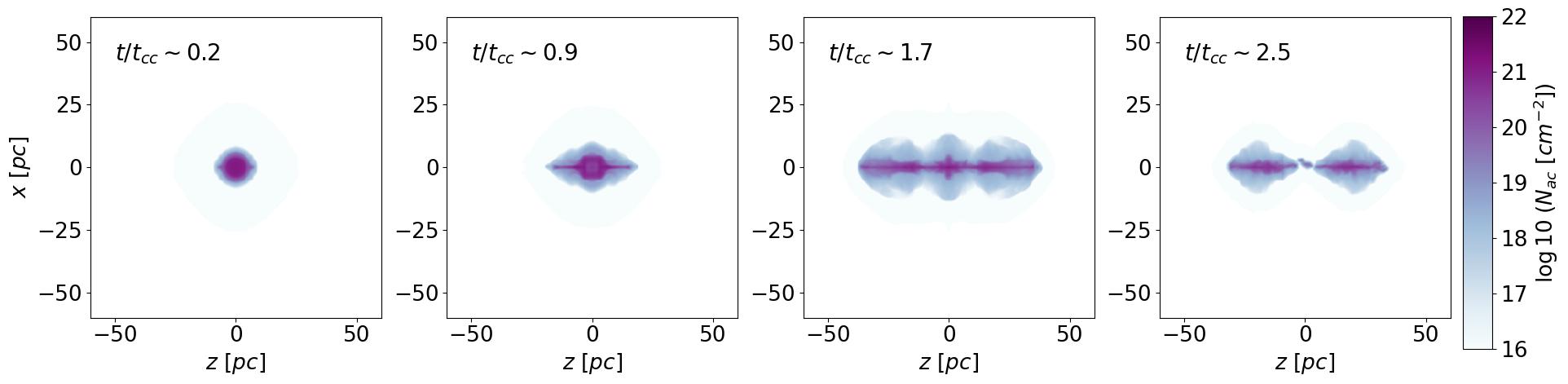}\vspace{-0.1cm}\\
    \includegraphics[width=0.85\textwidth]{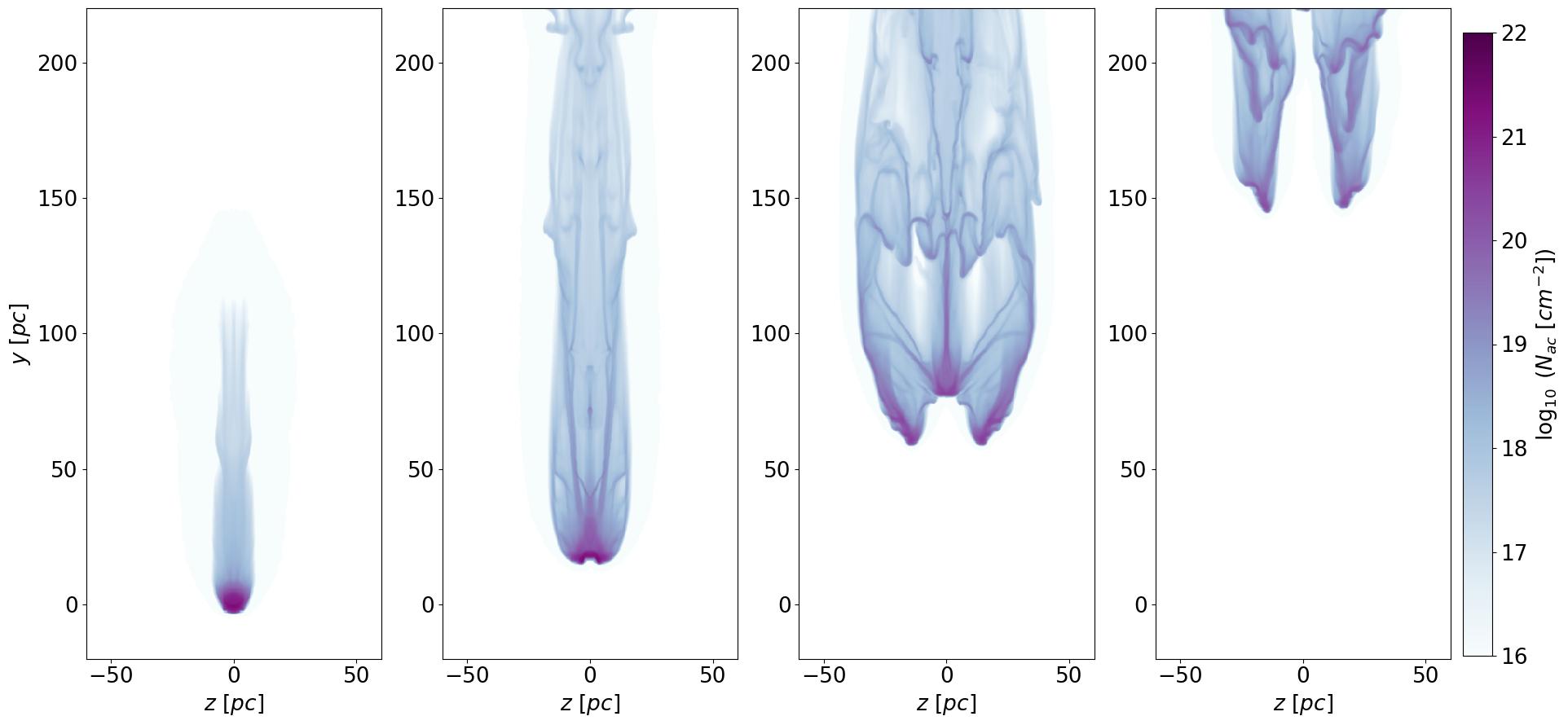}\vspace{-0.12cm}\\
  \end{tabular}
  \caption{2D maps showing the column densities of cloud gas. Panel 1a shows the down-the-barrel (top) and transverse (bottom) projections for model R32-AL, and panel 1b shows the same for model R32-TR. The morphological differences of the cloud and filaments, associated with the different initial field configurations, in both models is evident. From left to right, the plots display column densities at $t = 0.2, 1.2, 2.2$ and $3.2$ Myr (or $t/t_{\rm cc} = 0.2, 0.9, 1.7, 2.5$) respectively. The morphological evolution of the cloud is highly affected by the orientation of magnetic field. Especially, in R32-TR the draping effect caused by the transverse magnetic field prevents the clouds from expanding symmetrically in the $xz$ plane. A carbuncle effect can also be noticed in the models, so for our spectral analysis we pick sight lines at different radii from the centre.} 
\label{Figure1}
\end{center}
\end{figure*}

For the simulations reported in this paper (see Fig. \ref{Figure1}) we use a customised version of the PLUTOv4.3 code (see \citealt{2007ApJS..170..228M}) to solve the equations for mass, momentum, energy conservation, i.e.,
\begin{equation}
\frac{\partial \rho}{\partial t}+\bm{\nabla\cdot}\left[{\rho \bm{v}}\right]=0,
\label{eq:MassConservation}
\end{equation}

\begin{equation}
\frac{\partial \left[\rho \bm{v}\right]}{\partial t}+\bm{\nabla\cdot}\left[{\rho\bm{v}\bm{v}}-{\bm{B}\bm{B}}+{\bm{I}}P\right]=0,
\label{eq:MomentumConservation}
\end{equation}

\begin{equation}
\frac{\partial E}{\partial t}+\bm{\nabla\cdot}\left[\left(E+P\right)\bm{v}-\bm{B}\left(\bm{v}\bm{\cdot B}\right)\right]=\Gamma-\Lambda,
\label{eq:EnergyConservation}
\end{equation}

\noindent jointly with the magnetic induction equation,
\begin{equation}
\frac{\partial \bm{B}}{\partial t}-\bm{\nabla\times}\left(\bm{v}\bm{\times B}\right)=0,
\label{eq:induction}
\end{equation}

\noindent the solenoidal condition,
\begin{equation}
\bm{\nabla\cdot B}=0,
\label{eq:solenoidal}
\end{equation}

\noindent an additional advection equation of the form,
\begin{equation}
\frac{\partial\left[\rho C\right]}{\partial t}+\bm{\nabla\cdot}\left[{\rho C \bm{v}}\right]=0,
\label{eq:tracer}
\end{equation}

\noindent and an ideal equation of state
\begin{equation}
P_{\rm {th}}=P_{\rm th}(\rho,\epsilon)=\left(\gamma-1\right)\rho\epsilon.
\label{eq:EOS}
\end{equation}

\noindent In the equations above $\rho$ is the mass density, $\bm{v}$ is the velocity, $\bm{B}$ is the magnetic field\footnote{The factor $1/\sqrt{4\pi}$ is included into the definition of magnetic field.}, $P=P_{\rm th}+P_{\rm mag}$ is the total pressure (i.e., thermal plus magnetic: $P_{\rm mag}=\frac{1}{2}|\bm{B}|^2$), $E=\rho\epsilon+\frac{1}{2}\rho\bm{v^2}+\frac{1}{2}|\bm{B}|^2$ is the total energy density, $\epsilon$ is the specific internal energy, $\gamma=5/3$ is the ratio of the specific heat capacities at constant pressure and volume, and $C$ is a Lagrangian scalar used to track the evolution of gas initially contained in the cloud ($C=1$ in the cloud, and $C=0$ elsewhere, i.e. in the wind).\par

\subsection{Radiative processes}
\label{subsec:cooling}

Our simulations include a temperature- and density-dependent, customised cooling and heating function. The cooling rates were calculated using CLOUDY (see \citealt{1998PASP..110..761F}) for a solar mix at redshift zero. Close-to-solar metallicities have been inferred in observational studies of the MW halo (\citealt{2001ApJ...549..281R}). Our tabulated cooling function, $\Lambda$, includes cooling rates from atomic species in the temperature range relevant for the CGM, from $10^4\,\rm K$ to $10^9\,\rm K$ (see \citealt{2008A&A...488..429T}). The rates are computed in units of $\rm erg\,cm^3\,s^{-1}$, so $\Lambda=[\rho/(\mu\,m_u)]^2\,\tilde{\Lambda}=n^2\tilde{\Lambda}$. A density-weighted heating rate that mimics metagalactic heating (\citealt{2012ApJ...746..125H}) is also applied to the gas, but cooling dominates the evolution of the gas at the explored temperatures. Additionally, cooling is switched off for gas at temperatures below the cooling floor, $T=10^4\,\rm K$. Applying a cooling floor of $10^4\,\rm K$ to the gas in our simulations mimics the effect of gas heating caused by background UV radiation from the central starburst, which we did not explicitly include in our cooling and heating function. In this context, \cite{2017ApJ...837...28S} shows that photo-heating rates for the number densities relevant for the clouds we model here is much more significant for temperatures below $10^4\,\rm K$.

\subsection{Numerical solver}
\label{subsec:solver}

The above system of equations is numerically solved in a 3D Cartesian coordinate system $(x,y,z)$ using an unsplit \verb#TVD# Lax-Friedrichs solver \citep{2007ApJS..170..228M} with parabolic spatial reconstruction and a \verb#RK3# time-marching algorithm. The solenoidal conditions for the magnetic field, $\bm{ \nabla\cdot B}$ is enforced by using a hyperbolic divergence cleaning algorithm (see \citealt{2002JCoPh.175..645D}). This choice of algorithms provides adequate computational performance and numerical stability.\par

\subsection{3D domain and resolution}
\label{subsec:3Ddomainandresolution}

Our 3D clouds are centred on the origin $(0,0,0)$ of the computational domain, which consists of a prism with a spatial range $-6\,r_{\rm cloud}\leq X_1\leq6\,r_{\rm cloud}$, $-2\,r_{\rm cloud}\leq X_2\leq22\,r_{\rm cloud}$, and $-6\,r_{\rm cloud}\leq X_3\leq6\,r_{\rm cloud}$, where $r_{\rm cloud}=10\,\rm pc$ is the initial cloud radius. The physical domain then has dimensions $(120\times 240 \times 120)\,\rm pc$ in all models. The standard numerical resolution is $R_{\rm 32}$ (i.e., $32$ cells cover the cloud radius), which corresponds to a uniform grid resolution of $(N_{\rm X_{1}}\times N_{\rm X_{2}}\times N_{\rm X_{3}})=(384\times768\times384)$. This resolution is adequate to describe the overall evolution of 3D clouds as shown in \cite{2018MNRAS.473.3454B} for similar configurations. A numerical resolution analysis in presented in Appendix \ref{AppA}. In addition, outflow boundary conditions are imposed on all sides of the computational domain.\par

\subsection{Wind-cloud models}
\label{subsec:Wind-Cloud-Models}

Our simulation set includes five 3D wind-cloud models in total (see table \ref{Table1}). The models are based on earlier adiabatic models reported in \cite{2016MNRAS.455.1309B}, but they include radiative processes. The clouds have uniform density distributions with smoothed edges and are spherical with radius $r_{\rm cloud}=10\,\rm pc$. The function describing the radial density gradient is:
\begin{equation}
\rho(r)=\rho_{\rm wind} + \frac{\rho_{\rm core}-\rho_{\rm wind}}{1+\left(r/r_{\rm core}\right)^N},
\label{eq:DensityProfile}
\end{equation}

\noindent where $\rho_{\rm core}$ is the core density of the cloud with radius $r_{\rm core}=5\,\rm pc$, $\rho_{\rm wind}$ is the density of the wind, and $N=10$ controls the steepness of the curve describing the density gradient. To ensure a smooth transition into the background gas, the density at the boundary of the cloud ($r_{\rm cloud}$) is $2.0\rho_{\rm wind}$ for all the simulations reported here. The mass-weighted average number density and temperature of the cloud is $\langle n_{\rm cloud}\rangle=6.9\times 10^{-1}\,\rm cm^{-3}$ and $\langle T_{\rm cloud}\rangle=5.5\times 10^3\,\rm K$. The clouds are embedded in a supersonic wind with number density, $\bar{n}_{\rm wind}=10^{-3}\,\rm cm^{-3}$, and Mach number, ${\cal M_{\rm wind}}=|\bm{v_{\rm wind}}|/c_{\rm wind}=4$, where $|{\bm{v_{\rm wind}}}|\equiv v_{\rm wind}=5.2\times 10^2\,\rm km\,s^{-1}$ and $c_{\rm wind}=\sqrt{\gamma P/\rho_{\rm wind}}=1.3\times 10^2\,\rm km\,s^{-1}$ are the speed and sound speed of the wind, respectively. The initial density contrast between the cloud and the wind ranges from $\chi=2$ at the cloud envelope to $\chi=10^3$ at the cloud core.

\begin{table}
\caption{Initial conditions for the 3D wind-cloud models} 
\begin{adjustbox}{width=0.48\textwidth}
\begin{tabular}{c c c c c c}
\hline
\textbf{(1)} & \textbf{(2)} & \textbf{(3)} & \textbf{(4)} & \textbf{(5)} & \textbf{(6)}\\
\textbf{Model}~$\rightarrow$ & R32-HD & R32-AL & R32-TR & R16-AL & R16-TR\\
\textbf{Parameter}~$\downarrow$ &  &  &  & & \\\hline 
Cooling & Yes & Yes & Yes & Yes & Yes \\
$\bf{\rm cells/r_{\rm cloud}}$ & $32$ & $32$ & $32$ & $16$ & $16$\\
Resolution & $0.3\,\rm pc$ & $0.3\,\rm pc$ & $0.3\,\rm pc$  & $0.6\,\rm pc$ & $0.6\,\rm pc$\\
\textbf{$\overrightarrow{\bm B}$} orientation & -- & $B_{\rm y}$ & $B_{\rm x}$ & $B_{\rm y}$ & $B_{\rm x}$\\
\textbf{$|\overrightarrow{\bm B}|$} & -- & $0.2\,\rm\mu G$ & $0.2\,\rm\mu G$ & $0.2\,\rm\mu G$ & $0.2\,\rm\mu G$\\
\textbf{$\beta$} & $\infty$ & $100$ & $100$ & $100$ & $100$\\\hline
$\bf{r_{\rm cloud}}$ & \multicolumn{5}{c}{$10\,\rm pc$} \\
$\bf{T_{\rm cloud}}$ & \multicolumn{5}{c}{$8.1\times 10^2$ -- $4\times 10^5\,\rm K$} \\
$\bf{\overline{T}_{\rm cloud}}$ & \multicolumn{5}{c}{$1.3 \times 10^5\,\rm K$} \\
$\bf{\langle T_{\rm cloud}\rangle}$ & \multicolumn{5}{c}{$5.5 \times 10^3\,\rm K$} \\
$\bf{n_{\rm cloud}}$ & \multicolumn{5}{c}{$2\times 10^{-3}$ -- $1\,\rm cm^{-3}$} \\
$\bf{\overline{n}_{\rm cloud}}$ & \multicolumn{5}{c}{$1.5\times 10^{-1}\,\rm cm^{-3}$} \\
$\bf{\langle n_{\rm cloud}\rangle}$ & \multicolumn{5}{c}{$6.9\times 10^{-1}\,\rm cm^{-3}$} \\
$\bf{v_{\rm cloud}}$ & \multicolumn{5}{c}{$0\,\rm km\,s^{-1}$} \\
$\bf{T_{\rm wind}}$ & \multicolumn{5}{c}{$8.1\times 10^5\,\rm K$}\\
$\bf{n_{\rm wind}}$ & \multicolumn{5}{c}{$10^{-3}\,\rm cm^{-3}$}\\
$\bf{v_{\rm wind}}$ & \multicolumn{5}{c}{$5.2\times 10^2\,\rm km\,s^{-1}$}\\
$\bf{c_{\rm wind}}$ & \multicolumn{5}{c}{$1.3\times 10^2\,\rm km\,s^{-1}$}\\
$\bf{{\cal M}_{\rm wind}}$ & \multicolumn{5}{c}{$4$}\\
$\bf{\chi}$ & \multicolumn{5}{c}{$2$ -- $10^3$}\\
$\bf{t_{\rm cc}}$ & \multicolumn{5}{c}{$1.2\,\rm Myr$}\\\hline
\end{tabular}
\end{adjustbox}
\tablefoot{Column 1 shows the parameter name. Columns 2 through 5 indicate the model names (`R' stands for resolution in number of cells per cloud radius, `AL' and `TR' stand for magnetic fields aligned with and transverse to the wind direction, respectively. Rows show the initial conditions of each parameter for each model. Quantities with overlines are volume-weighted averages, and quantities in between angle brackets are mass-weighted averages.}
\label{Table1}
\end{table} 

Our purely hydrodynamic model, R32-HD, is used as a control run. In addition, uniform magnetic fields are incorporated into our MHD simulations via the so-called plasma beta of the gas, $\beta=P_{\rm th}/P_{\rm mag}=2P_{\rm th}/|\bm{B}|^2=100$, which is a dimensionless number that relates the thermal pressure, $P_{\rm th}$, to the magnetic pressure, $P_{\rm mag}=|{\bf B}|^2/2$. The magnetic field is uniformly distributed in the simulation domain at $t=0$ and has a strength of $0.2\,\rm\mu G$ (which is on the lower end of the values predicted by e.g. \citealt{2012ApJ...757...14J,2023A&A...670L..23H} for CGM environments). In model R32-AL the magnetic field is aligned to the direction of the wind (i.e., $\overrightarrow{\bf B}=\overrightarrow{\bf B_{\rm y}}$), while in model R32-TR the magnetic field is transverse to the direction of the wind (i.e., $\overrightarrow{\bf B}=\overrightarrow{\bf B_{\rm x}}$).\par


Our initial conditions for the wind and the cloud reflect the environmental conditions expected within the inner volumes of CGM winds. In particular, the chosen parameters are motivated by the environmental conditions of the CGM in the MW (see \citealt{2017A&A...607A..48R} for a survey on absorbers and also \citealt{2012MNRAS.423.3512C,2023A&A...674L..15V,2023MNRAS.524.1258N} for descriptions of the MW nuclear wind).

\subsection{Analysis tools and diagnostics}
\label{subsec:analysis}
To perform the analysis of our wind-cloud models, we define some diagnostics following the same convention as \cite{2016MNRAS.455.1309B}. First, we write the mass-weighted volume average of any variable as:
\begin{equation}
    \left \langle {\cal G}  \right \rangle = \frac{\int {\cal{G}} \rho \, C \mathrm{dV} }{\int\rho \, C \mathrm{dV}}.
    \label{mass-weighted-volume}
\end{equation}

\noindent From this equation, it is possible to define the average cloud\footnote{Throughout this paper, we use `cloud' to refer to the original cloud material traced by our scalar $C$. This is an appropriate choice in these models as most dense gas at late times is recondensed (original) cloud material as we do not capture mass growth.} extension along each axis as $\left \langle X_{j} \right \rangle$, with $j = x,y,z$, and its mean square as $\left \langle X_{j}^2 \right \rangle$. From these quantities, the effective radius along each axis is (\citealt{1994ApJ...433..757M}):
\begin{equation}
    \iota_{j} = \left [ 5\left ( \left \langle X_{j}^{2} \right \rangle - \left \langle X_{j} \right \rangle^2 \right ) \right ]^{1/2}.
\end{equation}

\noindent Other relevant diagnostics for the analysis are the cloud number density and temperature, measured as:
\begin{equation}
    n_{\rm cloud} = \frac{\rho_{\rm cloud}}{\mu m_p} = \frac{\rho \,C}{\mu m_p},
\end{equation}
\noindent and 
\begin{equation}
    T_{\rm cloud} = \frac{P_{\rm cloud}}{n_{\rm cloud} \, k_{\rm B}} = \frac{P_{\rm cloud} \, \mu m_p}{\rho_{\rm cloud} \, k_{\rm B}},
\end{equation}

\noindent where $\mu$ is the mean particle mass, $m_p$ the proton mass, $k_B$ is the Boltzmann constant, $\rho_{\rm cloud}$ is the mass density of the cloud and $P_{\rm cloud}=P_{\rm th}\,C$ is the cloud pressure. Additionally, we estimate the mixing fraction between the cloud and the wind as in \cite{1995ApJ...454..172X}:
\begin{equation}
    f_{\rm mix} = \frac{\int {\rho\,\cal C}_{\alpha} \mathrm{dV} }{M_{\rm cloud,0}},
\end{equation}

\noindent where $C_{\alpha}=C$ when $0.01 < C < 0.99$ and $M_{\rm cloud,0}$ is the initial mass of the cloud.

\noindent Regarding time-scales, the cloud-crushing time (as defined in \citealt{1996ApJ...473..365J}) is
\begin{equation}
t_{\rm cc}=\frac{2r_{\rm cloud}}{v_{\rm ts}}=\chi^{\frac{1}{2}}\frac{2r_{\rm cloud}}{{\cal M_{\rm wind}} c_{\rm wind}}=1.2\,\rm Myr,
\label{eq:CloudCrushing}
\end{equation}

\noindent where $v_{\rm ts}={\cal M_{\rm wind}} c_{\rm wind}/\chi^{\frac{1}{2}}$ is the speed of the shock transmitted from the wind to the cloud after the initial impact. The cloud cooling time is:
\begin{equation}
    t_{\rm cool}=\frac{3\,k_{\rm B}\rm \langle T_{\rm cloud}\rangle}{2\,\langle n_{\rm cloud}\rangle\,\Lambda}=6.6\,\rm kyr, 
\end{equation}

\noindent the mixed-gas cooling time (\citealt{2018MNRAS.480L.111G}) is:
\begin{equation}
    t_{\rm cool,mix}=\chi\frac{\Lambda(T_{\rm cloud})}{\Lambda(T_{\rm mix})}\,t_{\rm cool}=0.1\,\rm Myr, 
\end{equation}
\noindent where $T_{\rm mix}=\sqrt{T_{\rm wind}\,\langle T_{\rm cloud}\rangle}=6.7\times 10^4\,\rm K$. For these conditions, the critical radius for clouds to avoid disruption (see \citealt{2020MNRAS.492.1970G}) is $r_{\rm crit}=v_{\rm wind}\,t_{\rm cool, mix}/\chi^{1/2}=2.1\,\rm pc$\footnote{We note that this equation is for hydrodynamic models, so it should be taken only as a first approximation in MHD models, where using an effective density contrast, $\chi_{\rm eff}$ may be more appropriate. $\chi_{\rm eff}$ would be higher (and $r_{\rm crit}$ lower) in the transverse field case owing to shielding via draping.}. Since $t_{\rm cool,mix}<t_{\rm cc}/2$ and $r_{\rm cloud}>r_{\rm crit}$, we would expect mass growth in our models, but our computational domain is too small to capture the focusing effect described in earlier studies.

\subsection{The PLUTO-TRIDENT interface}
\label{subsec:ThePLUTO-TRIDENTinterface}

The column density maps and synthetic spectra are generated by TRIDENT (\citealt{2017ApJ...847...59H}), an extension of the yt analysis code (\citealt{2011ApJS..192....9T}). TRIDENT is a Python-based tool for post-processing hydrodynamical simulations to produce synthetic absorption spectra. It creates absorption-lines for any trajectory, operates across the ultraviolet, optical, and infrared bands, and generates column density maps for any specified ion (our target ions are listed in table \ref{Table2}). Specifically, we use the \cite{2017ApJS..230....8C} atomic data to choose the references for determining ion densities of: O\,{\sc vi}, N\,{\sc v} (\citealt{1988JPhB...21.3669P}), C\,{\sc ii} (\citealt{2004ADNDT..87....1F}), C\,{\sc iv} (\citealt{1998PhRvA..57.1652Y}), Si\,{\sc ii} (\citealt{2009A&A...508.1527B}) and Si\,{\sc iv} (\citealt{2006ADNDT..92..607F}). In TRIDENT, the number density of the ion $i$ for the species $X$ is derived from temperature, density, metallicity, ionisation fraction, and redshift by the \texttt{ion$\_$balance} module. We assume that the wind-cloud system is at redshift zero since our study simulates clouds in the Local Universe. This way $n_{X,i}$ is computed for a single cell as
\begin{equation}
    n _{X,i} = f_H\frac{\rho }{m_H}Z\left ( \frac{n_X}{n_H} \right )_{\odot}f_{X,i},
\end{equation}

\begin{table}\centering
\caption{Ion species mapped by our study.}
\begin{tabular}{c c c c c c c c c c}
\hline
\textbf{(1)} & \textbf{(2)} & \textbf{(3)} \\
\textbf{Phase} & \textbf{Ion} & \textbf{Temperature} \\ \hline
Cold phase & C\,{\sc ii}, Si\,{\sc ii} & $10^{4-4.5}$ K\\
Intermediate phase & C\,{\sc iv}, Si\,{\sc iv} & $10^{4.5-5}$ K\\
Warm phase & O\,{\sc vi}, N\,{\sc v} & $10^{5-5.5}$ K\\\hline
\end{tabular}
\label{Table2}
\end{table} 

\noindent where $f_H = 0.76$ is the mass fraction of primordial hydrogen, $\rho$ is the total gas density, $Z$ the metallicity, which we assume to be solar, $(n_X/n_H)_{\odot}$ is the solar abundance, and $f_{X_i}$ is the ionisation fraction, which is a function of the incident radiation.\par

By default, TRIDENT employs the \cite{2012ApJ...746..125H} UV metagalactic background. While this is appropriate for the study of the intergalactic medium, it is not suitable for generating observables in a nearby star-forming environment. For this purpose, we have generated a more realistic UV starburst background which is described in more detail in \cite{2023IAUS..362...56C}. The most relevant steps in creating the UV background are the choice of the Spectral Energy Distribution (SED) from STARBURST99 (\citealt{2017IAUS..316..359V}) that best represents the stellar population near the simulated atomic clouds. Specifically, we chose a SED assuming solar metallicity for consistency with the way $n_{X_i}$ is computed and a 3 Myr old starburst, which is in agreement with the average lifetime of atomic and molecular clouds detected, for example, in \cite{2013ApJ...770L...4M} and \cite{2017A&A...601A.146C}. We use a spectral resolution of $1$ km s$^{-1}$ in order to achieve high resolution in the spectral analysis combined with a reasonable computing time. We note that this resolution is a factor of $3$ times higher than the pixel spectral resolution of absorption spectra studies in our Galaxy (\citealt{2017A&A...607A..48R}). How our choice of spectral resolution influences the spectra is discussed in Appendix \ref{AppB}.

\section{Results}
\label{Results}

\subsection{Overall evolution of the wind-cloud system}
\label{General evolution}

Fig. \ref{Figure1} shows down-the-barrel and edge-on column density maps of cloud gas, in models R32-AL and R32-TR, at 4 different times. Panel 1a of this figure shows the evolution of model R32-AL, and panel 1b shows the model R32-TR. The disruption of wind-swept clouds embedded in supersonic winds occurs in four stages.\par 

During the first stage, the initial impact of the supersonic wind on the cloud triggers both reflected and transmitted shocks. The reflected shock creates a bow shock in front of the cloud while the transmitted shock travels through the cloud gas at a speed $\chi^{-0.5}v_{\rm wind}$. During the second stage, the cloud is subjected to shock heating, and pressure-gradient forces in the direction of streaming accelerate the cloud and stretch it. Shock heating increases the cloud temperature above the cooling floor, so cloud gas starts to lose thermal energy via radiative processes.\par

During the third stage, the acceleration continues and the cloud loses mass via the Kelvin-Helmholtz (KH) instabilities. The vorticity deposited by shear flows remove gas from the cloud and the wind moves it downstream, deforming the cloud into a filamentary shape. During the fourth stage, the cloud is accelerated sufficiently long that Rayleigh-Taylor (RT) instabilities at the front of the cloud become significant. As a result, the filament breaks up into smaller cloudlets as RT bubbles penetrate the cloud, causing it to expand further.\par

\subsection{On the effect of magnetic fields on wind-swept clouds}
\label{B-fields}

\begin{figure}
    \centering
    \hspace{-0.3cm}\includegraphics[width=0.5\textwidth]{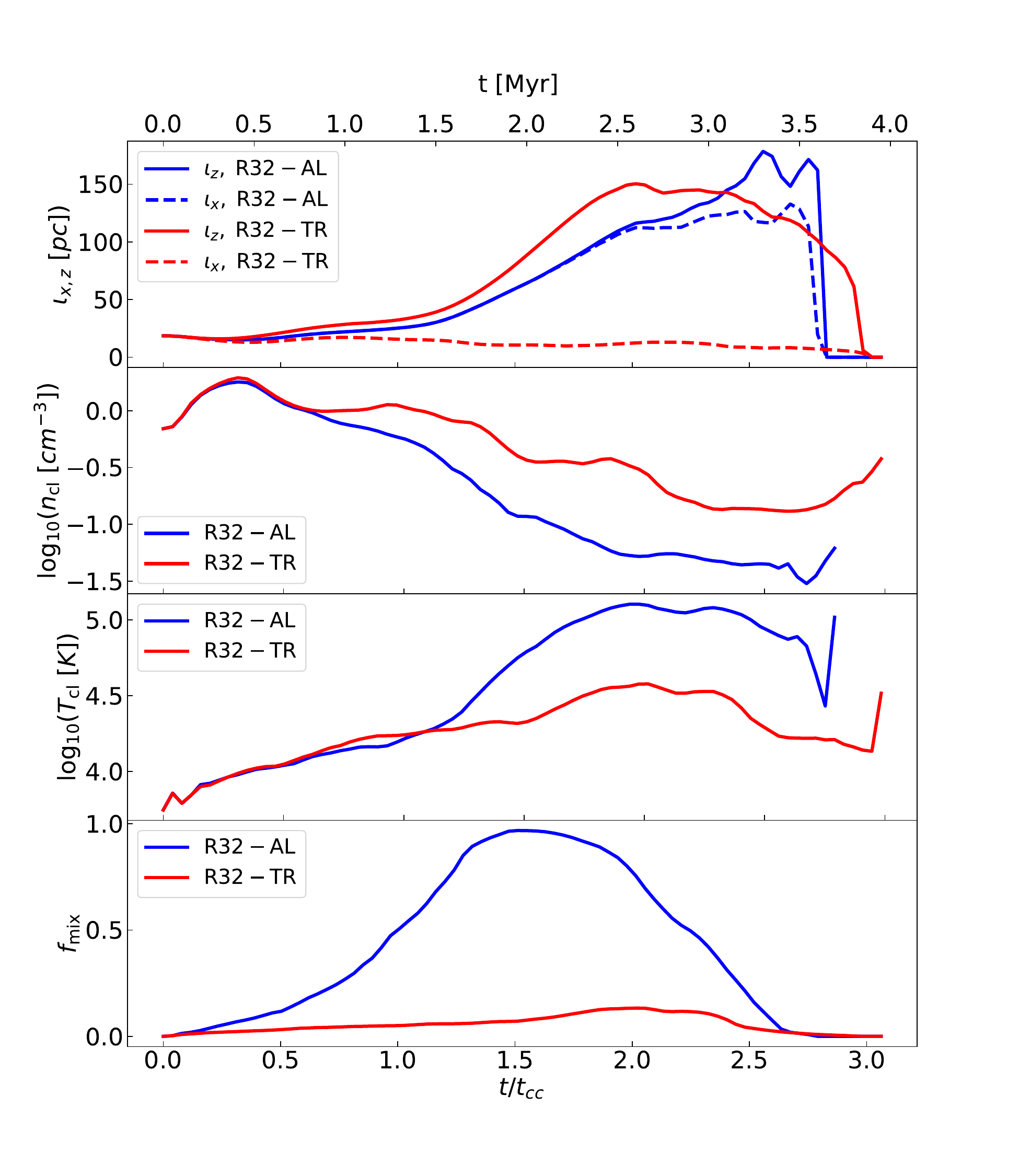}
    \caption{Evolution of several cloud properties in models with different initial magnetic field orientations (aligned in blue and transverse in red). The top panel shows the cloud elongations, the middle panels show the average cloud density and temperature, and the bottom panel shows the mixing fraction. Transverse fields cause cloud asymmetry, shield cold and dense gas, and reduce mixing.}
    \label{Figure2}
\end{figure}

Our simulations show that the initial orientation of the magnetic field has significant effects on several cloud properties, particularly on the morphology, density and temperature structure, and mixing profile of wind-swept clouds. Regarding morphology, in the model R32-AL the cloud develops a symmetric filamentary morphology containing a pressure-confined flux rope (see panel 1a of Fig. \ref{Figure1}), while in the model R32-TR the filament becomes anisotropic (see panel 1b) with a highly elongated morphology in the direction perpendicular to the $\bm B$ field and very narrow in the plane that contains $\bm B$. The top panel of Fig. \ref{Figure2} shows the transverse cloud elongations along $x$ and $z$-, which illustrate this effect.\par

\begin{figure*}
\begin{center}
  \begin{tabular}{c}
    3a. C\,{\sc ii} in model R32-AL\\
    \includegraphics[width=0.85\textwidth]{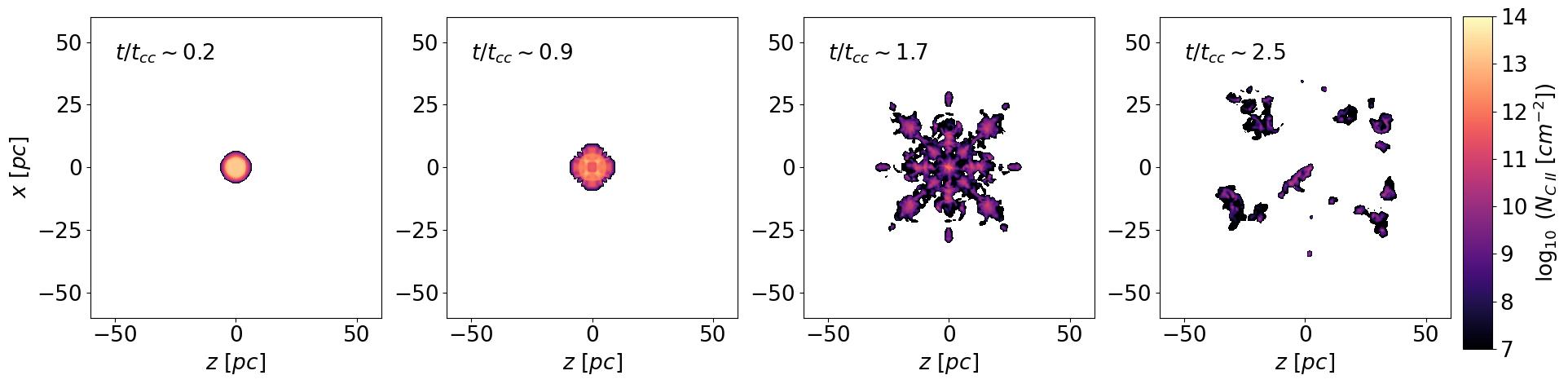}\vspace{-0.1cm}\\
    \includegraphics[width=0.85\textwidth]{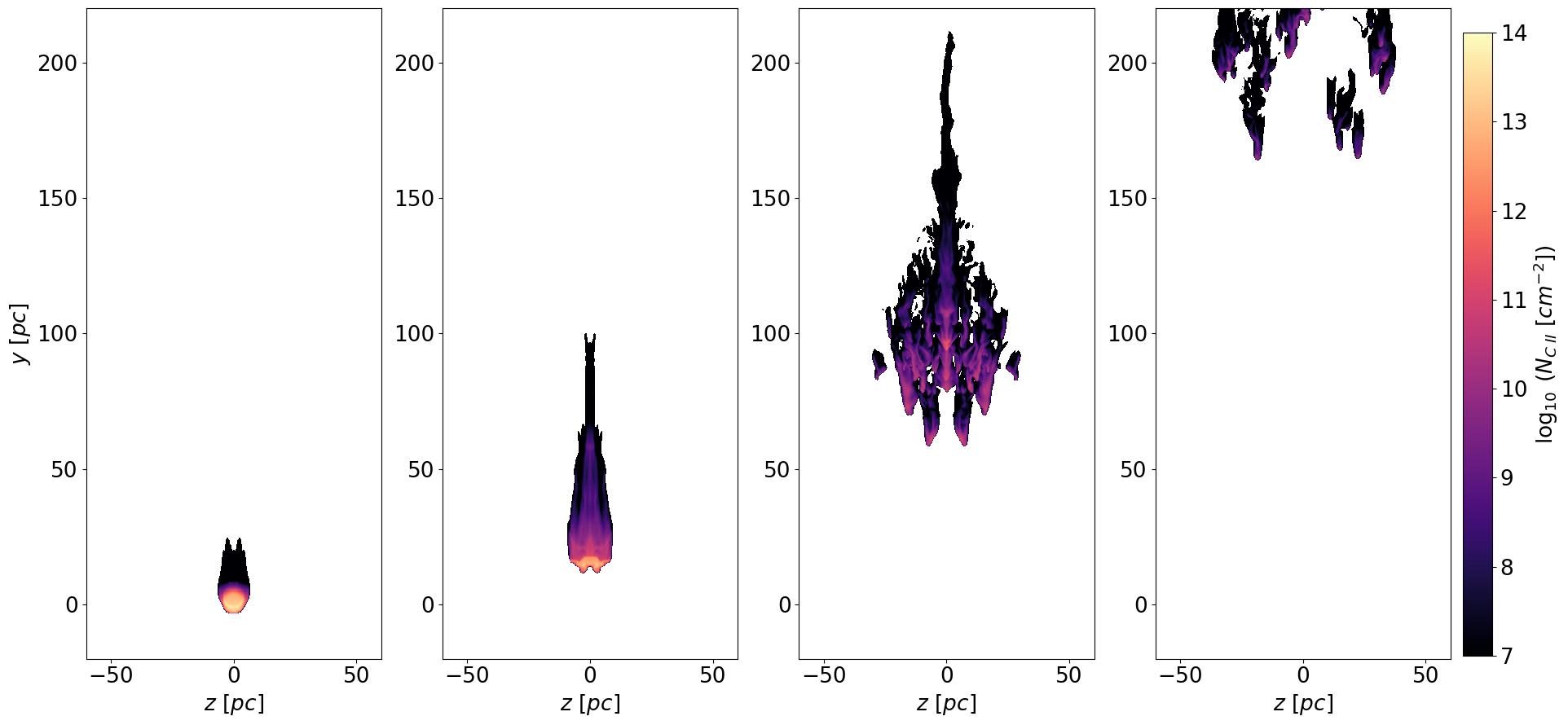}\\
    3b. C\,{\sc ii} in model R32-TR\\
    \includegraphics[width=0.85\textwidth]{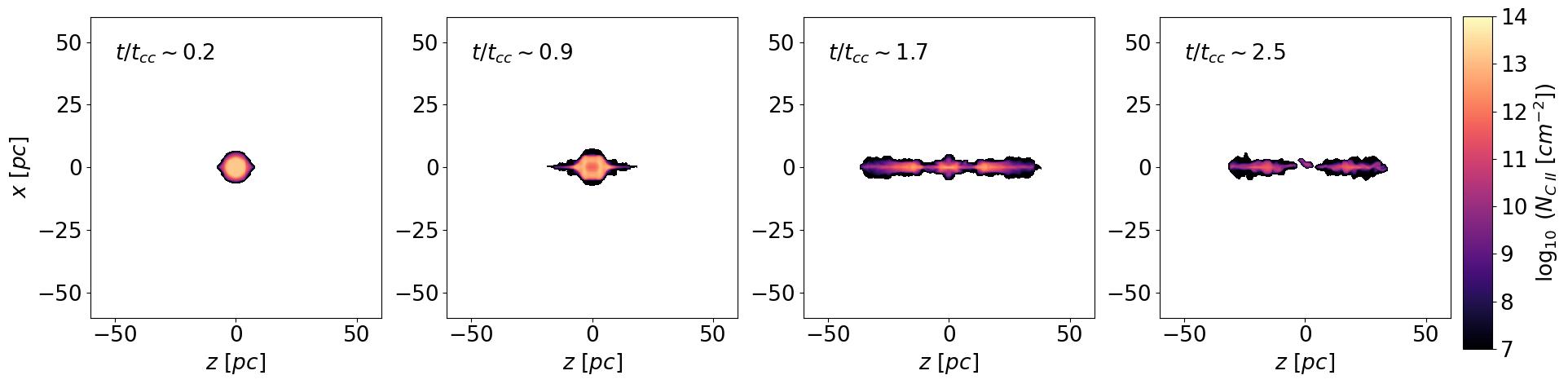}\vspace{-0.1cm}\\
    \includegraphics[width=0.85\textwidth]{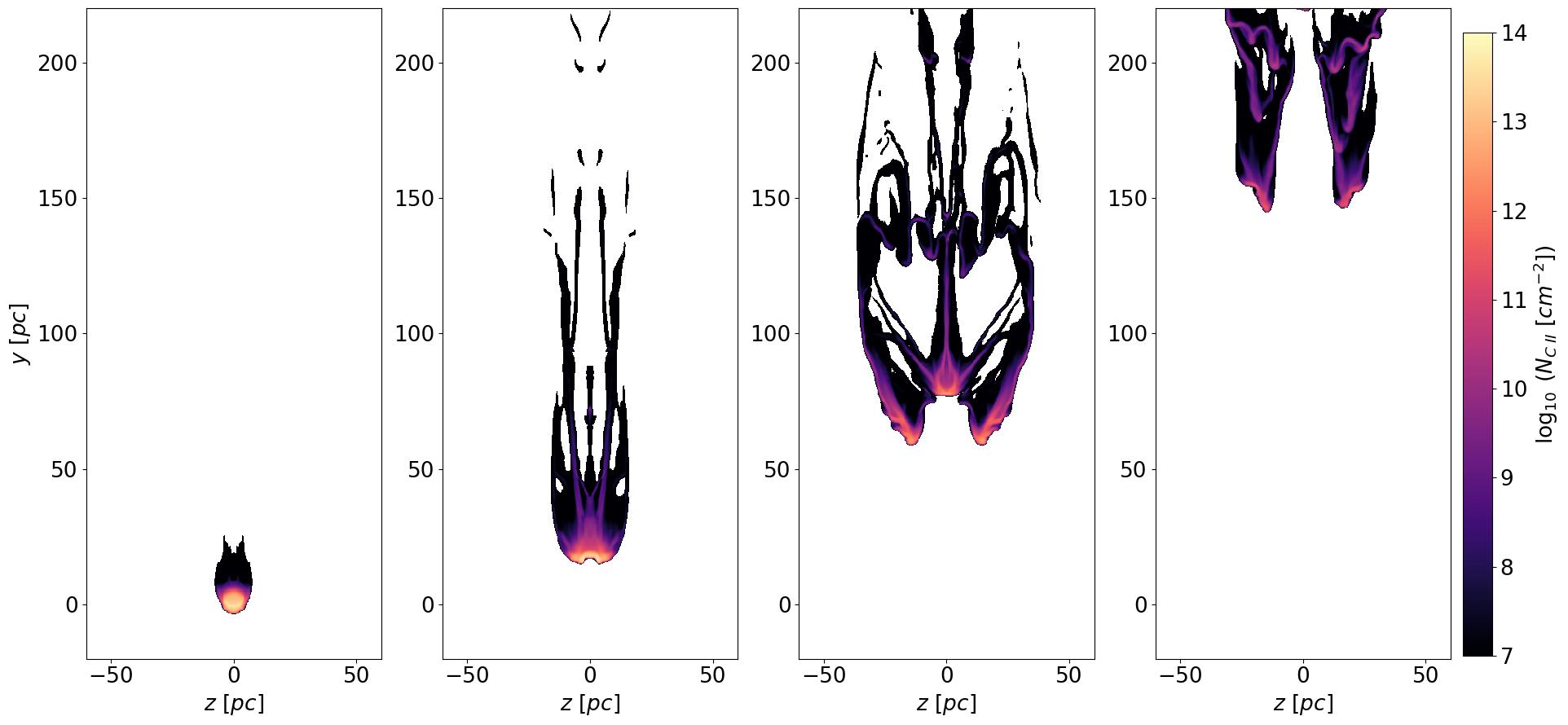}\\
  \end{tabular}
  \caption{2D maps showing the column densities of C\,{\sc ii}. Panel 3a shows the projections for R32-AL simulation along the $y$-axis for down-the-barrel images (top) and along the $x$-axis for edge-on (bottom). The same for the model R32-TR in panel 3b. From left to right, the plots display C\,{\sc ii} column densities at $t = 0.2, 1.2, 2.2$ and $3.2$ Myr (or $t/t_{\rm cc} = 0.2, 0.9, 1.7, 2.5$) respectively. We note how C\,{\sc ii}, tracer of the cold and dense gas, in the down-the-barrel projections of model R32-TR, survives up to $t/t_{\rm cc} \sim 2.5$ via magnetic shielding.} 
  \label{fig:cii_columndens}
\end{center}
\end{figure*}

\begin{figure*}
    \centering
    \includegraphics[width=\textwidth]{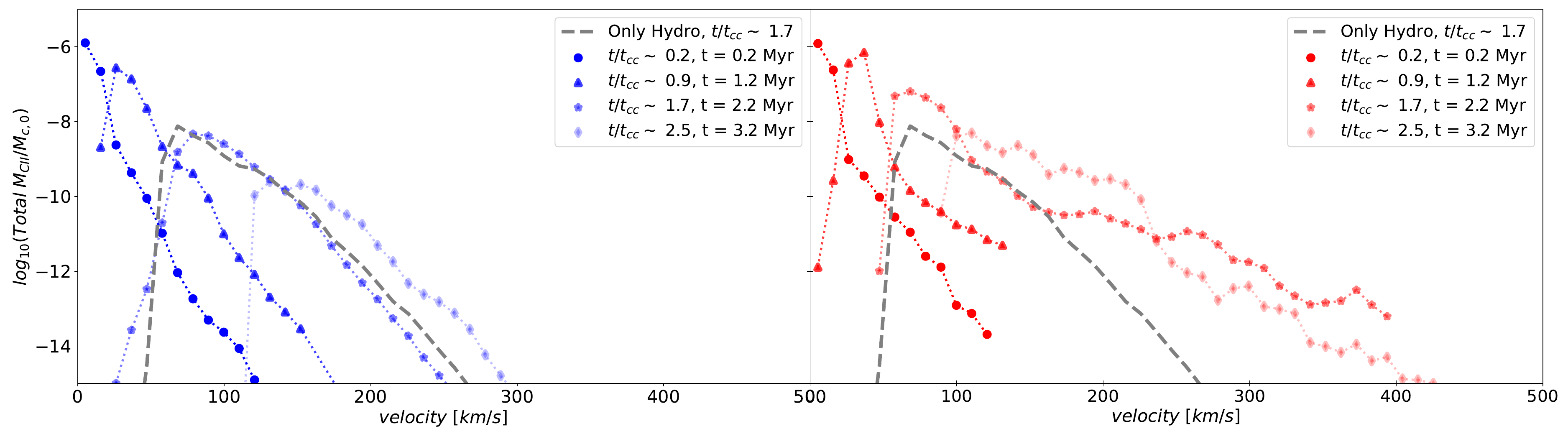}
    \caption{Histograms with velocities (in the $x$-axis) and the ratio between the total mass of C\,{\sc ii}, $M_{\rm CII}$, and the initial cloud mass, $M_{\rm cloud,0}$ (in the $y$-axis), at the same times as in Fig. \ref{fig:cii_columndens} , $t/t_{\rm cc} = 0.2$ (circles), $0.9$ (triangles), $1.7$ (stars), and $2.5$ (diamonds). We use the same colour coding of Fig. \ref{Figure2}, by showing in blue (left panel) the results from the simulation R32-AL and in red (right panel) the results from R32-TR. The dashed grey line refers to model R32-HD at $t/t_{\rm cc} = 1.7$. In model R32-TR, at $t/t_{\rm cc} > 1.7$, a larger fraction of C\,{\sc ii} mass reaches velocities $> 100$ km s$^{-1}$ than in R32-AL. Models R32-HD and R32-AL show similar trends, so transverse magnetic fields produce stronger kinematical effects on cold gas.}
    \label{fig:cii_hist}
\end{figure*}

The morphological difference observed in the clouds immersed in different fields is a well-established result in the literature (see e.g. \citealt{2008ApJ...680..336S,2018ApJ...865...64G,2020ApJ...892...59C,2024MNRAS.527..135H}) and it is due to the draping effect that the magnetic field has in the transverse model (\citealt{2008ApJ...677..993D}). The field lines wrap around the cloud along the $xy$ plane. This creates a net force pointing inwards, so the cloud gas moves in the direction perpendicular to the direction of wrapping. This elongates the cloud along the $z$-axis and promotes RT disruption (see insets in Fig. \ref{Figure1}).\par

Regarding the density and temperature structure of the cloud, the middle panels of Fig. \ref{Figure2} show that the transverse field can more effectively shield cold, dense gas. As a result, the cloud densities in the transverse field case are higher than in the aligned case, while temperatures show the opposite effect. This is also in agreement with previous findings, which show that the wrapping of a transverse field around the cloud triggers the emergence of a thin magnetic layer (\citealt{2000ApJ...543..775G,2007ApJ...670..221D}). Such a layer reduces the effects of dynamical instabilities, which are responsible for cloud disruption, and lengthens the lifetime of dense gas by reducing mixing. The latter effect is illustrated in the bottom panel of Fig. \ref{Figure2}. The question is now if these clear and significant differences seen in the morphology, density and temperature structure, and mixing profile of wind-swept clouds also leave imprints on the resulting ion distribution, kinematics, and spectra.

\subsection{Ion column densities and kinematics}
\label{Ion column densities and kinematics}

For each ion listed in Table \ref{Table2}, we create column density maps along the $x$-axis for an edge-on view, and along the $y$-axis for a down-the-barrel perspective. For each of the three gas phases, we report column density maps of one ion: C\,{\sc ii} for the cooler phase, C\,{\sc iv} for the intermediate-temperature gas, and O\,{\sc vi} for the warm medium. Next, we analyse the kinematics of the ions, reporting their velocity distributions at different times.

\subsubsection{\texorpdfstring{Cold phase: $10^{4-4.5}$ K}{Cold phase}}






Fig. \ref{fig:cii_columndens}\textcolor{blue}{.a} shows the C\,{\sc ii} column density in model R32-AL. In the upper panel, the down-the-barrel projections are displayed, while in the lower the edge-on ones are shown. Initially, C\,{\sc ii} is found in the central part of the cloud with a column density up to $10^{14}$ $\rm{cm}^{-2}$ at $t/t_{\rm cc} = 0.2$. After one Myr, in the down-the-barrel projection, the C\,{\sc ii} has not significantly changed its state, maintaining a diameter not exceeding $15$ pc and column densities greater than $10^{12}$ $\rm{cm}^{-2}$. However, in the edge-on plots, it is more evident that some of the C\,{\sc ii} is affected by losses via the KH instability and begins to spread along the downstream of the tail. At this point, the C\,{\sc ii} density drastically decreases when the cloud material becomes more exposed to the wind. At $t/t_{\rm cc} = 1.7$, RT instabilities drive cloud destruction into cloudlets, lowering the density. The gradual decrease of the C\,{\sc ii} column density is also due to the exposure of individual cloudlets to the high temperatures of the wind, which cause Carbon to reach higher levels of excitation than C\,{\sc ii}. Finally, at $t/t_{\rm cc} = 2.5$, few C\,{\sc ii} traces remain in the box. \par

In Fig. \ref{fig:cii_columndens}\textcolor{blue}{.b}, the evolution of the C\,{\sc ii} column density in model R32-TR is shown (down-the-barrel in the top four panels and edge-on in the bottom four). The transverse magnetic field shields the cloud and prevents it from expanding along the $x$-axis, as it is visible in the upper four figures. This phenomenon causes the cross-section of the cloud to remain more compact, and less likely to develop KH instabilities. In the top four panels of Fig. \ref{fig:cii_columndens}\textcolor{blue}{.b}, the C\,{\sc ii} never reaches extensions greater than 10 pc along the $x$-axis, and because it remains more compact than in the simulation with an aligned magnetic field, parts of the cloud with column densities greater than $10^{12}$ $\rm{cm}^{-2}$ are still present even at $t/t_{\rm cc} = 2.5$. The inability of the cloud to expand symmetrically because of magnetic draping, forces the C\,{\sc ii} to extend wider along the tail already at $t/t_{\rm cc} = 0.9$, but it develops fewer cloudlets at $t/t_{\rm cc} = 1.7$ Myr, where the cloud core splits into three cloudlets as a result of RT instabilities. \par

\begin{figure*}
\begin{center}
  \begin{tabular}{c}
    5a. C\,{\sc iv} in model R32-AL\\
    \includegraphics[width=0.85\textwidth]{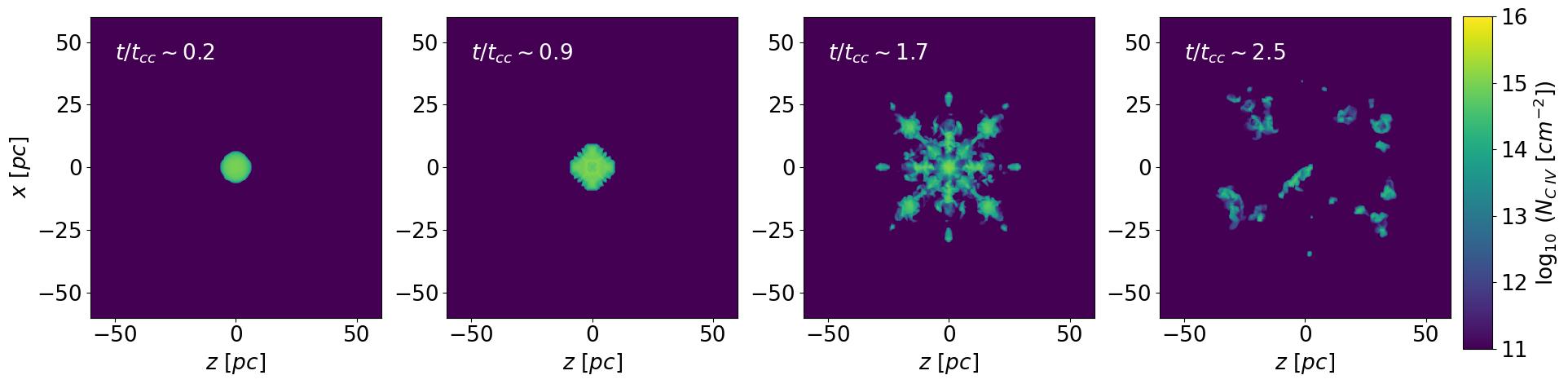}\vspace{-0.1cm}\\
    \includegraphics[width=0.85\textwidth]{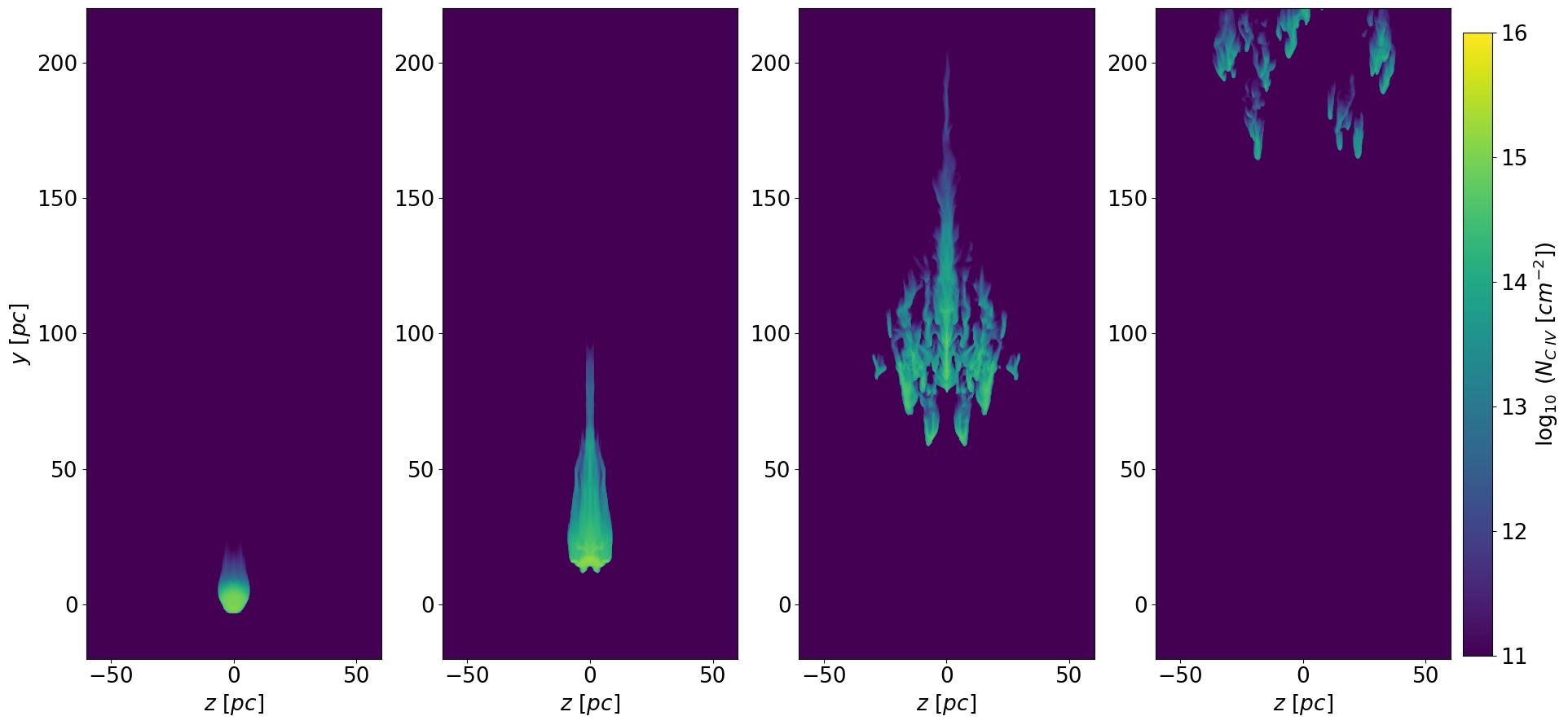}\\
    5b. C\,{\sc iv} in model R32-TR\\
    \includegraphics[width=0.85\textwidth]{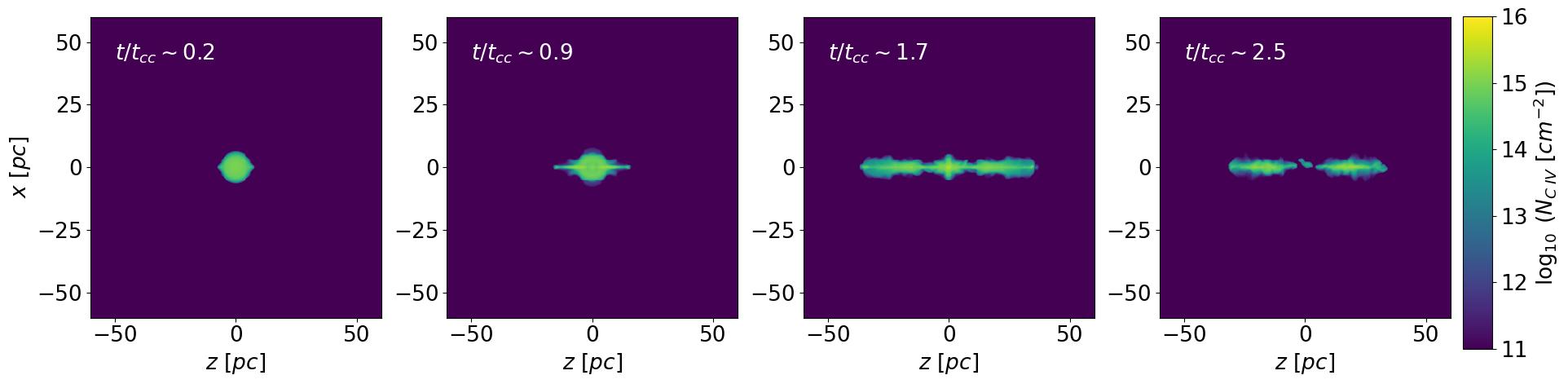}\vspace{-0.1cm}\\
    \includegraphics[width=0.85\textwidth]{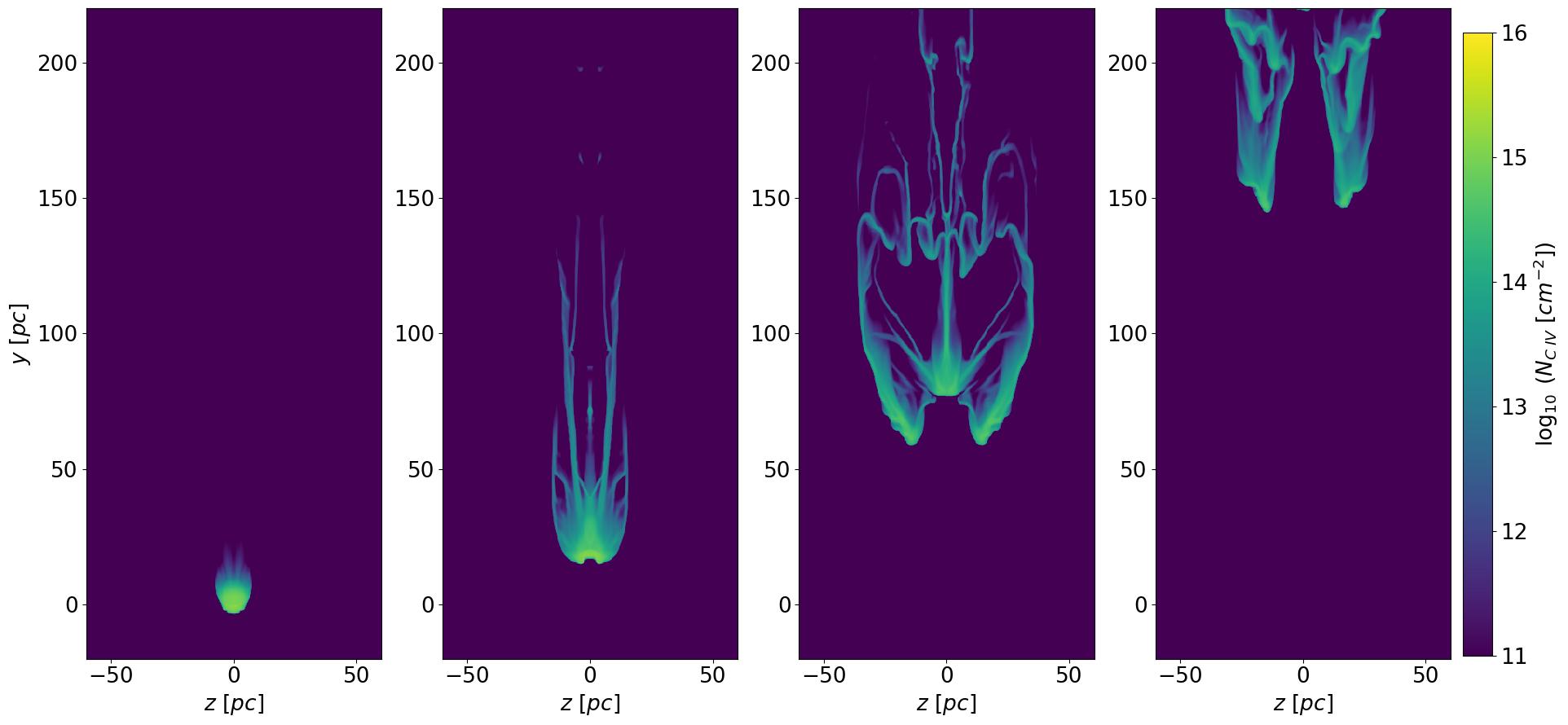}\\
  \end{tabular}
  \caption{2D maps showing the column densities of C\,{\sc iv}. Panel 3a shows the projections for R32-AL simulation along the $y$-axis for down-the-barrel images (top) and along the $x$-axis for edge-on (bottom). The same for the model R32-TR in panel 3b. From left to right, the plots display C\,{\sc iv} column densities at $t = 0.2, 1.2, 2.2$ and $3.2$ Myr (or $t/t_{\rm cc} = 0.2, 0.9, 1.7, 2.5$) respectively. We notice that C\,{\sc iv} follows a more extended volume of cloud material than C\,{\sc ii} (which mainly tracks the core), including both the cloud core and the tail. As for C\,{\sc ii}, also the C\,{\sc iv} column density morphology is affected by the different orientation of $\bm B$, maintaining an extension smaller that 10 pc along the $x$-axis in R32-TR due to draping. } 
  \label{fig:civ_columndens}
\end{center}
\end{figure*}

\begin{figure*}
    \centering
    \includegraphics[width=\textwidth]{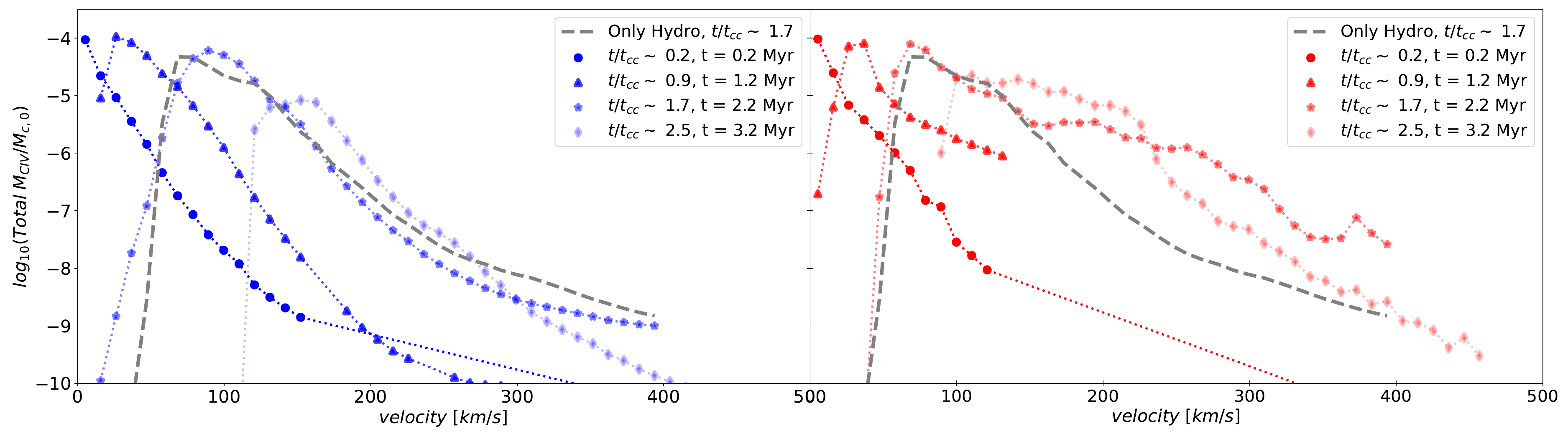}
    \caption{Histograms with velocities (in the $x$-axis) and the ratio between the total mass of C\,{\sc iv}, $M_{\rm CIV}$, and the initial cloud mass, $M_{\rm cloud,0}$ (in the $y$-axis), at the same times as in Fig. \ref{fig:civ_columndens}, $t/t_{\rm cc} = 0.2$ (circles), $0.9$ (triangles), $1.7$ (stars), and $2.5$ (diamonds). We show in blue (left panel) the results from the simulation R32-AL and in red (right panel) the results from R32-TR. The dashed grey line refers to the R32-HD at $t/t_{\rm cc} = 1.7$. We highlight that C\,{\sc iv} covers higher mass fractions than C\,{\sc ii} and O\,{\sc vi} (Figs. \ref{fig:cii_hist} and \ref{fig:ovi_hist}, respectively). Also, for $t/t_{\rm cc} \geq 1.7$, model R32-TR shows larger mass fractions of C\,{\sc iv} at velocities $> 200$ km s$^{-1}$ than model R32-AL, indicating a transverse magnetic field favours a more kinematically rich intermediate phase than an aligned field.}
    \label{fig:civ_hist}
\end{figure*}

The evolution of the mass fraction of C\,{\sc ii}, $M_{\rm C II}$, over the initial cloud mass $M_{\rm cloud,0}$ as a function of velocity is shown in Fig. \ref{fig:cii_hist} for both magnetic field directions. Up to $t/t_{\rm cc} = 0.9$, no major differences in the C\,{\sc ii} velocity distribution are observable in the simulations. In both cases, most of the C\,{\sc ii} maintains velocities below $100$ $\rm km\,s^{-1}$, since most of the cold gas is still confined to the cloud core. At $t/t_{\rm cc} = 1.7$, in the left panel the C\,{\sc ii} mass fraction peaks at $\sim 100$ $\rm km\,s^{-1}$. This reflects that the cloud at this stage gains speed as it expands, acquiring more momentum from the wind. However, in the right panel, at $t/t_{\rm cc} = 1.7$ a larger fraction of  C\,{\sc ii} reaches $v > 100$ $\rm km\,s^{-1}$. The transverse magnetic field envelops and shields the cold gas, squeezing it along the $x$-axis and expanding it along the $z$-axis. The change in the magnetic field topology results in a net magnetic force pointing inwards, which also drives cold gas along the current sheet that forms at the tail of the cloud. While the core regions in both models, R32-AL and R32-TR, have the same overall velocities, the tails in the R32-TR run cover a wider range of velocities. At this time, it is possible to appreciate the contribution of C\,{\sc ii} belonging to the tail, which reaches velocities up to $400$ $\rm km\,s^{-1}$. Finally, at $t/t_{\rm cc} = 2.5$ there is little trace of C\,{\sc ii} in both simulations, which nevertheless reaches velocities of no more than $300$ $\rm km\,s^{-1}$ in model R32-AL and up to $400$ $\rm km\,s^{-1}$ in model R32-TR.\par

For completeness, we show with a grey dashed line the trend of the hydrodynamic model without magnetic fields (model R32-HD), at $t/t_{\rm cc} \sim 1.7$. Comparing this trend with that of R32-AL, we note that the mass fraction of C\,{\sc ii} at different velocities is very similar between the two models. This result indicates that a magnetic field aligned with the wind direction does not have a strong effect on the cold phase of the gas. Instead, the comparison with R32-TR clearly shows how the transverse magnetic field produces stronger kinematical effects on cold gas and allows more C\,{\sc ii} mass to survive at $t = 2.2$ Myr via shielding. This phenomenon is visible from the general placement of the points representing $M_{\rm CII}/M_{\rm cloud,0}$ of model R32-TR at $t/t_{\rm cc} \sim 1.7$, which shows higher mass fractions than the grey dashed line of model R32-HD, especially at speeds above 200 $\rm km\,s^{-1}$. The main reason for this discrepancy is similar to the one previously reported to justify the difference between R32-AL and R32-TR.

\subsubsection{\texorpdfstring{Intermediate temperature phase: $10^{4.5-5}$ K}{Intermediate temperature phase}}




For the intermediate phase, we report the time evolution of the C\,{\sc iv} column density for the wind-cloud simulations with aligned and transverse magnetic field, respectively, in panels \textcolor{blue}{a} and \textcolor{blue}{b} of Fig. \ref{fig:civ_columndens}. An interesting difference between C\,{\sc iv} and C\,{\sc ii} is that C\,{\sc iv} is also present in the wind material, which is too hot to contain C\,{\sc ii}. C\,{\sc iv} would seem to be a better candidate for tracking the evolution of the atomic cloud than C\,{\sc ii}, since its column density reaches values above $10^{14} \; \rm{cm}^{-2}$, especially in the central regions of the cloud. Moreover, the ion also traces well the evolution of the tail that the cloud develops by maintaining a high contrast that reaches even four orders of magnitude compared with $N_{\rm CVI} \sim 10^{11} \; \rm{cm}^{-2}$ in the wind. The upper four panels of Fig.\ \ref{fig:civ_columndens}\textcolor{blue}{.a} indicate that C\,{\sc iv} follows an evolution similar to that reported for C\,{\sc ii}. Up to $t/t_{\rm cc} = 0.9$ the spatial distribution of C\,{\sc iv} in the $xz$ plane is still confined. At $t/t_{\rm cc} \geq 1.7$ the cloud is affected by the onset of RT instabilities and splits into cloudlets. In this case, however, in contrast to C\,{\sc ii}, C\,{\sc iv} is significantly present not only in the individual cores of the cloudlets where the gas is denser, but also in the material connecting these cloudlets. The bottom four panels of Fig. \ref{fig:civ_columndens}\textcolor{blue}{.a} show the evolution of C\,{\sc iv} edge-on. At $t/t_{\rm cc} = 0.9$ and $1.7$, C\,{\sc iv} is present both in the front of the cloud and also in the tail, where the gas reaches the highest velocities. As for C\,{\sc ii}, after $t/t_{\rm cc} = 2.5$ Myr, there are no longer many C\,{\sc iv} traces in the box.\par

In Fig.\ \ref{fig:civ_columndens}\textcolor{blue}{.b}, the evolution of the C\,{\sc iv} is affected by the different orientation of the magnetic field. In all of the four upper panels, the C\,{\sc iv} is prevented from expanding along the $x$-axis by maintaining along this direction an extension of less than $10$ pc. Again, as with the C\,{\sc ii}, at $t/t_{\rm cc} = 1.7$ the formation of $3$ cloudlets can be discerned, which remain visible until $t/t_{\rm cc} = 2.5$. In the lower four panels, the C\,{\sc iv} reaches $N_{\rm CIV} \geq 10^{13} \,\rm cm^{-2}$ both in the cloud core and in the tail up to $t/t_{\rm cc} = 0.9$. Then, it also keeps similar column densities in the front parts of the cloudlets and in the material connecting them at $t/t_{\rm cc} \geq 1.7$. The possibility of observing C\,{\sc iv} so abundantly, compared to other ions, in different parts of the cloud in a wide velocity range, can make it a good candidate for studying magnetic field effects in wind-cloud systems.\par

Fig. \ref{fig:civ_hist} shows the mass fraction of C\,{\sc iv} over $M_{\rm cloud,0}$ as a function of velocity at the same times as in Fig.\ \ref{fig:civ_columndens}, for R32-AL (left) and R32-TR (right). C\,{\sc iv} covers a higher mass fraction than C\,{\sc ii}, which suggests that ions tracing gas in a temperature range between $10^{4.5 -5}$ K might be the best approach for studying the effects of the magnetic fields on wind-swept clouds. In both simulations, at $t/t_{\rm cc} = 0.2$ most of the C\,{\sc iv} is in the cloud core with $v \leq 25$ $\rm km\,s^{-1}$. After one Myr in the left panel, it can be seen that a tail that reaches speeds $\gtrsim 200$ km s$^{-1}$ has already developed. In the right panel of Fig.\ \ref{fig:civ_hist}, however, the shielding effect of the transverse magnetic field prevents the cloud material from mixing with the external environment. Consequently, it is not possible to observe C\,{\sc iv} reaching $v \gtrsim 140$ $\rm km\,s^{-1}$ at early times. At $t/t_{\rm cc} \geq 1.7$, $M_{\rm CIV}/M_{\rm c,0}$ progressively decreases in the R32-AL simulation. The atomic cloud has already exchanged enough material with the environment, which can easily reach $T > 10^5$ K and thus is no longer detectable with C\,{\sc iv}. On the other hand, the right panel of Fig.\ \ref{fig:civ_hist} shows that at $t/t_{\rm cc} \geq 1.7$, the inability of the gas to mix with the wind material keeps it at $T = 10^{4.5 -5}$ K and, at the same time, the gas moving along the current sheet in the cloud tail reaches velocities up to $400$ $\rm km\,s^{-1}$ or higher.\par

As also mentioned for C\,{\sc ii}, we show with a grey dashed line the comparison with the R32-HD run at $t/t_{\rm cc} \sim 1.7$. Even for the gas at the intermediate temperature phase, the hydrodynamic model produces very similar results to R32-AL. This result again indicates that a directionally aligned magnetic field has no strong effect on the temporal evolution of the C\,{\sc iv}, either around the cloud or in the tail. As for the comparison with R32-TR, we note that there is a good overlap between the two models up to $v \sim 150$ $\rm km\,s^{-1}$. However, at higher velocities, the transverse magnetic field preserves the C\,{\sc iv}, producing a more filamentary structure (see Fig. \ref{fig:civ_columndens}) which can reaches speeds of up to 400 $\rm km\,s^{-1}$. Overall, model R32-TR favours a more kinematically rich intermediate phase than models R32-AL and R32-HD.

\subsubsection{\texorpdfstring{Warm phase: $10^{5-5.5}$ K}{Warm phase}}

\begin{figure*}
\begin{center}
  \begin{tabular}{c}
    7a. O\,{\sc vi} in model R32-AL\\
    \includegraphics[width=0.85\textwidth]{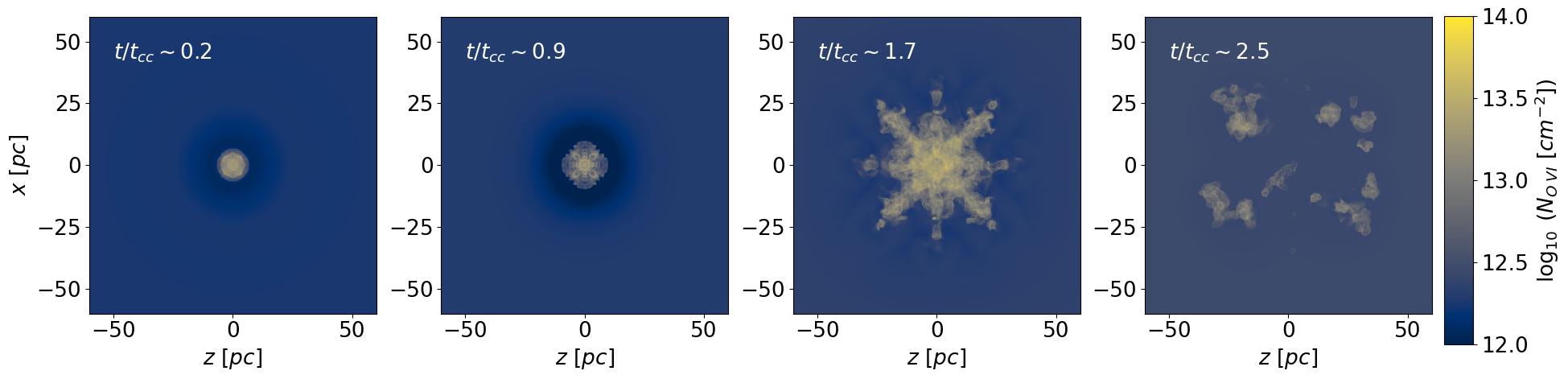}\vspace{-0.1cm}\\
    \includegraphics[width=0.85\textwidth]{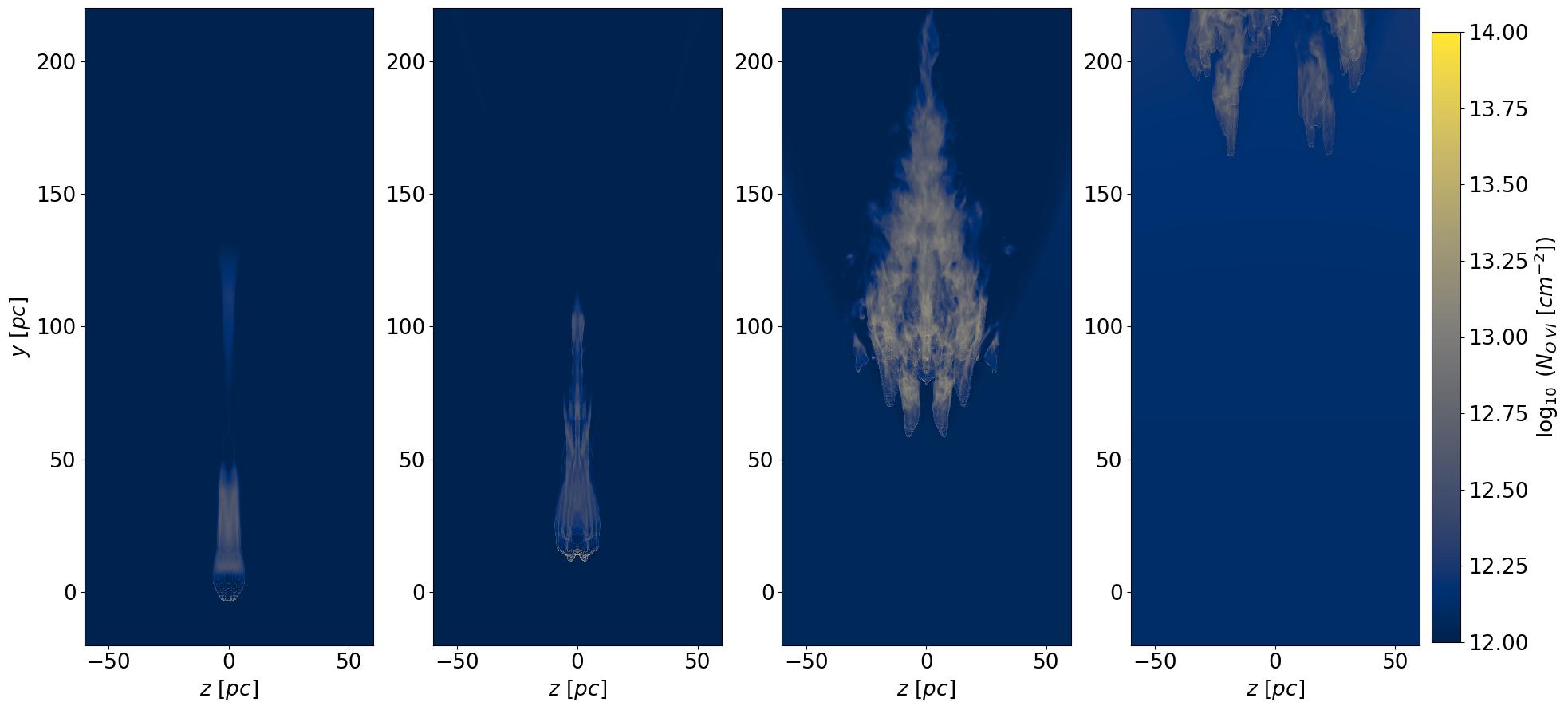}\\
    7b. O\,{\sc vi} in model R32-TR\\
    \includegraphics[width=0.85\textwidth]{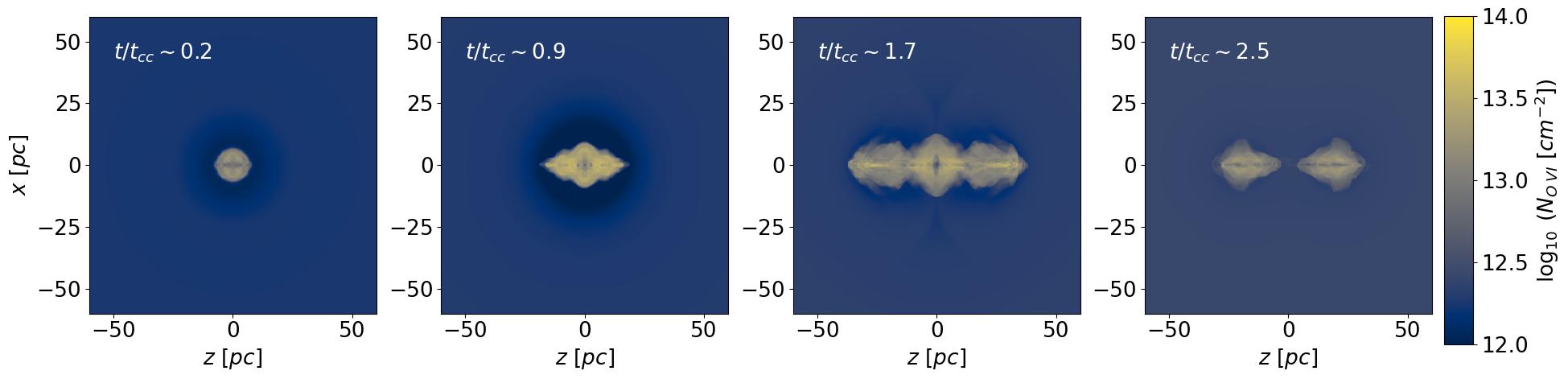}\vspace{-0.1cm}\\
    \includegraphics[width=0.85\textwidth]{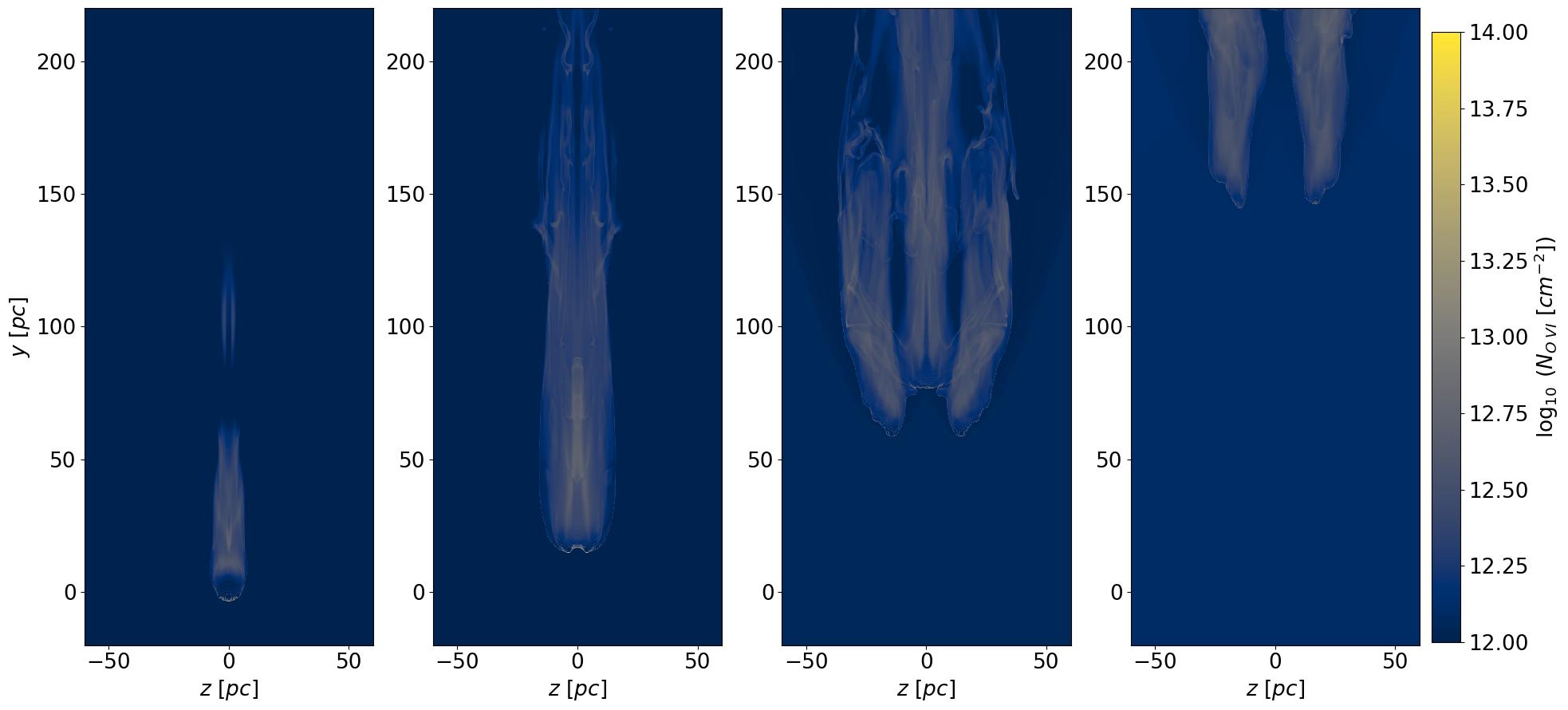}\\
  \end{tabular}
  \caption{2D maps showing the column densities of O\,{\sc vi}. Panel 3a shows the projections for the R32-AL simulation along the $y$-axis for down-the-barrel images (top) and along the $x$-axis for edge-on (bottom). The same for the model R32-TR in panel 3b. From left to right, the plots display C\,{\sc ii} column densities at $t = 0.2, 1.2, 2.2$ and $3.2$ Myr (or $t/t_{\rm cc} = 0.2, 0.9, 1.7, 2.5$) respectively. We see that the contrast among cloud and wind material is less for O\,{\sc vi} than C\,{\sc ii} and C\,{\sc iv}. O\,{\sc vi} traces the outer layers of the cloud, which are well mixed with the wind, so it also covers a more extended area in both down-the-barrel and edge-on projections than the cold and intermediate ions.} 
  \label{fig:ovi_columndens}
\end{center}
\end{figure*}

\begin{figure*}
    \centering
    \includegraphics[width=\textwidth]{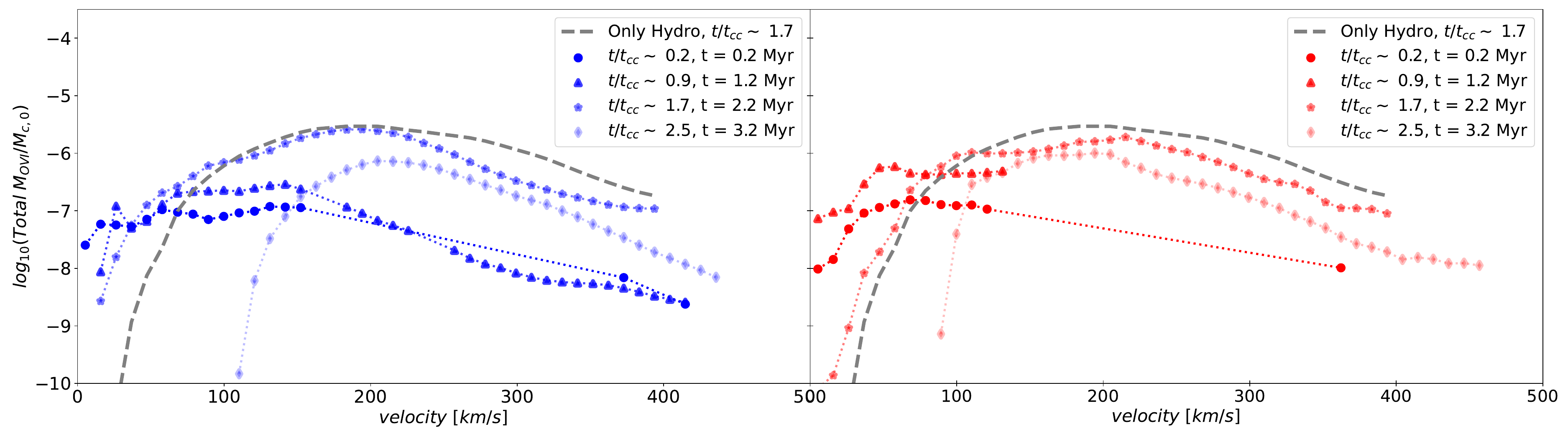}
    \caption{Histograms with velocities (in the $x$-axis) and the ratio between the total mass of O\,{\sc iv}, $M_{\rm OVI}$, and the initial cloud mass, $M_{\rm cloud,0}$ (in the $y$-axis), at the same times as in Fig. \ref{fig:ovi_columndens}, $t/t_{\rm cc} = 0.2$ (circles), $0.9$ (triangles), $1.7$ (stars), and $2.5$ (diamonds). We show in blue (left panel) the results from the simulation R32-AL and in red (right) the results from R32-TR. The dashed grey line refers to the R32-HD at $t/t_{\rm cc} = 1.7$. We see that O\,{\sc vi} spans a wider range of velocities both in R32-AL and R32-TR (at all times), compared to cold and intermediate ions. This ion's kinematics is more complex than the ones shown in Figs. \ref{fig:cii_hist} and \ref{fig:civ_hist} and less sensitive to observe differences caused by the magnetic field orientation.}
    \label{fig:ovi_hist}
\end{figure*}





Finally, in Fig. \ref{fig:ovi_columndens} we show the column density of O\,{\sc vi}, a tracer of the warm phase of the gas. In all panels O\,{\sc vi} is also present in the wind, which reduces the O\,{\sc vi} contrast between the cloud and the wind, compared to colder phases. In fact, distinguishing cloud and wind material in Fig. \ref{fig:ovi_columndens} is accomplished by choosing a range of $N_{\rm OVI}$ in the colour bar that spans only two orders of magnitude. In Fig. \ref{fig:ovi_columndens}\textcolor{blue}{.a}, in the four panels representing $N_{\rm OVI}$ down-the-barrel, O\,{\sc vi} belonging to the cloud has a more homogeneous and extended distribution than C\,{\sc ii} and C\,{\sc iv}; especially at $t/t_{\rm cc} \geq 1.7$. This is because the outer layers of the cloud are well mixed with the surrounding medium. The lower four panels indicate that O\,{\sc vi} is always absent in the cloud core, and only exists in the tail. At $t/t_{\rm cc} = 1.7$ Myr the RT bubbles penetrate the cloud forming cloudlets and expose more of the cloud surface to the galactic wind. As the cloud becomes more exposed to the wind, it heats up and mixes, thus, increasing $N_{\rm OVI}$.\par

In the upper panels of Fig. \ref{fig:ovi_columndens}\textcolor{blue}{.b} the effect of the transverse magnetic field is visible already at $t/t_{\rm cc} = 0.9$, where the expansion of the O\,{\sc vi} belonging to the cloud has as its preferred direction the one along the $z$ axis. In these images, the expansion along the $x$ axis is limited, but higher than in C\,{\sc ii} and C\,{\sc iv}, confirming that O\,{\sc vi} traces only the outer, more diffuse layers of the cloud. Finally, in the lower four panels of Fig. \ref{fig:ovi_columndens}\textcolor{blue}{.b}, the contrast between O\,{\sc vi} tracking cloud and wind material is more tenuous than in the down-the-barrel projection, with column densities only 1 dex higher than those present in the wind.\par

O\,{\sc vi} being present mostly in the wind-shocked cloud material, which then forms the tail, is also visible in Fig. \ref{fig:ovi_hist}. The two panels represent $M_{\rm OVI}/M_{\rm c,0}$ as a function of velocity for the R32-AL simulation in blue and R32-TR in red. At $t/t_{\rm cc} = 0.2$, the fraction of O\,{\sc vi} in the box does not reach $v > 150$ $\rm km\,s^{-1}$ in either case. At $t/t_{\rm cc} > 0.9$, in the right panel, $M_{\rm OVI}/M_{\rm c,0}$ increases progressively and the peak of the distribution shifts to higher speeds, up to $200$ $\rm km\,s^{-1}$ at $t/t_{\rm cc} = 2.5$. The same trend is also visible in the right plot, however, in this case, as the cloud remains more compact, the formation of the tail in the distribution at high speeds is delayed. Indeed, only from $t/t_{\rm cc} = 1.7$ it is possible to observe gas with $v>150$ $\rm km\,s^{-1}$.\par 

At the end, in both simulations, the outflow of the cloud from the box causes a lowering of the peak of the distribution at $t/t_{\rm cc} = 2.5$. Comparing the trends of R32-AL and R32-TR at $t/t_{\rm cc} \sim 1.7$ with the run without magnetic field (R32-HD, shown with a dashed grey line), we see that the trend of $M_{\rm OVI}/M_{\rm cloud,0}$ is similar in both cases, making O\,{\sc vi} a weak candidate for investigating the effects of magnetic fields. This is reasonable as the presence of this ion is regulated by mixing, which occurs in the outermost cloud layers where turbulence dominates (while magnetic fields have more influence on the colder phases).


\subsection{Spectral analysis}
\label{subsec:spectral_analysis}

\begin{figure}
    \centering
    \includegraphics[width=0.5\textwidth]{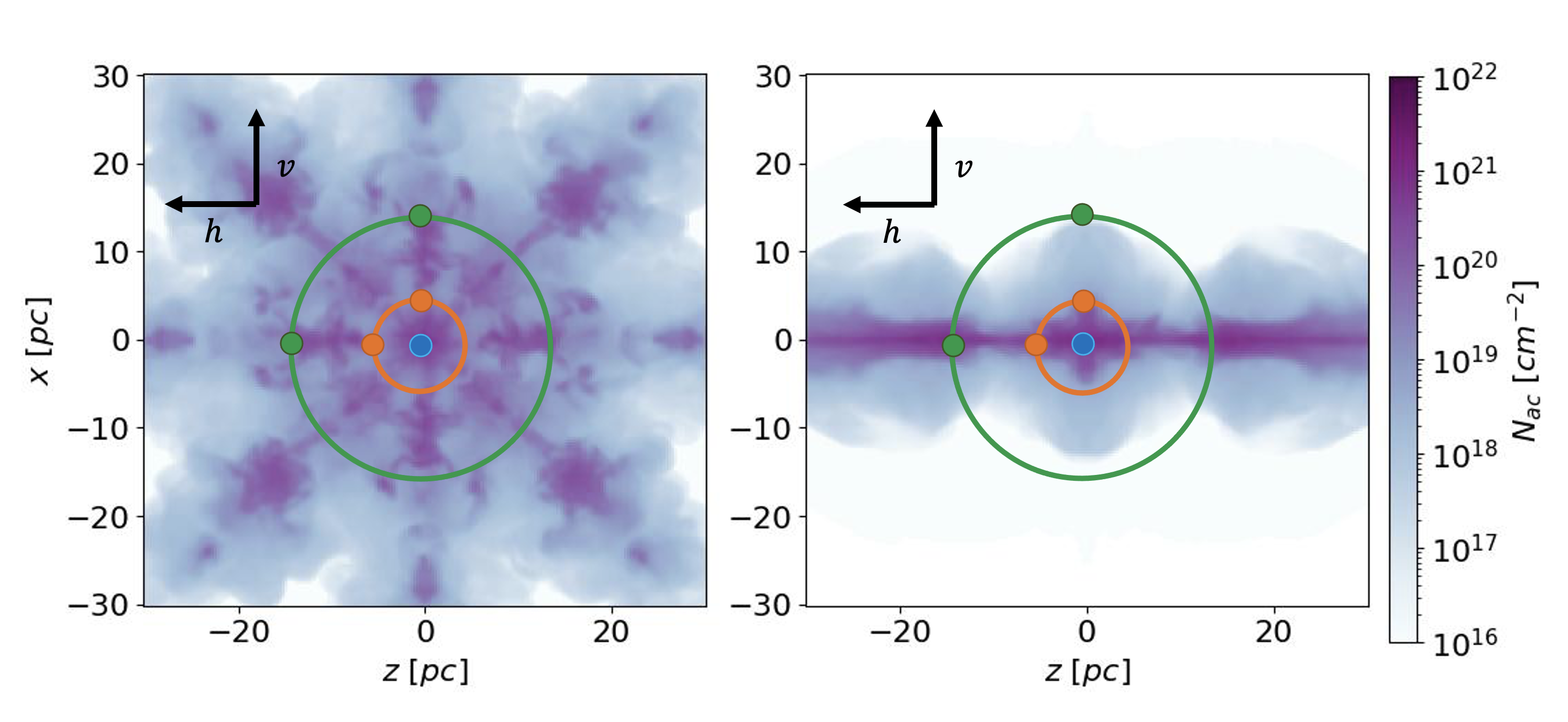}
    \caption{Zoom-in of the down-the-barrel column density map of cloud material in simulation R32-AL (left) and R32-TR (right) at $t/t_{\rm cc} = 1.7$. This figure shows on the $xz$ plane the points at which the spectra shown in Fig. \ref{fig:spectra_coldgas}, \ref{fig:spectra_intergas} and \ref{fig:spectra_warmgas} are computed. The blue dot marks the centre of the cloud, the orange circle the region at 5 pc from the centre and the green one the region at 15 pc. In both plots, the top left-hand corner indicates the '$v$' direction, which we refer to as $x$-axis, and the '$h$' direction, which we refer to as $z$-axis.}
    \label{fig:beam_direction}
\end{figure}

We produce down-the-barrel absorption lines of the 6 ions listed in Table \ref{Table2} using TRIDENT, in order to understand how the different orientations of the magnetic field affect their spectral line profiles. We choose to focus on the time $t/t_{\rm cc} = 1.7$ since this time represents a sufficiently evolved state that captures the mixing between the cold and hot gas. As shown in Section \ref{Spectral_time_evolution}, the differences between the absorption lines in R32-AL and R32-TR are most pronounced after $t/t_{\rm cc} \sim 0.9$. Nevertheless, it is important to choose a time earlier than $t/t_{\rm cc} \sim 2.5$, the moment when most of the cloud material has moved out of the box. Our analysis is based on creating lines in absorption using five down-the-barrel light rays: one passing through the cloud centre, and four passing at $5$ and $15$ pc from the centre, along the $x$- and $z$-axis (see Fig. \ref{fig:beam_direction}).\par

We produce down-the-barrel spectra in order to capture the largest column densities, but our method can be extended to other lines of sight. This first choice is a simplification, so we will investigate other orientations in future work. Fig. \ref{fig:beam_direction} shows the down-the-barrel column densities in the two simulations at $t/t_{\rm cc} = 1.7$ and also shows the points whose spectra we display in this section. The rosette-like pattern we see in our models, more evident in model R32-AL, is due to the carbuncle effect, caused by the uniform grid. While the carbuncle effect can be mitigated by adding a log-normal density distribution or a random turbulent velocity field at $t=0$, we opted to keep the clouds idealised to prevent the added turbulence from altering the spectral lines. Our previous cloud-wind simulations (see \citealt{2018MNRAS.473.3454B,2019MNRAS.486.4526B}) show that turbulent density and velocity fields can alter the cloud morphology, so, while more realistic and carbuncle effect-free, the spectral lines produced in such models deserve careful analysis in future work. However, we note that to study the direction dependency in our spectral analysis, we pick the $5$ reference points mentioned above. We also emphasise that when we pick points along (off-axis) diagonal directions, model R32-AL always has larger ion column densities than model R32-TR, so all the spectral analysis results we show below for the $x$-axis points are representative of such diagonal points.

\subsubsection{\texorpdfstring{Cold phase: C\,{\sc ii} and Si\,{\sc ii} absorption spectra}{Cold spectra}}

\begin{figure*}
\begin{center}
  \begin{tabular}{c c}
    \includegraphics[width=\textwidth]{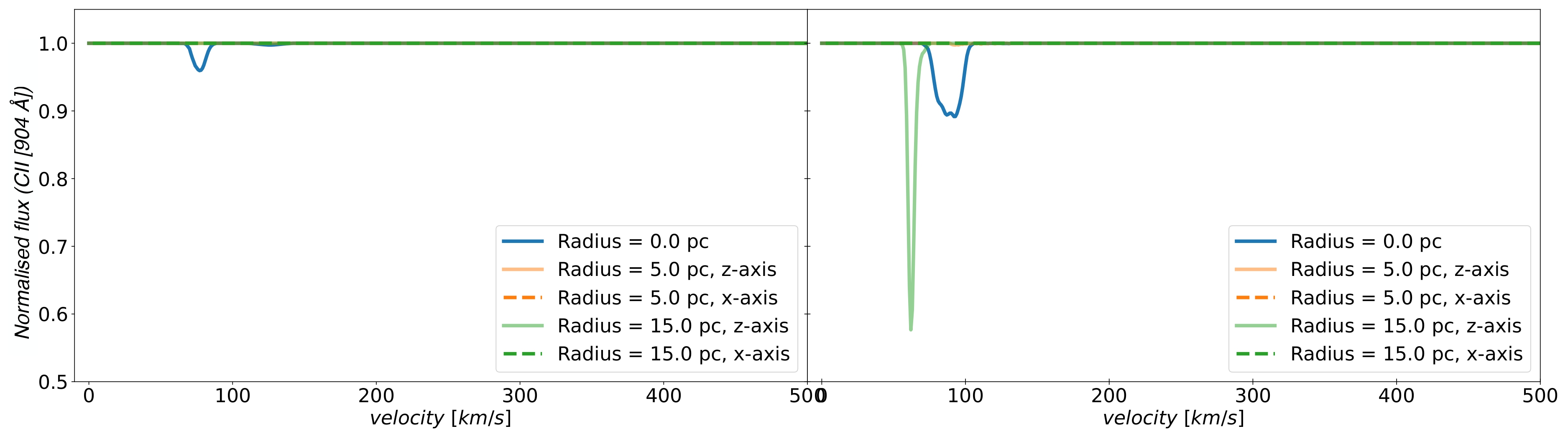}\\
    \includegraphics[width=\textwidth]{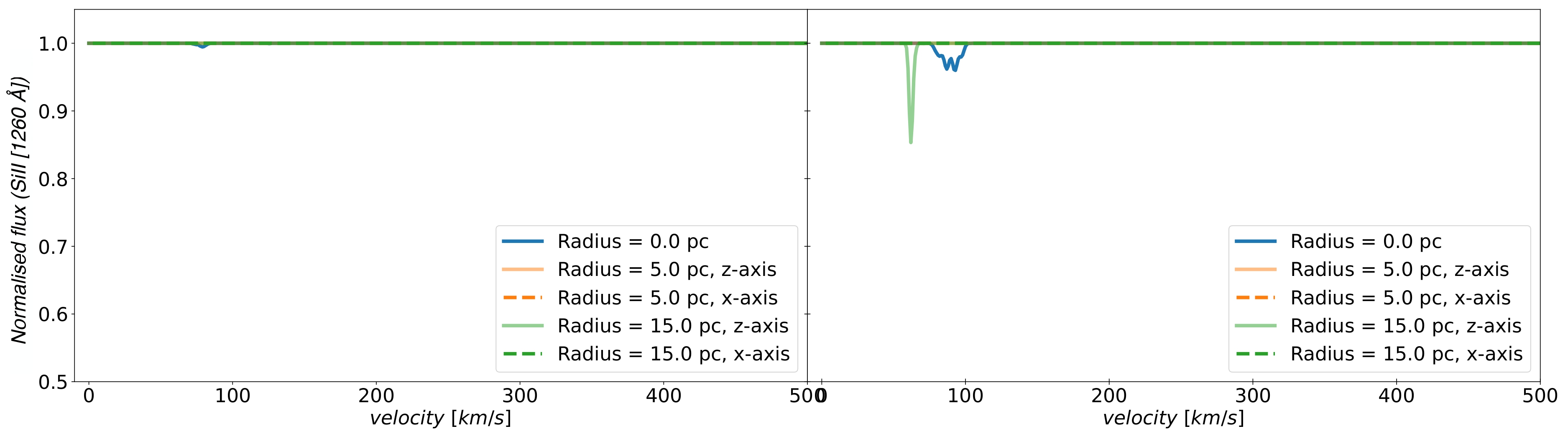}\\
  \end{tabular}
  \caption{Absorption spectra at $t/t_{\rm cc} = 1.7$ of ions tracing the cold phase of the gas generated with light rays passing through the centre (in blue), 5 pc from the centre along the $x$- (dashed orange line) and $z$-axis (solid orange line), and 15 pc from the centre along the $x$- (dashed green line) and $z$-axis (solid green line). In this figure, we display in the upper left panel the absorption profiles of the C\,{\sc ii} in the simulation R32-AL, in the upper right panel the C\,{\sc ii} in R32-TR, while the lower panels show the lines of the Si\,{\sc ii} in the simulation R32-AL (left) and R32-TR (right). We show that C\,{\sc ii} and Si\,{\sc ii} are tracers of the denser part of the cloud. These ions produce deeper absorption lines in the presence of a transverse magnetic field (see also Fig. \ref{fig:spectra_coldgas2} for $t/t_{\rm cc} = 1.2$). At $t/t_{\rm cc} = 1.7$, the differences are evident along the lines of sight observing the cloud centre and at 15 pc from the centre in the direction of the magnetic field, while at $t/t_{\rm cc} = 1.2$ the differences can be seen along the cloud centre and at 5 pc from it.} 
  \label{fig:spectra_coldgas}
\end{center}
\end{figure*}

Fig. \ref{fig:spectra_coldgas} shows the absorption lines of C\,{\sc ii} (upper panels) and Si\,{\sc ii} (bottom panels), for both simulations. At $t/t_{\rm cc} = 1.7$, in the aligned case, the gas reaches temperatures and densities such that the cold phase of the gas cannot produce any absorption lines, even in the central part of the cloud. On the other hand, when the magnetic field is transverse, both the C\,{\sc ii} and Si\,{\sc ii} located in the central part of the cloud produce deep lines. At $15$ pc from the centre, we can observe the effects in absorption of C\,{\sc ii} and Si\,{\sc ii} only along the $z$-axis. Along this direction, the cloud is not shielded, but rather becomes squeezed, by the magnetic field and is free to expand. For both ions, at this distance from the cloud centre the absorption line is less deep and formed by gas that is already accelerated by the wind at $v \sim 100$ $\rm km\,s^{-1}$, and more exposed to the hot wind. We can conclude that the effects of the transverse magnetic field on the state of the atomic cloud that we analysed in Sections \ref{General evolution}, \ref{B-fields} and \ref{Ion column densities and kinematics} also have a strong effect on the absorption spectra.\par

This result is confirmed by Fig. \ref{fig:spectra_coldgas2}, which shows similar absorption line trends for an earlier time ($t/t_{\rm cc} = 1.2$). Once an atomic cloud at T $\sim 10^4$ K interacts with a galactic wind with temperatures of $\sim 10^6$ K, observing the cold phase of the gas is only possible if the magnetic field is transverse to the direction of the wind. In addition, if these ions are detected, they are tracing only the densest and coldest regions of the wind-swept cloud.

\subsubsection{\texorpdfstring{Intermediate phase: C\,{\sc iv} and Si\,{\sc iv} absorption spectra}{Intermediate spectra}}

\begin{figure*}
\begin{center}
  \begin{tabular}{c c}
    \includegraphics[width=\textwidth]{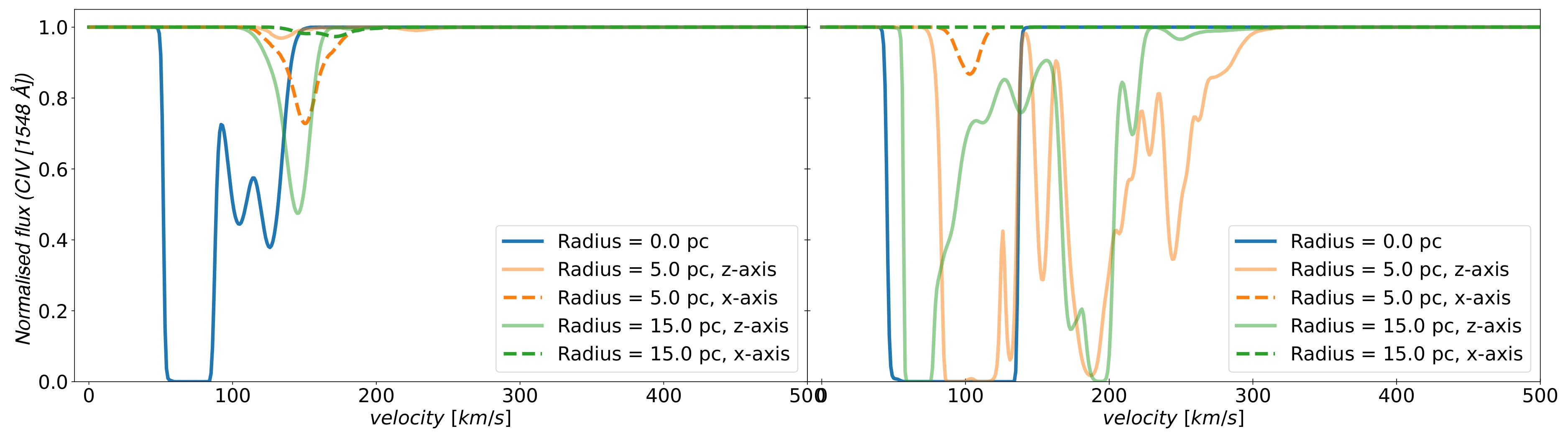}\\
    \includegraphics[width=\textwidth]{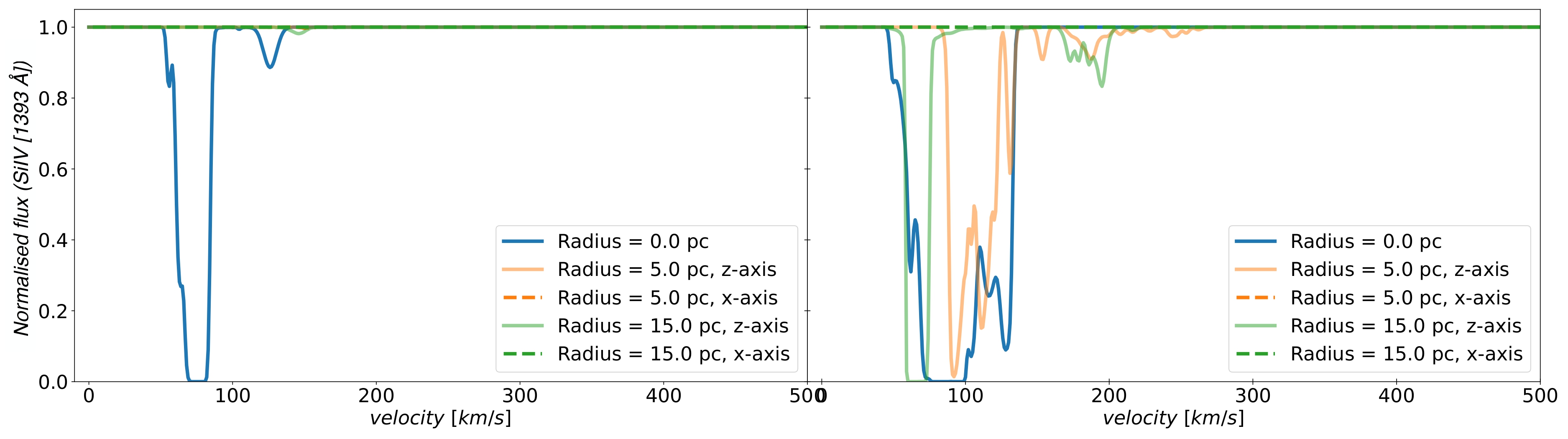}\\
  \end{tabular}
  \caption{Absorption spectra at $t/t_{\rm cc} = 1.7$ of ions tracing the intermediate-temperature gas, generated with light rays passing through the centre (in blue), 5 pc from the centre along the $x$- (dashed orange line) and $z$-axis (solid orange line), and 15 pc from the centre along the $x$- (dashed green line) and $z$-axis (solid green line). In this figure, we display in the upper left panel the absorption profiles of the C\,{\sc iv} in the simulation R32-AL, in the upper right panel the C\,{\sc iv} in R32-TR, while the lower panels show the lines of the Si\,{\sc iv} in the simulation R32-AL (left) and R32-TR (right). We note that a transverse magnetic field favours the production of broader and deeper lines of C\,{\sc iv} and Si\,{\sc iv}, compared to R32-AL. Fig. \ref{fig:spectra_intergas2} shows the same overall effects at an earlier time ($t/t_{\rm cc} = 1.2$), so intermediate ions are suitable candidates for tracing the effects of the underlying magnetic field morphology.} 
  \label{fig:spectra_intergas}
\end{center}
\end{figure*}

In Fig. \ref{fig:spectra_intergas} we plot spectra for the ions tracing the gas phase at $T=10^{4.5-5}$ K, showing the absorption lines of C\,{\sc iv} and Si\,{\sc iv}. In R32-AL, the evolution of the cloud is symmetric for both ions up to $15$ pc from the centre. Furthermore, in the left panels of Fig. \ref{fig:spectra_intergas}, both the C\,{\sc iv} and Si\,{\sc iv} create broader lines as the ray passes through the centre of the cloud in the $xz$ plane. Then, they become gradually narrower as one observes the outermost regions at $5$ and $15$ pc distances from the cloud centre. The broadening of the line generated by the ray passing through the centre is because we are observing gas belonging to both the cloud core and tail, which have high column densities and different velocities.\par

On the other hand, in R32-TR the evolution of the C\,{\sc iv} and Si\,{\sc iv} absorption lines is strongly dependent on the direction in which it is observed. If we focus on the right upper panel of Fig.\ \ref{fig:spectra_intergas}, when we observe at the centre of the cloud, the C\,{\sc iv} line is broad, saturated, and does not show substantial differences compared to R32-AL. However, if we move to the $5$ and $15$ pc points, the shape of the line is strongly affected by the different direction of the magnetic field. At $5$ pc from the centre the line generated by the gas along the $x$-axis is much weaker than along the $z$-axis. Along the $z$ direction, not only the line is saturated, but it is also very broad since it is generated by material spanning across a wide range of velocities. This occurs because we are observing different gas parcels, along the ray, tracing cloudlets moving along the squeezed tail of the cloud. The magnetic field squeezes the gas and accelerates it inwards, which consequently generates RT instabilities that create a significant dispersion in the velocity distribution of the cloud material. The same effect is more evident at $15$ pc, where along the $x$-axis the C\,{\sc iv} produces no profile in absorption, but along the $z$-axis it reaches velocities that span over $400$ $\rm km\,s^{-1}$.\par

In the bottom right panel of Fig. \ref{fig:spectra_intergas}, the Si\,{\sc iv} absorption lines are displayed for the R32-TR simulation. If we focus on the lines produced at $5$ and $15$ pc from the centre, we see that the gas traced by this ion produces a profile in absorption only along the $z$ direction. Si\,{\sc iv} traces slightly cooler and denser gas than C\,{\sc iv} consequently the effects described for C\,{\sc iv} are slightly less pronounced for this ion. This points towards a progressive effect of the transverse magnetic field on the ions with increasing temperature. As the temperature of the phase increases via wind-driven shock heating, the absorption lines in the direction of cloud expansion become increasingly deeper and broader. Overall, we find that model R32-TR favours the production of much broader and deeper lines of C\,{\sc iv} and Si\,{\sc iv} along the $z$-axis, compared to R32-AL (see also Fig. \ref{fig:spectra_intergas2} for $t/t_{\rm cc} = 1.2$). Thus, intermediate ions are suitable candidates for tracing the effects of differently oriented magnetic fields. 

\subsubsection{\texorpdfstring{Warm phase: N\,{\sc v} and O\,{\sc vi} absorption spectra}{Warm spectra}}

\begin{figure*}
\begin{center}
  \begin{tabular}{c c}
    \includegraphics[width=\textwidth]{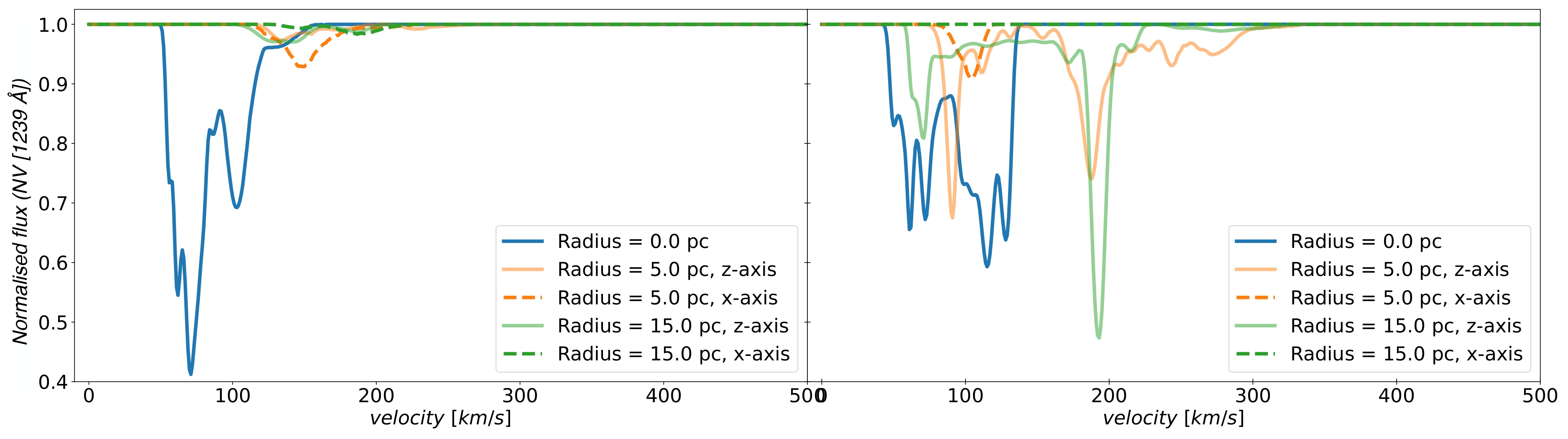}\\
    \includegraphics[width=\textwidth]{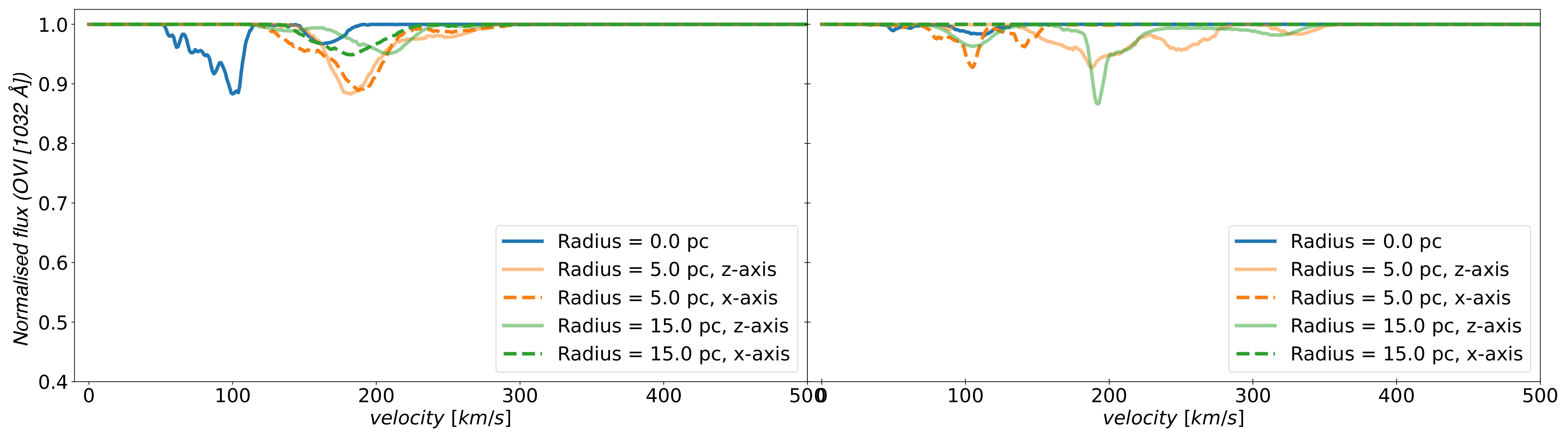}\\
  \end{tabular}
  \caption{Absorption spectra at $t/t_{\rm cc} = 1.7$ of ions tracing the warm phase of the gas generated with light rays passing through the centre (in blue), at 5 pc from the centre along the $x$- (dashed orange line) and $z$-axis (solid orange line), at 15 pc from the centre along the $x$- (dashed green line) and $z$-axis (solid green line). In this figure, we display in the upper left panel the absorption profiles of the N\,{\sc v} in the simulation R32-AL, in the upper right panel the N\,{\sc v} in R32-TR. while the lower panels show the lines of the O\,{\sc vi} in the simulation R32-AL (left) and R32-TR (right). We highlight that N\,{\sc v} and O\,{\sc vi} absorption lines are more complex than the other gas phases (also at an earlier time, see Fig. \ref{fig:spectra_warmgas2} for $t/t_{\rm cc} = 1.2$). Warm ions trace the outer and most turbulent layers of the cloud and wind material, so they show deeper lines in R32-AL (whose gas is more mixed and turbulent) than in R32-TR, but they are less sensitive to the $\bf{B}$ orientation.} 
  \label{fig:spectra_warmgas}
\end{center}
\end{figure*}

Fig. \ref{fig:spectra_warmgas} shows the absorption profiles of N\,{\sc v} in the top two panels and O\,{\sc vi} in the bottom two panels, again the left plots refer to the R32-AL simulation and the right ones to R32-TR. For both ions, the spectral line profile is much more complex than in the other phases. The reason for this is that, N\,{\sc v} and O\,{\sc vi} are tracers of highly mixed and turbulent gas at the interface between the cloud outer layers and the hot wind. Thus, we see lines of varying widths and depths at a wide range of velocities. This makes isolating strong features associated with the orientation of the magnetic field harder than in the other phases.\par 

However, one interesting feature we can infer from observing  N\,{\sc v} is that when the magnetic field is aligned with the wind direction, the cloud is more substantially mixed than in the transverse case. This is why, in the case of a spectra made with the ray passing through the centre, we observe a deeper line in the left plot compared to the right one. The same effect is also visible for O\,{\sc vi} albeit in a smaller way, since O\,{\sc vi} is also very much present in the wind and has smaller column densities. As a result, we conclude that it is more difficult to isolate the effects of the magnetic field with the warm phase tracers, N\,{\sc v} and O\,{\sc vi}. However, both ions are sensitive to mixing fraction and turbulence, which is more significant in the aligned field model. For this reason, they tend to produce deeper lines when the magnetic field is aligned, compared to the case with a transverse field. When looking into an earlier time, $t/t_{\rm cc} = 1.2$,  we see the same effects described above (see Fig. \ref{fig:spectra_warmgas2}).













\section{Discussion}
\label{Discussion}

\subsection{Spectral time evolution}
\label{Spectral_time_evolution}



\begin{figure*}
\begin{center}
  \begin{tabular}{c c}
    \includegraphics[width=\textwidth]{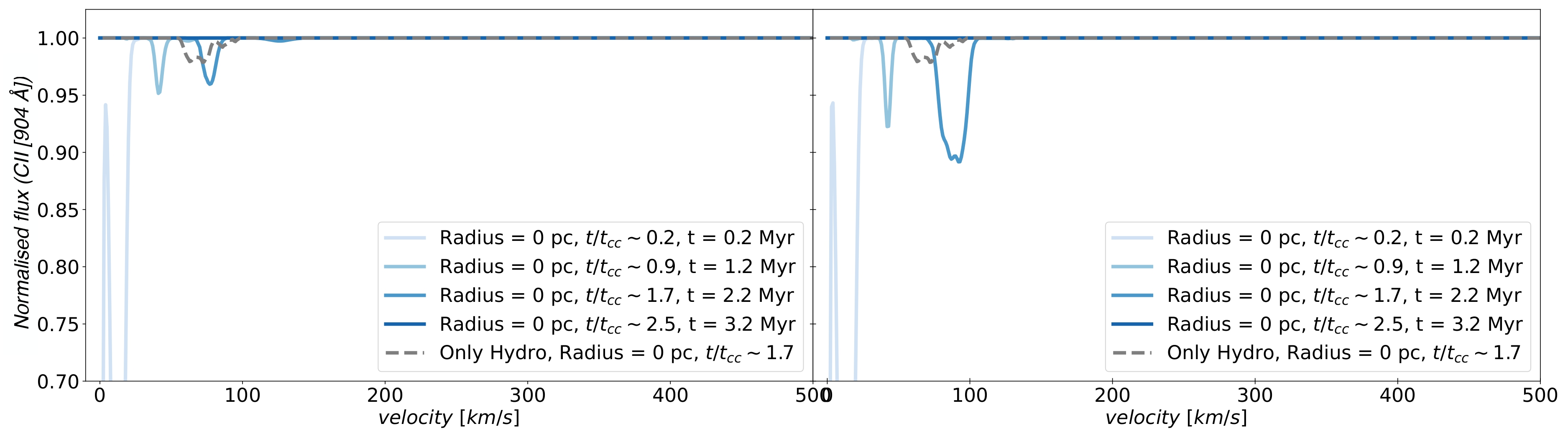}\\
    \includegraphics[width=\textwidth]{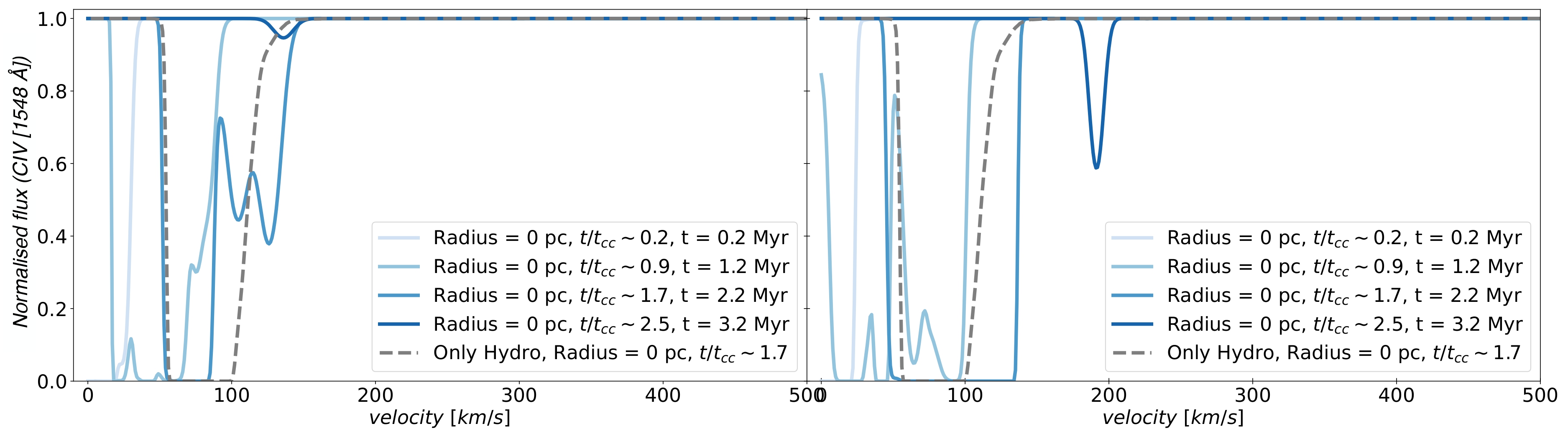}\\
    \includegraphics[width=\textwidth]{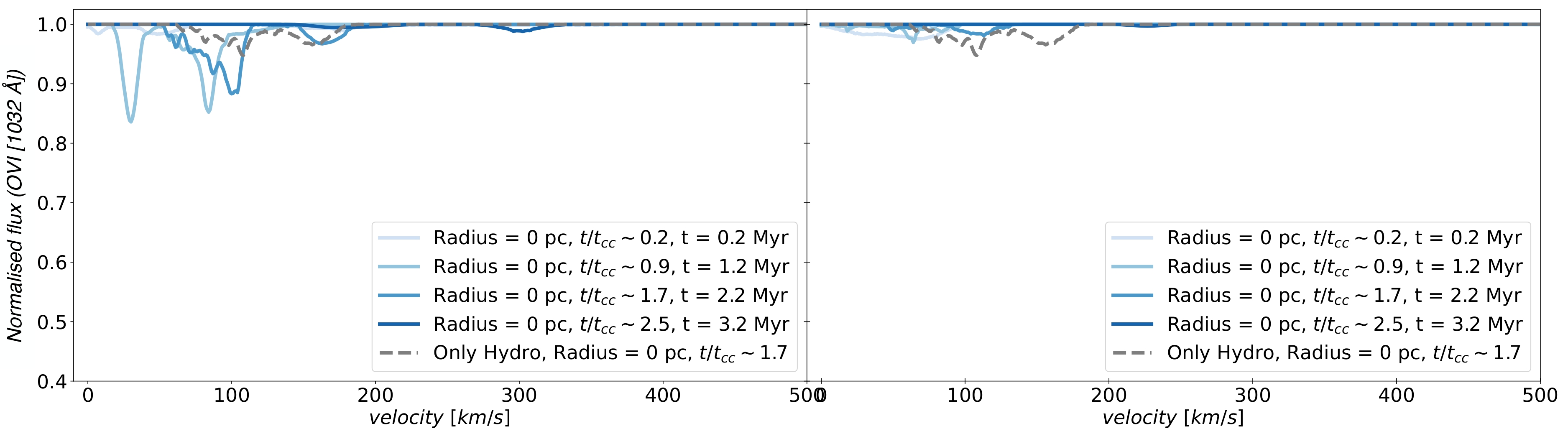}\\
  \end{tabular}
  \caption{Temporal evolution of the absorption spectra of one ion for each gas phase, generated with a light ray passing through the centre of the cloud. The absorption lines are generated at $t = 0.2, 1.2, 2.2, 3.2$ Myr (or $t/t_{\rm cc} = 0.2, 0.9, 1.7, 2.5$) respectively. We display C\,{\sc ii} as representative ion for the cold phase (top), C\,{\sc iv} for the intermediate temperature one (middle), and O\,{\sc vi} for the warm gas (bottom). As in Section \ref{subsec:spectral_analysis}, on the left side we show the simulation R32-AL, with an aligned magnetic field and on the right R32-TR, with a transverse magnetic field. We see that in R32-TR the gas reaches overall higher velocities than R32-AL at all times, particularly for the intermediate ions, which show velocities of $100-200$ km s$^{-1}$ at $t/t_{\rm cc} = 1.7$. We also stress that the intermediate gas phase is the one that shows broader and deeper lines at all times when the magnetic field is transverse.}
  \label{fig:spectra_timeev}
\end{center}
\end{figure*}


In Fig. \ref{fig:spectra_timeev} we display the time evolution of the spectra at the same times reported in Section \ref{Results}, for the three different gas phases: C\,{\sc ii} for the cold one, C\,{\sc iv} for the intermediate phase, and O\,{\sc vi} for the warm gas. In particular, this is the time evolution of the spectrum made with a light ray passing through the cloud centre. This direction is the one that shows better the differences among the two models, since it is the one where we expect to find the majority of the cloud material. Overall, in all the panels it is possible to see that every ion is accelerated during the wind-cloud interaction and reaches velocities of $100-200\,\rm km\,s^{-1}$ at $t/t_{\rm cc} \sim 1.7$. Moreover, it is at $t/t_{\rm cc} \sim 1.7$ that the difference among the two magnetic field configurations is more visible, in particular for C\,{\sc iv}. C\,{\sc ii} produces a strong absorption line in the initial stages, but already at $t/t_{\rm cc} \sim 0.9$ the absorption in the cloud centre is reduced. In both models with aligned and transverse magnetic fields, the C\,{\sc ii} ion reaches $\sim 100\,\rm km\,s^{-1}$, and it does not show an evident difference among the two models. Finally, the cold ion is not visible anymore at $t/t_{\rm cc} \sim 2.5$. C\,{\sc iv} is the ion that gets accelerated the most, as already shown also in Fig. \ref{fig:spectra_intergas}, and it is also the ion that shows the biggest difference among the two magnetic field orientations. Already at $t/t_{\rm cc} \sim 0.9$ the broadening of the absorption line is larger in R32-TR. Then, in the later stages C\,{\sc iv} is sped up to 200 $\rm km\,s^{-1}$ and the absorption line is still visible at $t/t_{\rm cc} \sim 2.5$. This proves how a transverse magnetic field prevents gas at $10^{4.5-5}$ K from full disruption and makes it detectable even after a few $\rm Myr$ from the beginning of the wind-cloud interaction. O\,{\sc vi} does not produce deep absorption lines during the entire cloud evolution. Even if, when the magnetic field is aligned with the direction of the wind, the lines are slightly deeper. Overall, it is difficult to draw major conclusions from this ion.\par

In each panel we show in grey the comparison with the only hydrodynamical simulatio (R32-HD) at $t/t_{\rm cc} \sim 1.7$. The major differences are the ones arising from the comparison with R32-TR, especially for C\,{\sc ii} and C\,{\sc iv}. R32-TR produces deeper and broader lines. This effect can be explained by magnetic field draping, which shields cold gas and takes intermediate temperature gas along with it. Generally the run without magnetic fields produces absorption lines very similar to R32-AL for C\,{\sc ii} and O\,{\sc vi}. This result enforces that an aligned magnetic field has a weak influence on the cloud evolution, especially for gas in the cold and warm phases.

\subsection{Implications to CGM observations}

Our analysis demonstrates that magnetic fields have distinct effects on different gas phases, such as, in the cold phase, we find no signature of C\,{\sc ii} and Si\,{\sc ii} when the field is aligned; in the intermediate phase, traced by C\,{\sc iv} and Si\,{\sc iv}, we find broader lines in the transverse case; and in the warm phase, we find deeper lines for O\,{\sc vi} and N\,{\sc v} in the aligned case (albeit, the lines are less sensitive overall to the field orientation). The main questions are now whether these findings can help explain the presence or lack of ions in particular observed winds and whether the imprint of magnetic fields can be detected in observational work.\par

Given the complexity of the multi-phase gas in the CGM of galaxies and the many variables at play, one-to-one quantitative comparisons between our simulations and observational studies are not really possible. Instead, we focus on qualitative comparisons, particularly to those CGM environments that are better resolved by spectroscopy in our Galaxy and the local Universe. The initial conditions in our models are also mainly motivated by such environments, so we discuss our results in the context of ion detections in the MW halo via the Cosmic Origins Spectrograph (COS) onboard the $Hubble$ space telescope (e.g. \citealt{2017A&A...607A..48R}).\par

First, we compare the column number densities of our models with typical observed values. \cite{2017A&A...607A..48R} show column densities as low as $N_{\rm CII}\sim 10^{13}\,\rm cm^{-2}$ and $N_{\rm SiII}\sim10^{12}\,\rm cm^{-2}$ for low ions. We show that such values can certainly arise in CGM wind-cloud interactions, but are more likely to produce absorption lines at early stages of such interactions. We also find that such ions reside mainly in the vicinity of the cores of the clouds. Once cores are disrupted, column densities fall below detectable limits (e.g. $N_{\rm SiII}<10^{12.3}$, see table 2 in \citealt{2017A&A...607A..48R}) and produce weak lines. In terms of magnetic field effects, transverse $\bf B$ configurations protect better low ions than their aligned counterparts. For C\,{\sc iv} \cite{2017A&A...607A..48R} report column densities of $N_{\rm CIV}\sim10^{12.5-14.5}\,\rm cm^{-2}$ and for Si\,{\sc iv} \cite{2019ApJ...871...35Z} report $N_{\rm SiIV}\sim10^{13.5}\,\rm cm^{-2}$, which are also comparable to those found in this paper. Unlike low ions, our models show that the intermediate-phase initially produces strong lines, and it becomes quite long-lived as the wind-cloud interaction reaches the shredding phase. Magnetic fields also leave clearer imprints in the form of deeper and broader absorption lines when the $\bf B$ field is transverse to the wind. As for higher ions, \cite{2003ApJS..146....1W} reported detections of O\,{\sc vi} in the MW halo. They report column densities $N_{\rm OVI}\sim 10^{14}\,\rm cm^{-2}$, which are on the upper limits of what we detect in our models. This is reasonable as high ions in wind-cloud models reside in the envelopes of clouds, which are well mixed with the background wind. In terms of the $\bf B$ field orientation, slightly stronger lines are detected in the aligned case as that model shows a larger degree of mixing.\par

Second, we compare the velocities of different gas phases. For the MW CGM, \cite{2017A&A...607A..48R} report typical velocities between $-400$ and $400\,\rm km\,s^{-1}$ for low and intermediate ions (Si\,{\sc ii}, Si\,{\sc iii}, C\,{\sc ii}, and C\,{\sc iv}). Similarly, \cite{2003ApJS..146....1W} map O\,{\sc vi} absorption in a broad range of velocities ($\leq 1200\,\rm km\,s^{-1}$), but finds central velocities only between $\sim 100-400\,\rm km\,s^{-1}$. These ranges are qualitatively within the typical ion velocities that we observe in our simulations, but again it is very likely that observations trace several clouds along sight-lines and thus a much more complex flow than what our single-cloud simulations can capture. Despite this, the fact that the ion velocity ranges are similar indicates that our initial conditions are appropriate to describe such CGM clouds. One difference we notice is that \cite{2017A&A...607A..48R} report broader and deeper spectral lines for the low ions than for intermediate ions (see their Fig. 1). As discussed before, that only occurs in the early stages of wind-cloud evolution, when the denser gas is still largely intact. This may indicate that in the MW CGM, there is some kind of replenishment process in action, which is responsible for maintaining cold gas. Dense-gas re-condensation is definitely viable as recent simulation work suggests (\citealt{2020MNRAS.492.1970G}), provided that the mixed phase can cool down quickly behind the cloud. Studying spectral lines in mass-growth scenarios (\citealt{2010MNRAS.404.1464M,2021MNRAS.501.1143K}) and more complex multi-cloud systems (\citealt{2021MNRAS.506.5658B}) would be an interesting avenue to explore in future work.\par

Our work can also be qualitatively compared to extragalactic sources. For instance, \cite{2018MNRAS.474.1688C} studied O\,{\sc vi} absorption in 6 MEGASAURA galaxies (\citealt{2018AJ....155..104R}), including a $z=2.9$ lensed galaxy, SGAS J122651.3+215220. The detected O\,{\sc vi} spectral line in the CGM of this galaxy displays velocities in the range $200-500\,\rm km\,s^{-1}$, similar to our synthetic ion velocities, but much larger column densities, $N_{\rm OVI}\sim10^{15}\,\rm cm^{-2}$, than ours for this ion. This discrepancy points to a higher wind temperature for that system, which the authors estimate to be $T_{\rm wind}>10^7\,\rm K$. The authors of this paper also pointed out N\,{\sc v} absorption was absent from the observations (see also \citealt{2016MNRAS.457.3133C}), despite the fact that this ion resides in conditions similar to O\,{\sc vi}. Albeit for distinct wind conditions, we notice differences in the spectra for N\,{\sc v} absorption in our models. Overall the spectra of these ions are complex and time dependent. In the R32-AL model, N\,{\sc v} can only be clearly detected in direct pointings to the cloud centre at $t/t_{\rm cc}=1.7$, while in the R32-TR model this ion can be detected at different radii. This would imply that transverse fields would favour N\,{\sc v} conditions and make it more detectable, but at $t/t_{\rm cc}=1.2$ (see Appendix \ref{AppB}), N\,{\sc v} has much deeper lines in the aligned case. Thus, additional simulations with higher wind velocities and temperatures are needed to study this ion in a wider parameter space.\par

Another extragalactic study that reports ion column densities and velocities is \cite{2015ApJ...809..147H}, who used stacking techniques to study C\,{\sc ii}, Si\,{\sc ii}, Si\,{\sc iii}, Si\,{\sc iv}, and N\,{\sc ii} in a sample of low-redshift starburst galaxies. They estimated column densities of $N \sim 10^{13-16}\,\rm cm^{-2}$ and outflow speeds of $v\sim 100-500\,\rm km\,s^{-1}$ for these ions. While direct one-to-one comparisons cannot be established given the vastly different scales between the observations and our models, we notice that their observed column densities are a little higher than in our models. This is reasonable as they are detecting the contributions of much more complex multicloud outflows. In this context, MHD wind-multicloud models would reveal substantially more features, than single-cloud models, for extragalactic sources. Going back to our questions, our models show that magnetic fields have tangible effects on high-resolution absorption spectra. As they alter the morphology of wind-swept clouds, they favour stronger lines in transverse-field models and can make other lines weaker, which all adds up to the complexity of galactic wind spectroscopy.


\subsection{Comparison to previous numerical work}

This paper investigates, for the first time, the effect of magnetic fields and their orientation on spectral lines produced by wind-cloud models. Other authors have already simulated synthetic spectra of atomic clouds interacting with galactic winds, but only in hydrodynamic cases. Comparison with other numerical work is then complicated as the initial conditions, numerical codes, and the way the spectra are generated are different. Despite this, we think it is important to comment on how our analysis compares to others in the literature. Some examples are \citealt{2021ApJ...919..112D} and \citealt{2018ApJ...864...96C}, who present synthetic absorption lines and velocity distributions of the same ions used for our analysis.\par

In \citealt{2021ApJ...919..112D}, the authors perform a simulation of a cloud with turbulent mass distribution and radiative cooling with a cooling floor at $\sim 10\,\rm K$. They report velocity distributions and covering fractions of the same ions investigated in our paper, at a single intermediate time between the moment of maximum cloud compression and complete destruction. By comparing Figures \ref{fig:cii_hist}, \ref{fig:civ_hist} and \ref{fig:ovi_hist} with their Table \ref{Table2}, we see that results show general agreement with their trends. The C\,{\sc ii} velocity histograms peak at lower velocities than the ions tracing the hottest gas phases. They also find that the ions representing the warm phase of the gas, such as N\,{\sc v} and O\,{\sc vi}, peak at higher velocities, almost 200 $\rm km\,s^{-1}$, as in the later stages of our clouds. This result indicates that even in absence of magnetic fields, if the gas can cool down more than $10^4$ K, the coldest ions remain the most difficult to accelerate. In addition, the covering fractions found by us are larger than theirs, due to our larger cloud size compared to the box dimension.\par

Finally, \citealt{2018ApJ...864...96C} uses a technique based on TRIDENT to calculate how thermal conduction affects the absorption lines of ions such as: H\,{\sc i}, Mg\,{\sc ii}, C\,{\sc ii}, C\,{\sc iii}, C\,{\sc iv}, Si\,{\sc iii}, Si\,{\sc iv}, N\,{\sc v} and O\,{\sc vi}. They have a cooling floor set at $10^4$ K and use a UV background consistent with the cloud's proximity to a star-forming region. Accordingly to them, O\,{\sc vi} is tracing both the cloud and the wind material, making its absorption profile broad and shallow. Overall, their absorption lines are deeper and broader than ours, since they generate spectra by combining information from the cells with higher densities, instead of looking at single lines of sight as we do. For this reason, although we do not reach their speeds in the synthetic spectra, our velocities in Figures \ref{fig:cii_hist}, \ref{fig:civ_hist} and \ref{fig:ovi_hist} seem to agree with theirs, especially in the model R32-TR.

\subsection{Caveats and limitations}
\label{subsec:caveats}
While our simulations show new important insights into the role of magnetic fields in shaping the physics and observables of CGM clouds, we discuss some caveats and limitations of our work below.\par

First, we have not studied models with thermal conduction (e.g. see \citealt{2021MNRAS.502.1263K,2023ApJ...951..113B}). Previous hydrodynamical simulations with conduction have showed that in high-column density clouds, like the ones we model here, thermal conduction leads to very thin, cold filaments that seem detached from the outside ambient gas (\citealt{2016ApJ...822...31B}). The outer layers of the clouds evaporate, but not the cores which manage to radiate the conducted energy away. In such scenarios, we would then expect low ions to display stronger absorption lines while high ions may be highly absent as they normally reside in $>10^{5.5}\,\rm K$ gas (which would not be present). Second, we have not explored the effects of stronger magnetic fields (i.e. lower plasma betas) on the ion spectral lines. \cite{2016MNRAS.455.1309B} showed that reducing the plasma beta from $\beta=100$ to $10$ in oblique field models have two main effects. On the one hand, a stronger field reduces the erosive effects of KH instabilities, leading to enhanced shielding and a more laminar flows around the cloud (see also \citealt{2018ApJ...865...64G}), while on the other hand it promotes RT instabilities, leading to an earlier cloud break-up (see also \citealt{2020ApJ...892...59C}). Considering these effects, we would then expect deeper spectral lines, particularly for the low ions, in strong-field cases.\par

Other limitations of our simulations, which we plan to systematically address in the future, include studying winds with higher winds and temperatures (\citealt{2018ApJ...864...96C}), clouds with more realistic initial density distributions (\citealt{2009ApJ...703..330C,2019MNRAS.486.4526B}), and projection effects. In our current simulations, our wind and clouds have fixed parameters, and we only produce down-the-barrel spectra, but there is of course room for extending the parameter space, the number of sight-lines, and the view angles. The reader is referred to Appendix \ref{AppA} for details on numerical resolution effects, Appendix \ref{AppB} for spectral resolution effects, and Appendix \ref{AppC} for the time dependence of spectra.


\section{Conclusions}
\label{Conclusions}


We find that the initial orientation of magnetic fields have significant effects on the evolution of wind-swept clouds. To study the influence of two orientations of the magnetic fields (aligned and transverse), we carried out a suite of 3D magnetohydrodynamical simulations of wind-cloud systems. To extract synthetic spectra from such simulations, we developed a framework that links PLUTO simulations to TRIDENT via the yt-package infrastructure. We also incorporated a UV background that takes into account stellar feedback using the STARBURST99 and CLOUDY packages.\par

We investigated the kinematics and spectra of ions tracing three different phases of the gas (cold via C\,{\sc ii}  and Si\,{\sc ii}, intermediate via C\,{\sc iv}  and Si\,{\sc iv}, warm via O\,{\sc vi} and N\,{\sc v}), and we found the following:

\begin{enumerate}
    \item Influence of magnetic fields on wind-swept clouds: We find that the initial transverse magnetic field makes the cloud asymmetric, shields and protects dense cold gas, and reduces mixing fractions compared to the aligned case. This configuration favours magnetic draping, which squeezes the cloud in one direction, expands it perpendicularly, and makes the overall flow more laminar.
    \item Imprints of magnetic fields on ion kinematics: By studying the velocity distributions of different ions, we find that a) the warmer the phase, the higher velocities the gas reaches, b) gas  can reach overall higher velocities in the transverse field case, and c) the latter effect is more visible in the intermediate phase.
    \item Magnetic field signatures on the spectra of the cold phase: We find that the cold phase is only detectable when the magnetic field is transverse to the wind as this configuration protects the core of the cloud via shielding. We find no signature of C\,{\sc ii} and Si\,{\sc ii} when the field is aligned.
    \item Magnetic field signatures on the spectra of the intermediate phase: Overall we find that the transverse field configuration favours the emergence of deeper and broader spectral lines compared to the aligned case. The deeper lines are associated with shielding effects, while the broader lines are due to gas escaping along the squeezed tail owing to Rayleigh-Taylor instabilities. C\,{\sc iv} and Si\,{\sc iv} are good indicators of the magnetic field morphology.
    \item Magnetic field signatures on the spectra of the warm phase: This phase traces the outer layers of the clouds, which are dominated by turbulence generated by Kelvin-Helmholtz instabilities. This makes the spectra much more complex and time dependent than in the other phases, so these ions are not optimal tracers of the underlying magnetic fields. A distinctive feature that we find is that N\,{\sc v} has overall deeper lines in the aligned case, but can only be detected along the cloud centre at late times.
\end{enumerate}

In summary, we find that magnetic fields have clear observational signatures on down-the-barrel absorption spectra. To disentangle these imprints, intermediate ions show the clearest effects. We thus believe that pursuing future studies on detailed comparisons between observations and synthetic spectra obtained from simulations are essential to improve our understanding on the CGM. In future work, we will look into studying other magnetic field configurations, spectra in wind-multicloud models, the effects of different UV backgrounds, and mapping more ions to the sample.

\begin{acknowledgements}
We thank the anonymous referee for a constructive report. The authors gratefully acknowledge the Gauss Centre for Supercomputing e.V. (\url{www.gauss-centre.eu}) for funding this project by providing computing time (via grant pn34qu) on the GCS Supercomputer SuperMUC-NG at the Leibniz Supercomputing Centre (\url{www.lrz.de}). WEBB is supported by the National Secretariat of Higher Education, Science, Technology, and Innovation of Ecuador, SENESCYT.  We also thank the developers of the PLUTO code, the TRIDENT, CLOUDY, CLOUDY cooling tools, and the STARBURST99 packages for making them available to the community and the members of the FOGGIE collaboration for providing us with an updated version of ion fraction table. ES was supported in part by NASA grant 80NSSC23K0646.
\end{acknowledgements}

   


\bibliographystyle{aa}
\bibliography{wbandabarragan}

\begin{thebibliography}{103}
\expandafter\ifx\csname natexlab\endcsname\relax\def\natexlab#1{#1}\fi

\bibitem[{{Banda-Barrag{\'a}n} {et~al.}(2020){Banda-Barrag{\'a}n}, {Br{\"u}ggen}, {Federrath}, {Wagner}, {Scannapieco}, \& {Cottle}}]{2020MNRAS.499.2173B}
{Banda-Barrag{\'a}n}, W.~E., {Br{\"u}ggen}, M., {Federrath}, C., {et~al.} 2020, \mnras, 499, 2173

\bibitem[{{Banda-Barrag{\'a}n} {et~al.}(2021){Banda-Barrag{\'a}n}, {Br{\"u}ggen}, {Heesen}, {Scannapieco}, {Cottle}, {Federrath}, \& {Wagner}}]{2021MNRAS.506.5658B}
{Banda-Barrag{\'a}n}, W.~E., {Br{\"u}ggen}, M., {Heesen}, V., {et~al.} 2021, \mnras, 506, 5658

\bibitem[{{Banda-Barrag{\'a}n} {et~al.}(2018){Banda-Barrag{\'a}n}, {Federrath}, {Crocker}, \& {Bicknell}}]{2018MNRAS.473.3454B}
{Banda-Barrag{\'a}n}, W.~E., {Federrath}, C., {Crocker}, R.~M., \& {Bicknell}, G.~V. 2018, \mnras, 473, 3454

\bibitem[{{Banda-Barrag{\'a}n} {et~al.}(2016){Banda-Barrag{\'a}n}, {Parkin}, {Federrath}, {Crocker}, \& {Bicknell}}]{2016MNRAS.455.1309B}
{Banda-Barrag{\'a}n}, W.~E., {Parkin}, E.~R., {Federrath}, C., {Crocker}, R.~M., \& {Bicknell}, G.~V. 2016, \mnras, 455, 1309

\bibitem[{{Banda-Barrag{\'a}n} {et~al.}(2019){Banda-Barrag{\'a}n}, {Zertuche}, {Federrath}, {Garc{\'\i}a Del Valle}, {Br{\"u}ggen}, \& {Wagner}}]{2019MNRAS.486.4526B}
{Banda-Barrag{\'a}n}, W.~E., {Zertuche}, F.~J., {Federrath}, C., {et~al.} 2019, \mnras, 486, 4526

\bibitem[{{Bautista} {et~al.}(2009){Bautista}, {Quinet}, {Palmeri}, {Badnell}, {Dunn}, \& {Arav}}]{2009A&A...508.1527B}
{Bautista}, M.~A., {Quinet}, P., {Palmeri}, P., {et~al.} 2009, \aap, 508, 1527

\bibitem[{{Ben Bekhti} {et~al.}(2008){Ben Bekhti}, {Richter}, {Westmeier}, \& {Murphy}}]{2008A&A...487..583B}
{Ben Bekhti}, N., {Richter}, P., {Westmeier}, T., \& {Murphy}, M.~T. 2008, \aap, 487, 583

\bibitem[{{Bland-Hawthorn} \& {Cohen}(2003)}]{2003ApJ...582..246B}
{Bland-Hawthorn}, J. \& {Cohen}, M. 2003, \apj, 582, 246

\bibitem[{{Bolatto} {et~al.}(2013){Bolatto}, {Warren}, {Leroy}, {Walter}, {Veilleux}, {Ostriker}, {Ott}, {Zwaan}, {Fisher}, {Weiss}, {Rosolowsky}, \& {Hodge}}]{2013Natur.499..450B}
{Bolatto}, A.~D., {Warren}, S.~R., {Leroy}, A.~K., {et~al.} 2013, \nat, 499, 450

\bibitem[{{Br{\"u}ggen} \& {Scannapieco}(2016)}]{2016ApJ...822...31B}
{Br{\"u}ggen}, M. \& {Scannapieco}, E. 2016, \apj, 822, 31

\bibitem[{{Br{\"u}ggen} \& {Scannapieco}(2020)}]{2020ApJ...905...19B}
{Br{\"u}ggen}, M. \& {Scannapieco}, E. 2020, \apj, 905, 19

\bibitem[{{Br{\"u}ggen} {et~al.}(2023){Br{\"u}ggen}, {Scannapieco}, \& {Grete}}]{2023ApJ...951..113B}
{Br{\"u}ggen}, M., {Scannapieco}, E., \& {Grete}, P. 2023, \apj, 951, 113

\bibitem[{{Carretti} {et~al.}(2013){Carretti}, {Crocker}, {Staveley-Smith}, {Haverkorn}, {Purcell}, {Gaensler}, {Bernardi}, {Kesteven}, \& {Poppi}}]{2013Natur.493...66C}
{Carretti}, E., {Crocker}, R.~M., {Staveley-Smith}, L., {et~al.} 2013, \nat, 493, 66

\bibitem[{{Casavecchia} {et~al.}(2023){Casavecchia}, {Banda-Barrag{\'a}n}, {Br{\"u}ggen}, \& {Brighenti}}]{2023IAUS..362...56C}
{Casavecchia}, B., {Banda-Barrag{\'a}n}, W.~E., {Br{\"u}ggen}, M., \& {Brighenti}, F. 2023, IAU Symposium, 362, 56

\bibitem[{{Cashman} {et~al.}(2017){Cashman}, {Kulkarni}, {Kisielius}, {Ferland}, \& {Bogdanovich}}]{2017ApJS..230....8C}
{Cashman}, F.~H., {Kulkarni}, V.~P., {Kisielius}, R., {Ferland}, G.~J., \& {Bogdanovich}, P. 2017, \apjs, 230, 8

\bibitem[{{Chisholm} {et~al.}(2018){Chisholm}, {Bordoloi}, {Rigby}, \& {Bayliss}}]{2018MNRAS.474.1688C}
{Chisholm}, J., {Bordoloi}, R., {Rigby}, J.~R., \& {Bayliss}, M. 2018, \mnras, 474, 1688

\bibitem[{{Chisholm} {et~al.}(2016){Chisholm}, {Tremonti}, {Leitherer}, {Chen}, \& {Wofford}}]{2016MNRAS.457.3133C}
{Chisholm}, J., {Tremonti}, C.~A., {Leitherer}, C., {Chen}, Y., \& {Wofford}, A. 2016, \mnras, 457, 3133

\bibitem[{{Cooper} {et~al.}(2008){Cooper}, {Bicknell}, {Sutherland}, \& {Bland-Hawthorn}}]{2008ApJ...674..157C}
{Cooper}, J.~L., {Bicknell}, G.~V., {Sutherland}, R.~S., \& {Bland-Hawthorn}, J. 2008, \apj, 674, 157

\bibitem[{{Cooper} {et~al.}(2009){Cooper}, {Bicknell}, {Sutherland}, \& {Bland-Hawthorn}}]{2009ApJ...703..330C}
{Cooper}, J.~L., {Bicknell}, G.~V., {Sutherland}, R.~S., \& {Bland-Hawthorn}, J. 2009, \apj, 703, 330

\bibitem[{{Corbelli} {et~al.}(2017){Corbelli}, {Braine}, {Bandiera}, {Brouillet}, {Combes}, {Druard}, {Gratier}, {Mata}, {Schuster}, {Xilouris}, \& {Palla}}]{2017A&A...601A.146C}
{Corbelli}, E., {Braine}, J., {Bandiera}, R., {et~al.} 2017, \aap, 601, A146

\bibitem[{{Cottle} {et~al.}(2018){Cottle}, {Scannapieco}, \& {Br{\"u}ggen}}]{2018ApJ...864...96C}
{Cottle}, J., {Scannapieco}, E., \& {Br{\"u}ggen}, M. 2018, \apj, 864, 96

\bibitem[{{Cottle} {et~al.}(2020){Cottle}, {Scannapieco}, {Br{\"u}ggen}, {Banda-Barrag{\'a}n}, \& {Federrath}}]{2020ApJ...892...59C}
{Cottle}, J., {Scannapieco}, E., {Br{\"u}ggen}, M., {Banda-Barrag{\'a}n}, W., \& {Federrath}, C. 2020, \apj, 892, 59

\bibitem[{{Crocker}(2012)}]{2012MNRAS.423.3512C}
{Crocker}, R.~M. 2012, \mnras, 423, 3512

\bibitem[{{Danehkar} {et~al.}(2021){Danehkar}, {Oey}, \& {Gray}}]{2021ApJ...921...91D}
{Danehkar}, A., {Oey}, M.~S., \& {Gray}, W.~J. 2021, \apj, 921, 91

\bibitem[{{Danehkar} {et~al.}(2022){Danehkar}, {Oey}, \& {Gray}}]{2022ApJ...937...68D}
{Danehkar}, A., {Oey}, M.~S., \& {Gray}, W.~J. 2022, \apj, 937, 68

\bibitem[{{Das} \& {Gronke}(2024)}]{2024MNRAS.527..991D}
{Das}, H.~K. \& {Gronke}, M. 2024, \mnras, 527, 991

\bibitem[{{de la Cruz} {et~al.}(2021){de la Cruz}, {Schneider}, \& {Ostriker}}]{2021ApJ...919..112D}
{de la Cruz}, L.~M., {Schneider}, E.~E., \& {Ostriker}, E.~C. 2021, \apj, 919, 112

\bibitem[{{Dedner} {et~al.}(2002){Dedner}, {Kemm}, {Kr{\"o}ner}, {Munz}, {Schnitzer}, \& {Wesenberg}}]{2002JCoPh.175..645D}
{Dedner}, A., {Kemm}, F., {Kr{\"o}ner}, D., {et~al.} 2002, Journal of Computational Physics, 175, 645

\bibitem[{{Di Teodoro} {et~al.}(2018){Di Teodoro}, {McClure-Griffiths}, {Lockman}, {Denbo}, {Endsley}, {Ford}, \& {Harrington}}]{2018ApJ...855...33D}
{Di Teodoro}, E.~M., {McClure-Griffiths}, N.~M., {Lockman}, F.~J., {et~al.} 2018, \apj, 855, 33

\bibitem[{{Dursi}(2007)}]{2007ApJ...670..221D}
{Dursi}, L.~J. 2007, \apj, 670, 221

\bibitem[{{Dursi} \& {Pfrommer}(2008)}]{2008ApJ...677..993D}
{Dursi}, L.~J. \& {Pfrommer}, C. 2008, \apj, 677, 993

\bibitem[{{Everett} {et~al.}(2008){Everett}, {Zweibel}, {Benjamin}, {McCammon}, {Rocks}, \& {Gallagher}}]{2008ApJ...674..258E}
{Everett}, J.~E., {Zweibel}, E.~G., {Benjamin}, R.~A., {et~al.} 2008, \apj, 674, 258

\bibitem[{{Faucher-Gigu{\`e}re} \& {Oh}(2023)}]{2023ARA&A..61..131F}
{Faucher-Gigu{\`e}re}, C.-A. \& {Oh}, S.~P. 2023, \araa, 61, 131

\bibitem[{{Ferland} {et~al.}(1998){Ferland}, {Korista}, {Verner}, {Ferguson}, {Kingdon}, \& {Verner}}]{1998PASP..110..761F}
{Ferland}, G.~J., {Korista}, K.~T., {Verner}, D.~A., {et~al.} 1998, \pasp, 110, 761

\bibitem[{{Fox} {et~al.}(2019){Fox}, {Richter}, {Ashley}, {Heckman}, {Lehner}, {Werk}, {Bordoloi}, \& {Peeples}}]{2019ApJ...884...53F}
{Fox}, A.~J., {Richter}, P., {Ashley}, T., {et~al.} 2019, \apj, 884, 53

\bibitem[{{Froese Fischer} \& {Tachiev}(2004)}]{2004ADNDT..87....1F}
{Froese Fischer}, C. \& {Tachiev}, G. 2004, Atomic Data and Nuclear Data Tables, 87, 1

\bibitem[{{Froese Fischer} {et~al.}(2006){Froese Fischer}, {Tachiev}, \& {Irimia}}]{2006ADNDT..92..607F}
{Froese Fischer}, C., {Tachiev}, G., \& {Irimia}, A. 2006, Atomic Data and Nuclear Data Tables, 92, 607

\bibitem[{{Gregori} {et~al.}(2000){Gregori}, {Miniati}, {Ryu}, \& {Jones}}]{2000ApJ...543..775G}
{Gregori}, G., {Miniati}, F., {Ryu}, D., \& {Jones}, T.~W. 2000, \apj, 543, 775

\bibitem[{{Gronke} \& {Oh}(2018)}]{2018MNRAS.480L.111G}
{Gronke}, M. \& {Oh}, S.~P. 2018, \mnras, 480, L111

\bibitem[{{Gronke} \& {Oh}(2020)}]{2020MNRAS.492.1970G}
{Gronke}, M. \& {Oh}, S.~P. 2020, \mnras, 492, 1970

\bibitem[{{Gr{\o}nnow} {et~al.}(2018){Gr{\o}nnow}, {Tepper-Garc{\'{\i}}a}, \& {Bland-Hawthorn}}]{2018ApJ...865...64G}
{Gr{\o}nnow}, A., {Tepper-Garc{\'{\i}}a}, T., \& {Bland-Hawthorn}, J. 2018, \apj, 865, 64

\bibitem[{{Haardt} \& {Madau}(2012)}]{2012ApJ...746..125H}
{Haardt}, F. \& {Madau}, P. 2012, \apj, 746, 125

\bibitem[{{Heckman} {et~al.}(2015){Heckman}, {Alexandroff}, {Borthakur}, {Overzier}, \& {Leitherer}}]{2015ApJ...809..147H}
{Heckman}, T.~M., {Alexandroff}, R.~M., {Borthakur}, S., {Overzier}, R., \& {Leitherer}, C. 2015, \apj, 809, 147

\bibitem[{{Heckman} \& {Thompson}(2017)}]{2017arXiv170109062H}
{Heckman}, T.~M. \& {Thompson}, T.~A. 2017, ArXiv e-prints [\eprint[arXiv]{1701.09062}]

\bibitem[{{Heesen} {et~al.}(2023){Heesen}, {O'Sullivan}, {Br{\"u}ggen}, {Basu}, {Beck}, {Seta}, {Carretti}, {Krause}, {Haverkorn}, {Hutschenreuter}, {Bracco}, {Stein}, {Bomans}, {Dettmar}, {Chy{\.z}y}, {Heald}, {Paladino}, \& {Horellou}}]{2023A&A...670L..23H}
{Heesen}, V., {O'Sullivan}, S.~P., {Br{\"u}ggen}, M., {et~al.} 2023, \aap, 670, L23

\bibitem[{{Heywood} {et~al.}(2019){Heywood}, {Camilo}, {Cotton}, {Yusef-Zadeh}, {Abbott}, {Adam}, {Aldera}, {Bauermeister}, {Booth}, {Botha}, {Botha}, {Brederode}, {Brits}, {Buchner}, {Burger}, {Chalmers}, {Cheetham}, {de Villiers}, {Dikgale-Mahlakoana}, {du Toit}, {Esterhuyse}, {Fanaroff}, {Foley}, {Fourie}, {Gamatham}, {Goedhart}, {Gounden}, {Hlakola}, {Hoek}, {Hokwana}, {Horn}, {Horrell}, {Hugo}, {Isaacson}, {Jonas}, {Jordaan}, {Joubert}, {J{\'o}zsa}, {Julie}, {Kapp}, {Kenyon}, {Kotz{\'e}}, {Kriel}, {Kusel}, {Lehmensiek}, {Liebenberg}, {Loots}, {Lord}, {Lunsky}, {Macfarlane}, {Magnus}, {Magozore}, {Mahgoub}, {Main}, {Malan}, {Malgas}, {Manley}, {Maree}, {Merry}, {Millenaar}, {Mnyandu}, {Moeng}, {Monama}, {Mphego}, {New}, {Ngcebetsha}, {Oozeer}, {Otto}, {Passmoor}, {Patel}, {Peens-Hough}, {Perkins}, {Ratcliffe}, {Renil}, {Rust}, {Salie}, {Schwardt}, {Serylak}, {Siebrits}, {Sirothia}, {Smirnov}, {Sofeya}, {Swart}, {Tasse}, {Taylor}, {Theron}, {Thorat}, {Tiplady}, {Tshongweni}, {van Balla}, {van der Byl},
  {van der Merwe}, {van Dyk}, {Van Rooyen}, {Van Tonder}, {Van Wyk}, {Wallace}, {Welz}, \& {Williams}}]{2019Natur.573..235H}
{Heywood}, I., {Camilo}, F., {Cotton}, W.~D., {et~al.} 2019, \nat, 573, 235

\bibitem[{{Hidalgo-Pineda} {et~al.}(2024){Hidalgo-Pineda}, {Farber}, \& {Gronke}}]{2024MNRAS.527..135H}
{Hidalgo-Pineda}, F., {Farber}, R.~J., \& {Gronke}, M. 2024, \mnras, 527, 135

\bibitem[{{Hopkins} {et~al.}(2012){Hopkins}, {Quataert}, \& {Murray}}]{2012MNRAS.421.3522H}
{Hopkins}, P.~F., {Quataert}, E., \& {Murray}, N. 2012, \mnras, 421, 3522

\bibitem[{{Hummels} {et~al.}(2017){Hummels}, {Smith}, \& {Silvia}}]{2017ApJ...847...59H}
{Hummels}, C.~B., {Smith}, B.~D., \& {Silvia}, D.~W. 2017, \apj, 847, 59

\bibitem[{{Jansson} \& {Farrar}(2012)}]{2012ApJ...757...14J}
{Jansson}, R. \& {Farrar}, G.~R. 2012, \apj, 757, 14

\bibitem[{{Jones} {et~al.}(1996){Jones}, {Ryu}, \& {Tregillis}}]{1996ApJ...473..365J}
{Jones}, T.~W., {Ryu}, D., \& {Tregillis}, I.~L. 1996, \apj, 473, 365

\bibitem[{{Kanjilal} {et~al.}(2021){Kanjilal}, {Dutta}, \& {Sharma}}]{2021MNRAS.501.1143K}
{Kanjilal}, V., {Dutta}, A., \& {Sharma}, P. 2021, \mnras, 501, 1143

\bibitem[{{Keeney} {et~al.}(2006){Keeney}, {Danforth}, {Stocke}, {Penton}, {Shull}, \& {Sembach}}]{2006ApJ...646..951K}
{Keeney}, B.~A., {Danforth}, C.~W., {Stocke}, J.~T., {et~al.} 2006, \apj, 646, 951

\bibitem[{{Kim} \& {Ostriker}(2018)}]{2018ApJ...853..173K}
{Kim}, C.-G. \& {Ostriker}, E.~C. 2018, \apj, 853, 173

\bibitem[{{Kooij} {et~al.}(2021){Kooij}, {Gr{\o}nnow}, \& {Fraternali}}]{2021MNRAS.502.1263K}
{Kooij}, R., {Gr{\o}nnow}, A., \& {Fraternali}, F. 2021, \mnras, 502, 1263

\bibitem[{{Lehner} {et~al.}(2015){Lehner}, {Howk}, \& {Wakker}}]{2015ApJ...804...79L}
{Lehner}, N., {Howk}, J.~C., \& {Wakker}, B.~P. 2015, \apj, 804, 79

\bibitem[{{Leroy} {et~al.}(2015){Leroy}, {Walter}, {Martini}, {Roussel}, {Sandstrom}, {Ott}, {Weiss}, {Bolatto}, {Schuster}, \& {Dessauges-Zavadsky}}]{2015ApJ...814...83L}
{Leroy}, A.~K., {Walter}, F., {Martini}, P., {et~al.} 2015, \apj, 814, 83

\bibitem[{{Liu} {et~al.}(2015){Liu}, {Gao}, {Isaak}, {Daddi}, {Yang}, {Lu}, \& {van der Werf}}]{2015ApJ...810L..14L}
{Liu}, D., {Gao}, Y., {Isaak}, K., {et~al.} 2015, \apjl, 810, L14

\bibitem[{{Lopez-Rodriguez} {et~al.}(2021){Lopez-Rodriguez}, {Guerra}, {Asgari-Targhi}, \& {Schmelz}}]{2021ApJ...914...24L}
{Lopez-Rodriguez}, E., {Guerra}, J.~A., {Asgari-Targhi}, M., \& {Schmelz}, J.~T. 2021, \apj, 914, 24

\bibitem[{{Mac Low} {et~al.}(1994){Mac Low}, {McKee}, {Klein}, {Stone}, \& {Norman}}]{1994ApJ...433..757M}
{Mac Low}, M.-M., {McKee}, C.~F., {Klein}, R.~I., {Stone}, J.~M., \& {Norman}, M.~L. 1994, \apj, 433, 757

\bibitem[{{Marinacci} {et~al.}(2010){Marinacci}, {Binney}, {Fraternali}, {Nipoti}, {Ciotti}, \& {Londrillo}}]{2010MNRAS.404.1464M}
{Marinacci}, F., {Binney}, J., {Fraternali}, F., {et~al.} 2010, \mnras, 404, 1464

\bibitem[{{McClure-Griffiths} {et~al.}(2013){McClure-Griffiths}, {Green}, {Hill}, {Lockman}, {Dickey}, {Gaensler}, \& {Green}}]{2013ApJ...770L...4M}
{McClure-Griffiths}, N.~M., {Green}, J.~A., {Hill}, A.~S., {et~al.} 2013, \apjl, 770, L4

\bibitem[{{McCourt} {et~al.}(2015){McCourt}, {O'Leary}, {Madigan}, \& {Quataert}}]{2015MNRAS.449....2M}
{McCourt}, M., {O'Leary}, R.~M., {Madigan}, A.-M., \& {Quataert}, E. 2015, \mnras, 449, 2

\bibitem[{{Mignone} {et~al.}(2007){Mignone}, {Bodo}, {Massaglia}, {Matsakos}, {Tesileanu}, {Zanni}, \& {Ferrari}}]{2007ApJS..170..228M}
{Mignone}, A., {Bodo}, G., {Massaglia}, S., {et~al.} 2007, \apjs, 170, 228

\bibitem[{{Murray} {et~al.}(2011){Murray}, {M{\'e}nard}, \& {Thompson}}]{2011ApJ...735...66M}
{Murray}, N., {M{\'e}nard}, B., \& {Thompson}, T.~A. 2011, \apj, 735, 66

\bibitem[{{Noon} {et~al.}(2023){Noon}, {Krumholz}, {Di Teodoro}, {McClure-Griffiths}, {Lockman}, \& {Armillotta}}]{2023MNRAS.524.1258N}
{Noon}, K.~A., {Krumholz}, M.~R., {Di Teodoro}, E.~M., {et~al.} 2023, \mnras, 524, 1258

\bibitem[{{Oppenheimer} {et~al.}(2010){Oppenheimer}, {Dav{\'e}}, {Kere{\v{s}}}, {Fardal}, {Katz}, {Kollmeier}, \& {Weinberg}}]{2010MNRAS.406.2325O}
{Oppenheimer}, B.~D., {Dav{\'e}}, R., {Kere{\v{s}}}, D., {et~al.} 2010, \mnras, 406, 2325

\bibitem[{{Peach} {et~al.}(1988){Peach}, {Saraph}, \& {Seaton}}]{1988JPhB...21.3669P}
{Peach}, G., {Saraph}, H.~E., \& {Seaton}, M.~J. 1988, Journal of Physics B Atomic Molecular Physics, 21, 3669

\bibitem[{{Richter} {et~al.}(2017){Richter}, {Nuza}, {Fox}, {Wakker}, {Lehner}, {Ben Bekhti}, {Fechner}, {Wendt}, {Howk}, {Muzahid}, {Ganguly}, \& {Charlton}}]{2017A&A...607A..48R}
{Richter}, P., {Nuza}, S.~E., {Fox}, A.~J., {et~al.} 2017, \aap, 607, A48

\bibitem[{{Richter} {et~al.}(2001){Richter}, {Savage}, {Wakker}, {Sembach}, \& {Kalberla}}]{2001ApJ...549..281R}
{Richter}, P., {Savage}, B.~D., {Wakker}, B.~P., {Sembach}, K.~R., \& {Kalberla}, P. M.~W. 2001, \apj, 549, 281

\bibitem[{{Rigby} {et~al.}(2018){Rigby}, {Bayliss}, {Sharon}, {Gladders}, {Chisholm}, {Dahle}, {Johnson}, {Paterno-Mahler}, {Wuyts}, \& {Kelson}}]{2018AJ....155..104R}
{Rigby}, J.~R., {Bayliss}, M.~B., {Sharon}, K., {et~al.} 2018, \aj, 155, 104

\bibitem[{{Rubin} {et~al.}(2022){Rubin}, {Juarez}, {Cooksey}, {Werk}, {Prochaska}, {O'Meara}, {Burchett}, {Rickards Vaught}, {Kulkarni}, \& {Straka}}]{2022ApJ...936..171R}
{Rubin}, K. H.~R., {Juarez}, C., {Cooksey}, K.~L., {et~al.} 2022, \apj, 936, 171

\bibitem[{{Rubin} {et~al.}(2014){Rubin}, {Prochaska}, {Koo}, {Phillips}, {Martin}, \& {Winstrom}}]{2014ApJ...794..156R}
{Rubin}, K. H.~R., {Prochaska}, J.~X., {Koo}, D.~C., {et~al.} 2014, \apj, 794, 156

\bibitem[{{Rupke} {et~al.}(2005){Rupke}, {Veilleux}, \& {Sanders}}]{2005ApJS..160..115R}
{Rupke}, D.~S., {Veilleux}, S., \& {Sanders}, D.~B. 2005, \apjs, 160, 115

\bibitem[{{Scannapieco}(2017)}]{2017ApJ...837...28S}
{Scannapieco}, E. 2017, \apj, 837, 28

\bibitem[{{Scannapieco} \& {Br{\"u}ggen}(2015)}]{2015ApJ...805..158S}
{Scannapieco}, E. \& {Br{\"u}ggen}, M. 2015, \apj, 805, 158

\bibitem[{{Scannapieco} {et~al.}(2001){Scannapieco}, {Thacker}, \& {Davis}}]{2001ApJ...557..605S}
{Scannapieco}, E., {Thacker}, R.~J., \& {Davis}, M. 2001, \apj, 557, 605

\bibitem[{{Schneider} {et~al.}(2018){Schneider}, {Robertson}, \& {Thompson}}]{2018ApJ...862...56S}
{Schneider}, E.~E., {Robertson}, B.~E., \& {Thompson}, T.~A. 2018, \apj, 862, 56

\bibitem[{{Shin} {et~al.}(2008){Shin}, {Stone}, \& {Snyder}}]{2008ApJ...680..336S}
{Shin}, M.-S., {Stone}, J.~M., \& {Snyder}, G.~F. 2008, \apj, 680, 336

\bibitem[{{Shopbell} \& {Bland-Hawthorn}(1998)}]{1998ApJ...493..129S}
{Shopbell}, P.~L. \& {Bland-Hawthorn}, J. 1998, \apj, 493, 129

\bibitem[{{Sparre} {et~al.}(2020){Sparre}, {Pfrommer}, \& {Ehlert}}]{2020MNRAS.499.4261S}
{Sparre}, M., {Pfrommer}, C., \& {Ehlert}, K. 2020, \mnras, 499, 4261

\bibitem[{{Sparre} {et~al.}(2019){Sparre}, {Pfrommer}, \& {Vogelsberger}}]{2019MNRAS.482.5401S}
{Sparre}, M., {Pfrommer}, C., \& {Vogelsberger}, M. 2019, \mnras, 482, 5401

\bibitem[{{Su} {et~al.}(2010){Su}, {Slatyer}, \& {Finkbeiner}}]{2010ApJ...724.1044S}
{Su}, M., {Slatyer}, T.~R., \& {Finkbeiner}, D.~P. 2010, \apj, 724, 1044

\bibitem[{{Tchernyshyov} {et~al.}(2022){Tchernyshyov}, {Werk}, {Wilde}, {Prochaska}, {Tripp}, {Burchett}, {Bordoloi}, {Howk}, {Lehner}, {O'Meara}, {Tejos}, \& {Tumlinson}}]{2022ApJ...927..147T}
{Tchernyshyov}, K., {Werk}, J.~K., {Wilde}, M.~C., {et~al.} 2022, \apj, 927, 147

\bibitem[{{Te{\c{s}}ileanu} {et~al.}(2008){Te{\c{s}}ileanu}, {Mignone}, \& {Massaglia}}]{2008A&A...488..429T}
{Te{\c{s}}ileanu}, O., {Mignone}, A., \& {Massaglia}, S. 2008, \aap, 488, 429

\bibitem[{{Thompson} {et~al.}(2016){Thompson}, {Quataert}, {Zhang}, \& {Weinberg}}]{2016MNRAS.455.1830T}
{Thompson}, T.~A., {Quataert}, E., {Zhang}, D., \& {Weinberg}, D.~H. 2016, \mnras, 455, 1830

\bibitem[{{Tremonti} {et~al.}(2004){Tremonti}, {Heckman}, {Kauffmann}, {Brinchmann}, {Charlot}, {White}, {Seibert}, {Peng}, {Schlegel}, {Uomoto}, {Fukugita}, \& {Brinkmann}}]{2004ApJ...613..898T}
{Tremonti}, C.~A., {Heckman}, T.~M., {Kauffmann}, G., {et~al.} 2004, \apj, 613, 898

\bibitem[{{T{\"u}llmann} {et~al.}(2006){T{\"u}llmann}, {Pietsch}, {Rossa}, {Breitschwerdt}, \& {Dettmar}}]{2006A&A...448...43T}
{T{\"u}llmann}, R., {Pietsch}, W., {Rossa}, J., {Breitschwerdt}, D., \& {Dettmar}, R.~J. 2006, \aap, 448, 43

\bibitem[{{Tumlinson} {et~al.}(2017){Tumlinson}, {Peeples}, \& {Werk}}]{2017ARA&A..55..389T}
{Tumlinson}, J., {Peeples}, M.~S., \& {Werk}, J.~K. 2017, \araa, 55, 389

\bibitem[{{Turk} {et~al.}(2011){Turk}, {Smith}, {Oishi}, {Skory}, {Skillman}, {Abel}, \& {Norman}}]{2011ApJS..192....9T}
{Turk}, M.~J., {Smith}, B.~D., {Oishi}, J.~S., {et~al.} 2011, The Astrophysical Journal Supplement Series, 192, 9

\bibitem[{{van Marle} \& {Keppens}(2012)}]{2012A&A...547A...3V}
{van Marle}, A.~J. \& {Keppens}, R. 2012, \aap, 547, A3

\bibitem[{{V{\'a}zquez} \& {Leitherer}(2017)}]{2017IAUS..316..359V}
{V{\'a}zquez}, G.~A. \& {Leitherer}, C. 2017, in Formation, Evolution, and Survival of Massive Star Clusters, ed. C.~{Charbonnel} \& A.~{Nota}, Vol. 316, 359--360

\bibitem[{{Veena} {et~al.}(2023){Veena}, {Riquelme}, {Kim}, {Menten}, {Schilke}, {Sormani}, {Banda-Barrag{\'a}n}, {Wyrowski}, {Fuller}, \& {Cheema}}]{2023A&A...674L..15V}
{Veena}, V.~S., {Riquelme}, D., {Kim}, W.~J., {et~al.} 2023, \aap, 674, L15

\bibitem[{{Veilleux} {et~al.}(2005){Veilleux}, {Cecil}, \& {Bland-Hawthorn}}]{2005ARA&A..43..769V}
{Veilleux}, S., {Cecil}, G., \& {Bland-Hawthorn}, J. 2005, \araa, 43, 769

\bibitem[{{Veilleux} {et~al.}(2020){Veilleux}, {Maiolino}, {Bolatto}, \& {Aalto}}]{2020A&ARv..28....2V}
{Veilleux}, S., {Maiolino}, R., {Bolatto}, A.~D., \& {Aalto}, S. 2020, \aapr, 28, 2

\bibitem[{{Villagran} {et~al.}(2020){Villagran}, {Vel{\'a}zquez}, {G{\'o}mez}, \& {Giacani}}]{2020MNRAS.491.2855V}
{Villagran}, M.~A., {Vel{\'a}zquez}, P.~F., {G{\'o}mez}, D.~O., \& {Giacani}, E.~B. 2020, \mnras, 491, 2855

\bibitem[{{Wakker} {et~al.}(2003){Wakker}, {Savage}, {Sembach}, {Richter}, {Meade}, {Jenkins}, {Shull}, {Ake}, {Blair}, {Dixon}, {Friedman}, {Green}, {Green}, {Kruk}, {Moos}, {Murphy}, {Oegerle}, {Sahnow}, {Sonneborn}, {Wilkinson}, \& {York}}]{2003ApJS..146....1W}
{Wakker}, B.~P., {Savage}, B.~D., {Sembach}, K.~R., {et~al.} 2003, \apjs, 146, 1

\bibitem[{{Walch} {et~al.}(2015){Walch}, {Girichidis}, {Naab}, {Gatto}, {Glover}, {W{\"u}nsch}, {Klessen}, {Clark}, {Peters}, {Derigs}, \& {Baczynski}}]{2015MNRAS.454..238W}
{Walch}, S., {Girichidis}, P., {Naab}, T., {et~al.} 2015, \mnras, 454, 238

\bibitem[{{Xu} \& {Stone}(1995)}]{1995ApJ...454..172X}
{Xu}, J. \& {Stone}, J.~M. 1995, \apj, 454, 172

\bibitem[{{Yan} {et~al.}(1998){Yan}, {Tambasco}, \& {Drake}}]{1998PhRvA..57.1652Y}
{Yan}, Z.-C., {Tambasco}, M., \& {Drake}, G.~W.~F. 1998, \pra, 57, 1652

\bibitem[{{Zhang}(2018)}]{2018Galax...6..114Z}
{Zhang}, D. 2018, Galaxies, 6, 114

\bibitem[{{Zhang} \& {Thompson}(2012)}]{2012MNRAS.424.1170Z}
{Zhang}, D. \& {Thompson}, T.~A. 2012, \mnras, 424, 1170

\bibitem[{{Zheng} {et~al.}(2019){Zheng}, {Peek}, {Putman}, \& {Werk}}]{2019ApJ...871...35Z}
{Zheng}, Y., {Peek}, J.~E.~G., {Putman}, M.~E., \& {Werk}, J.~K. 2019, \apj, 871, 35

\end{thebibliography}

\appendix{}
\section{A word on convergence}
\label{AppA}
\begin{figure*}
\begin{center}
  \begin{tabular}{c c}
    \includegraphics[width=\textwidth]{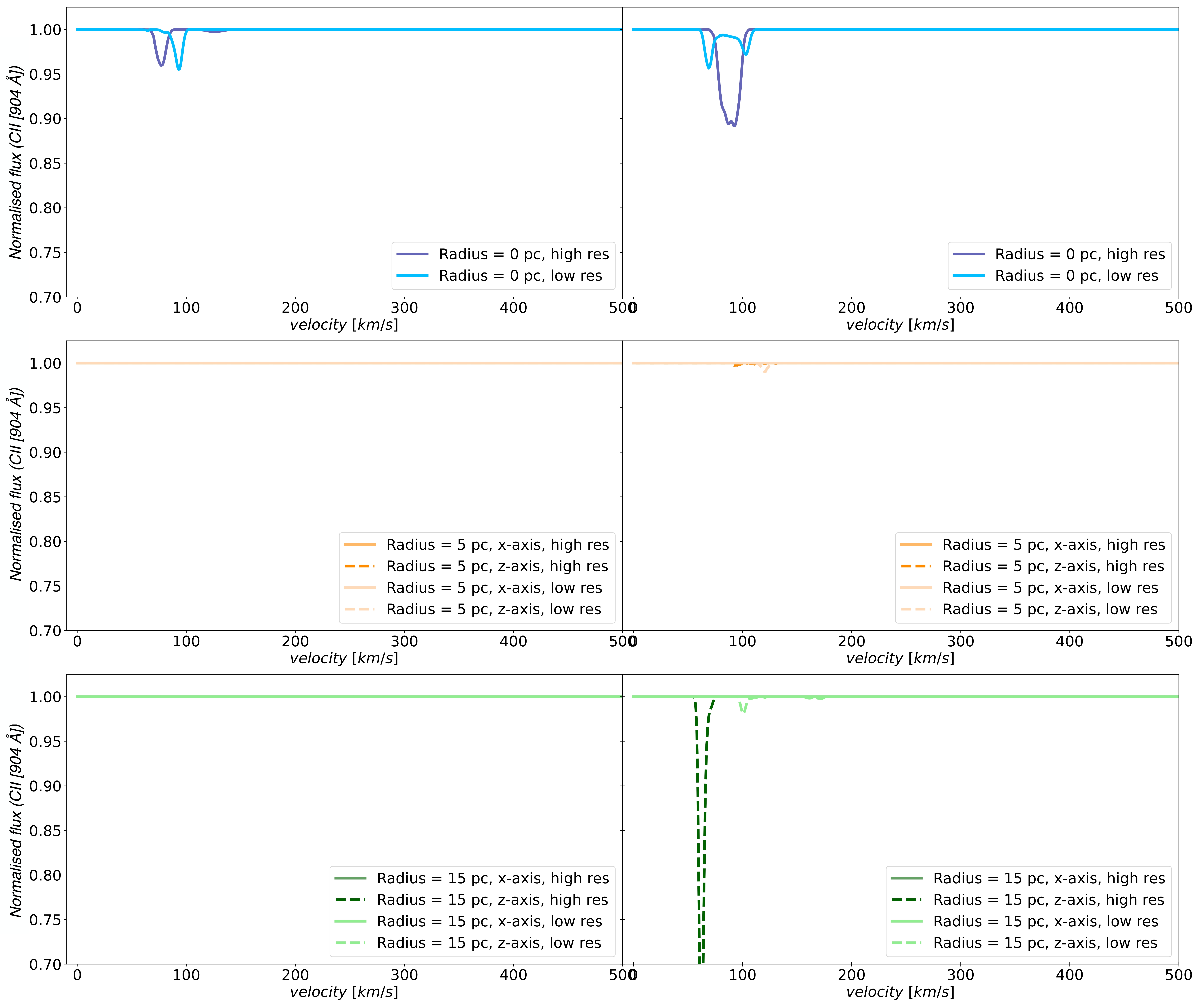}\\
  \end{tabular}
  \caption{Absorption spectra of C\,{\sc ii} generated with light rays passing through the centre (upper panels), at 5 pc from the centre (middle panels), and at 15 pc from the centre (bottom panels). In this figure, we display in the left panels the absorption profiles produced with R32-AL and R16-AL, both having a magnetic field aligned with the wind direction. On the right ones, the simulations displayed are R32-TR and R16-TR, with a transverse magnetic field. The higher resolution spectra are represented with darker colours: dark blue in the upper panels, red in the middle, and dark green in the bottom. While, the lower resolution ones are plotted in light blue (upper panels), orange (middle), and light green (bottom). Solid lines show the spectra generated with light rays passing at 5 and 15 pc along the $z$-axis, while dashed lines along the $x$-axis.} 
  \label{fig:spectra_convergCII}
\end{center}
\end{figure*}

\begin{figure*}
\begin{center}
  \begin{tabular}{c c}
    \includegraphics[width=\textwidth]{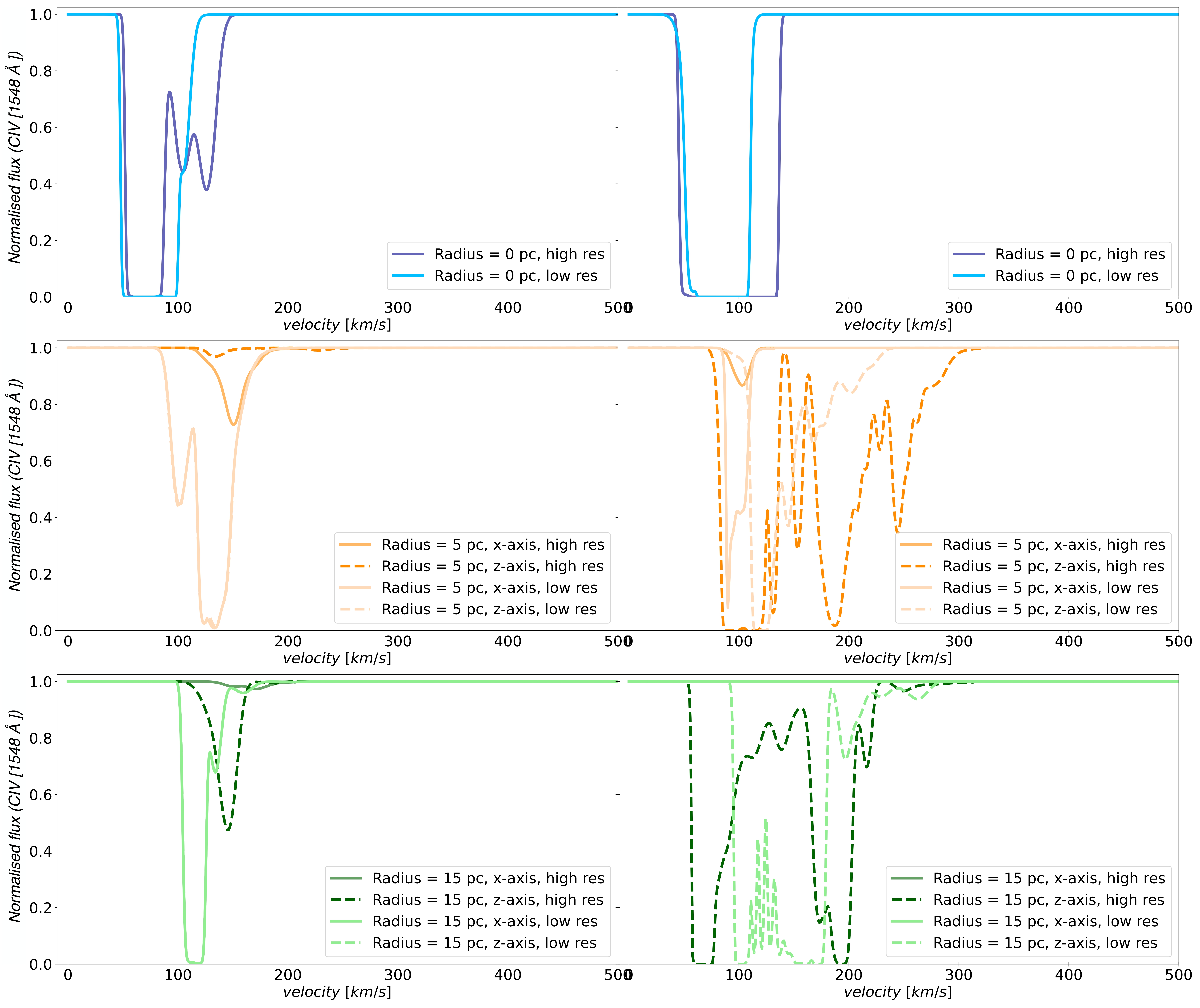}\\
  \end{tabular}
  \caption{Absorption spectra of C\,{\sc iv} generated with light rays passing through the centre (upper panels), at 5 pc from the centre (middle panels), and at 15 pc from the centre (bottom panels). In this figure, we display in the left panels the absorption profiles produced with R32-AL and R16-AL, both having a magnetic field aligned with the wind direction. On the right ones, the simulations displayed are R32-TR and R16-TR, with a transverse magnetic field. The higher resolution spectra are represented with darker colours: dark blue in the upper panels, red in the middle, and dark green in the bottom. While, the lower resolution ones are plotted in light blue (upper panels), orange (middle), and light green (bottom). Solid lines show the spectra generated with light rays passing at 5 and 15 pc along the $z$-axis, while dashed lines along the $x$-axis.} 
  \label{fig:spectra_convergCIV}
\end{center}
\end{figure*}

\begin{figure*}
\begin{center}
  \begin{tabular}{c c}
    \includegraphics[width=\textwidth]{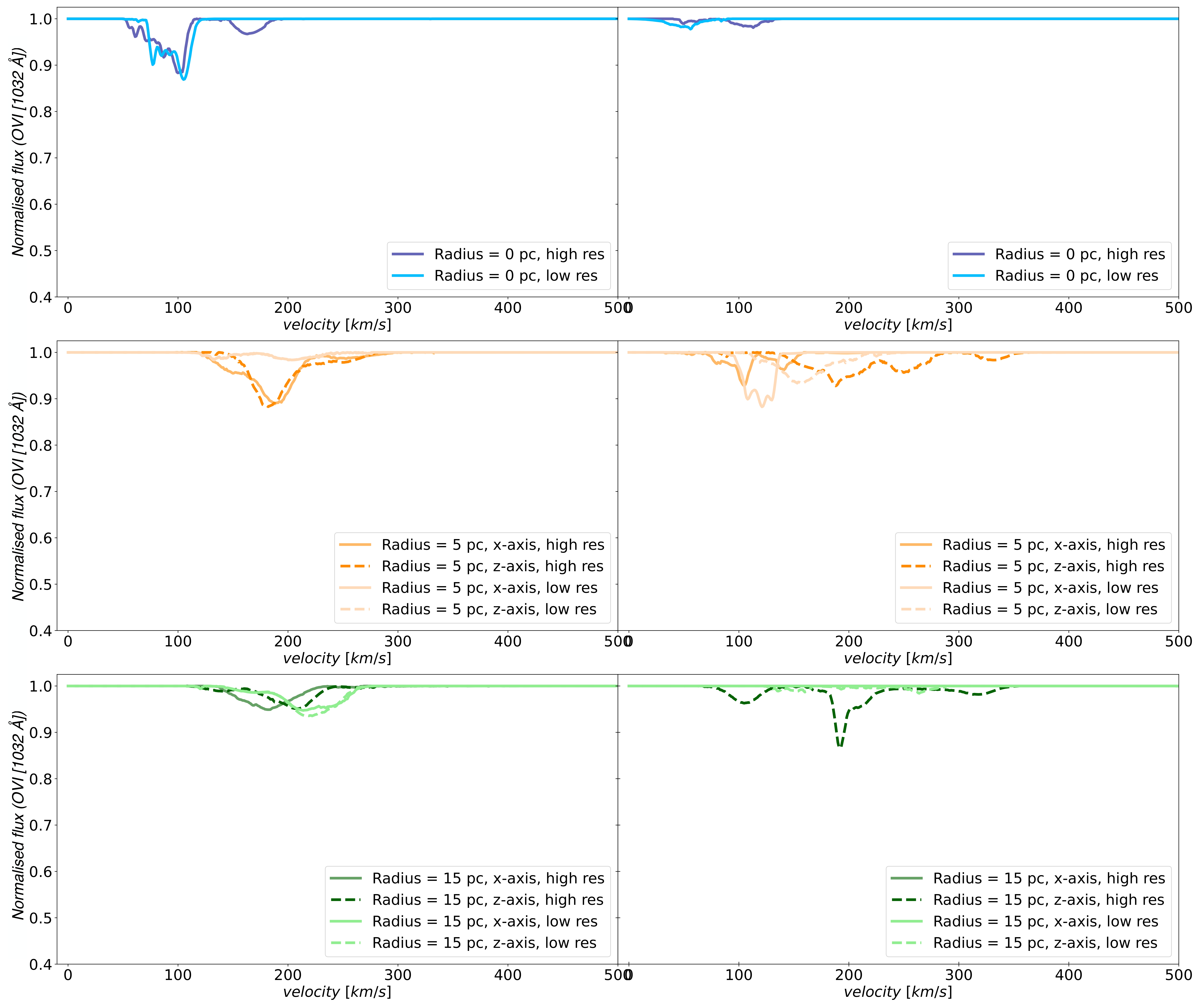}\\
  \end{tabular}
  \caption{Absorption spectra of O\,{\sc vi} generated with light rays passing through the centre (upper panels), at 5 pc from the centre (middle panels), and at 15 pc from the centre (bottom panels). In this figure, we display in the left panels the absorption profiles produced with R32-AL and R16-AL, both having a magnetic field aligned with the wind direction. On the right ones, the simulations displayed are R32-TR and R16-TR, with a transverse magnetic field. The higher resolution spectra are represented with darker colours: dark blue in the upper panels, red in the middle, and dark green in the bottom. While, the lower resolution ones are plotted in light blue (upper panels), orange (middle), and light green (bottom). Solid lines show the spectra generated with light rays passing at 5 and 15 pc along the $z$-axis, while dashed lines along the $x$-axis.} 
  \label{fig:spectra_convergOVI}
\end{center}
\end{figure*}

In Figs. \ref{fig:spectra_convergCII}, \ref{fig:spectra_convergCIV}, and \ref{fig:spectra_convergOVI} we display how using lower resolution simulations of a wind-cloud interactions (models R16-AL and R16-TR) affects the absorption spectra reported in the main body of the paper. These three figures show in order C\,{\sc ii}, C\,{\sc iv} and O\,{\sc vi} as representative of the cold, intermediate, and warm phase of the gas. The high resolution simulation is the one used for all our analysis. In these figures we compare the results with the simulations R16-AL and R16-TR, which have a computational domain consisting of a uniform grid of resolution ($N_{X_1} \times N_{X_2} \times N_{X_3} $) = ($192 \times 384 \times 192$). Overall, the higher resolution setup produces deeper, broader and more complex absorption lines.\par

By looking at individual ions, the C\,{\sc ii} absorption lines in Fig. \ref{fig:spectra_convergCII} are deeper when the resolution of the box is higher, especially in the centre of the cloud and at 15 pc. This means that higher resolution is needed in order to produce absorption features by gas with temperature below $10^{4.5}$ K, that otherwise are not even visible. C\,{\sc iv} is reported in Fig. \ref{fig:spectra_convergCIV} and it shows how in general the absorption lines are broader in the higher resolution case. However, also with the lower resolution setup, the difference among the two magnetic field orientations can be observed, especially in the middle and bottom panels. On the other hand, the middle left and bottom left panels show how the cloud is perfectly symmetric when the resolution is lower. Indeed, the absorption lines produced by a light ray passing at a distance of 5 and 15 pc from the centre, in the $z$- and $x$-axis (see Fig. \ref{fig:beam_direction}) are perfectly overlapping. Finally, Fig. \ref{fig:spectra_convergOVI} shows how O\,{\sc vi} is difficult to detect in both cases, with different resolutions.

\section{Spectral resolution analysis}
\label{AppB}
\begin{figure*}
\begin{center}
  \begin{tabular}{c c}
    \includegraphics[width=\textwidth]{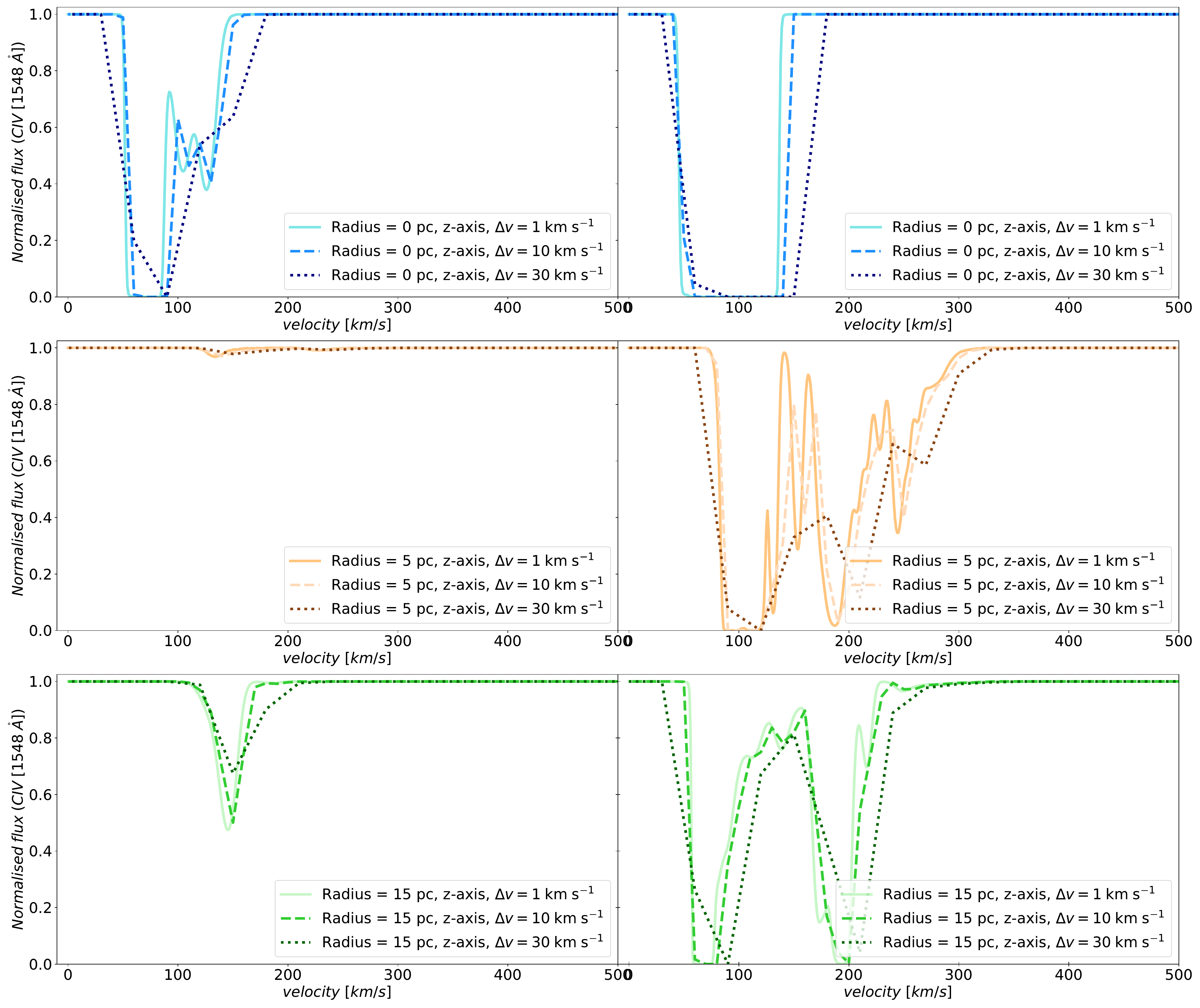}\\
  \end{tabular}
  \caption{Absorption spectra of C\,{\sc iv} generated with light rays passing through the centre (upper panels), at 5 pc from the centre (middle panels), and at 15 pc from the centre (bottom panels), all along the $z$-axis. In this figure, we display the absorption profiles produced with R32-AL on the left and R32-TR on the right. In the same panel, different lines have different spectral resolution: $1$ km s$^{-1}$ (solid), $10$ km s$^{-1}$ (dashed) and $30$ km s$^{-1}$ (dotted).} 
  \label{fig:spectra_convergCIV2}
\end{center}
\end{figure*}

In Fig. \ref{fig:spectra_convergCIV2} we show the effects of different spectroscopic resolution on the lines. To do so, we select the C\,{\sc iv} absorption line. We use this ion as it is the one affected the most by the different orientation of the initial magnetic field. The spectra are produced by varying the spectral resolution in the velocity field within TRIDENT. We observe that with a resolution of $10$ and $30$ km s$^{-1}$, the global width and depth of the lines is overall the same compared to the $1$ km s$^{-1}$ resolution case. This result proves that at lower spectral resolution, the effects of different magnetic field orientations are still visible for intermediate gas ions. The same result holds for the other ions we studied in this paper.

\section{Spectra at $t/t_{\rm cc} = 1.2$}
\label{AppC}
\begin{figure*}
\begin{center}
  \begin{tabular}{c c}
    \includegraphics[width=\textwidth]{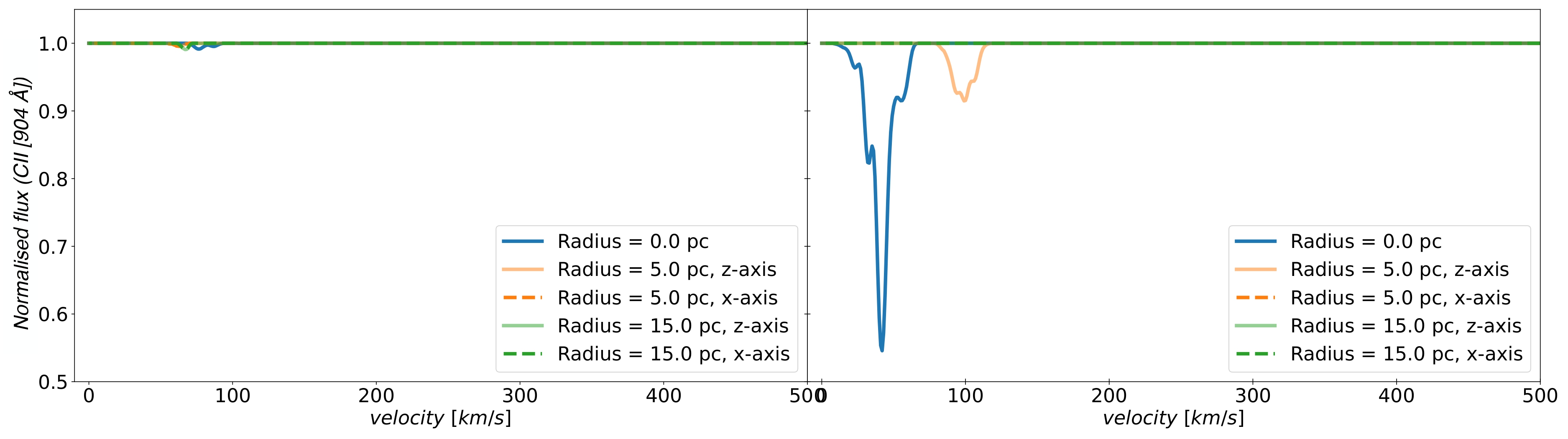}\\
    \includegraphics[width=\textwidth]{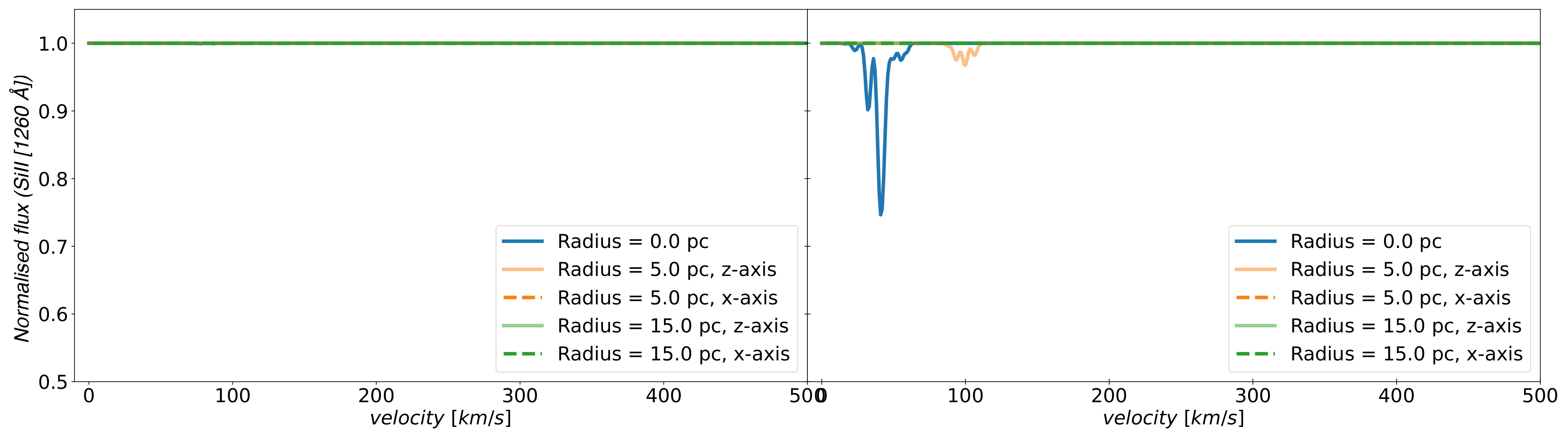}\\
  \end{tabular}
  \caption{Absorption spectra at $t/t_{\rm cc} = 1.2$ of ions tracing the cold phase of the gas generated with light rays passing through the centre (in blue), 5 pc from the centre along the $x$- (dashed orange line) and $z$-axis (solid orange line), and 15 pc from the centre along the $x$- (dashed green line) and $z$-axis (solid green line). In this figure, we display in the upper left panel the absorption profiles of the C\,{\sc ii} in the simulation R32-AL, in the upper right panel the C\,{\sc ii} in R32-TR, while the lower panels show the lines of the Si\,{\sc ii} in the simulation R32-AL (left) and R32-TR (right).} 
  \label{fig:spectra_coldgas2}
\end{center}
\end{figure*}

\begin{figure*}
\begin{center}
  \begin{tabular}{c c}
    \includegraphics[width=\textwidth]{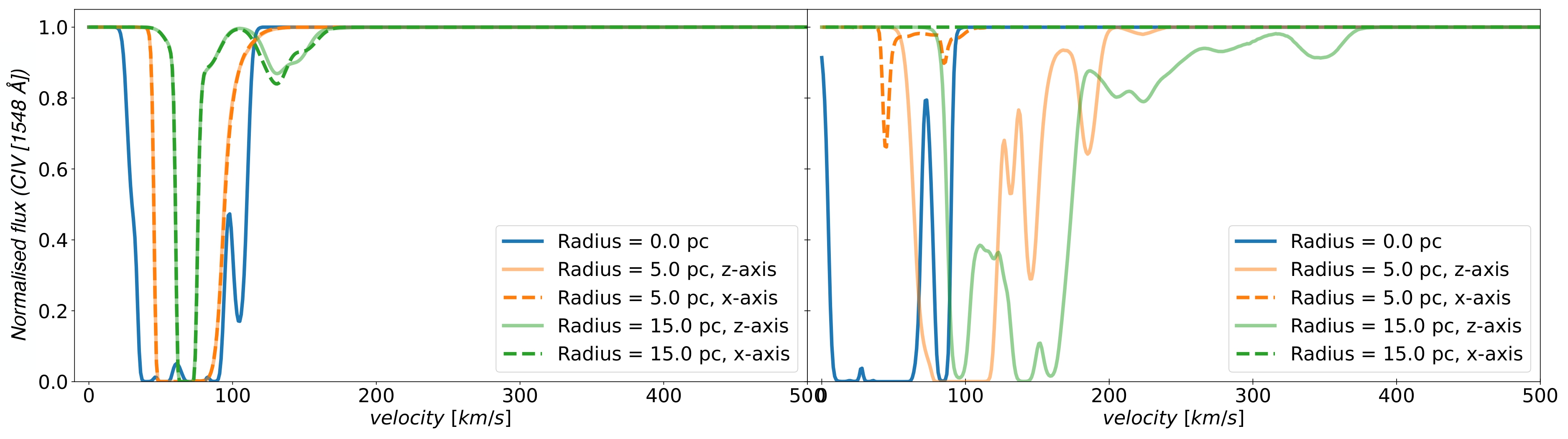}\\
    \includegraphics[width=\textwidth]{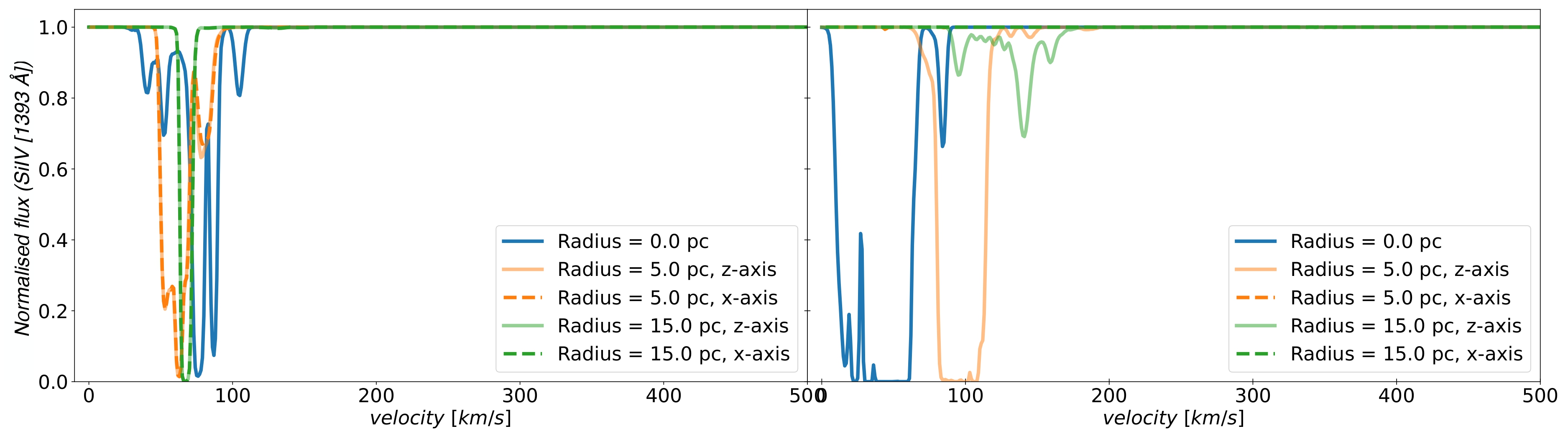}\\
  \end{tabular}
  \caption{Absorption spectra at $t/t_{\rm cc} = 1.2$ of ions tracing the intermediate-temperature gas, generated with light rays passing through the centre (in blue), 5 pc from the centre along the $x$- (dashed orange line) and $z$-axis (solid orange line), and 15 pc from the centre along the $x$- (dashed green line) and $z$-axis (solid green line). In this figure, we display in the upper left panel the absorption profiles of the C\,{\sc iv} in the simulation R32-AL, in the upper right panel the C\,{\sc iv} in R32-TR, while the lower panels show the lines of the Si\,{\sc iv} in the simulation R32-AL (left) and R32-TR (right).} 
  \label{fig:spectra_intergas2}
\end{center}
\end{figure*}

\begin{figure*}
\begin{center}
  \begin{tabular}{c c}
    \includegraphics[width=\textwidth]{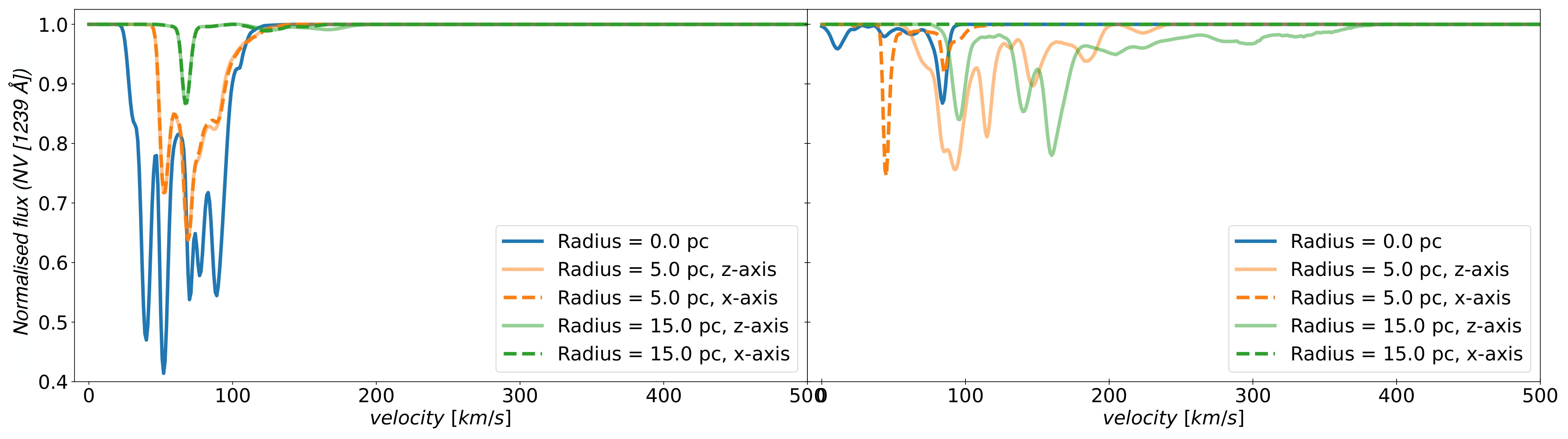}\\
    \includegraphics[width=\textwidth]{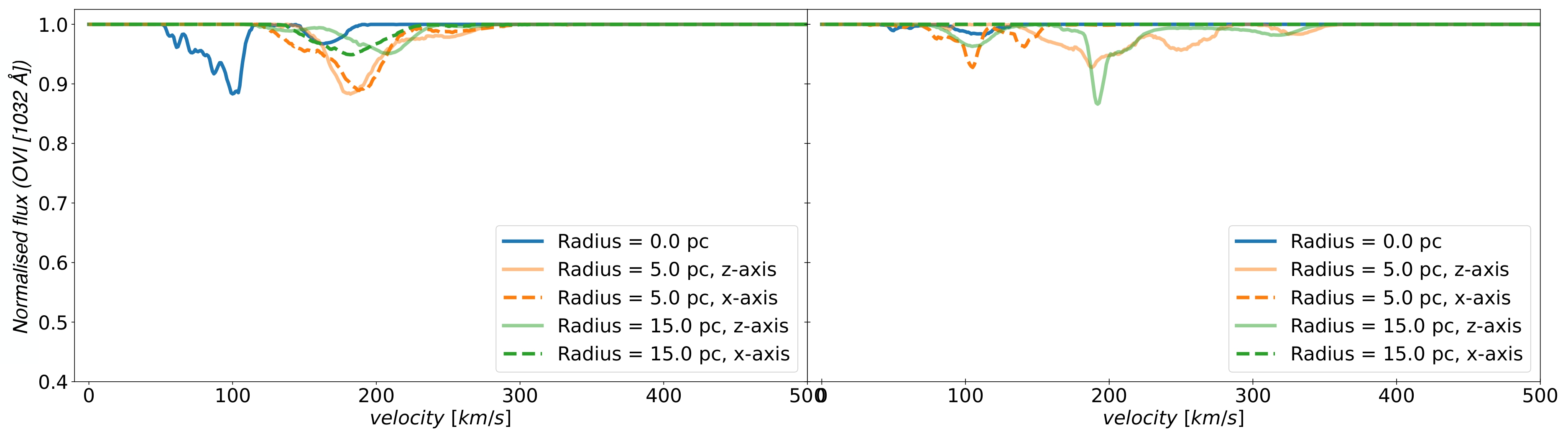}\\
  \end{tabular}
  \caption{Absorption spectra at $t/t_{\rm cc} = 1.2$ of ions tracing the warm phase of the gas generated with light rays passing through the centre (in blue), at 5 pc from the centre along the $x$- (dashed orange line) and $z$-axis (solid orange line), at 15 pc from the centre along the $x$- (dashed green line) and $z$-axis (solid green line). In this figure, we display in the upper left panel the absorption profiles of the N\,{\sc v} in the simulation R32-AL, in the upper right panel the N\,{\sc v} in R32-TR. while the lower panels show the lines of the O\,{\sc vi} in the simulation R32-AL (left) and R32-TR (right).} 
  \label{fig:spectra_warmgas2}
\end{center}
\end{figure*}

In Figs. \ref{fig:spectra_coldgas2}, \ref{fig:spectra_intergas2}, and \ref{fig:spectra_warmgas2} we display spectra for low, intermediate, and high ions at time $t/t_{\rm cc} = 1.2$. These figures are included for comparison with the spectra presented in the main body of the paper. They show that the results presented throughout the manuscript for the different ions hold for an earlier time.

\end{document}